\documentclass[useAMS,fleqn,usenatbib]{mnras}
\usepackage{hyperref}

\usepackage{graphicx}	
\usepackage{amsmath}	
\usepackage{amssymb}	

\usepackage{lscape}
\usepackage{color}
\usepackage{comment}
\usepackage{natbib, aas_macros}
\usepackage{bm}
\usepackage[dvipsnames,svgnames,x11names]{xcolor}
\newcommand{\orcid}[1]{\href{https://orcid.org/#1}{\includegraphics[width=12pt]{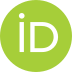}}}

\title[Active gas features revealed by joint MUSTANG-2 and XXL analysis]{
Active gas features in three HSC-SSP CAMIRA clusters revealed by high angular resolution analysis of MUSTANG-2 SZE and XXL X-ray observations\thanks{Based on data collected at Subaru Telescope, which is operated by the National Astronomical Observatory of Japan. Based on observations obtained with XMM-Newton, an ESA science mission with instruments and contributions directly funded by ESA Member States and NASA}}
\author[N. Okabe et al.]
{Nobuhiro Okabe$^{1,2,3,4}$\thanks{E-mail:okabe@hiroshima-u.ac.jp} \orcid{0000-0003-2898-0728},
Simon Dicker$^{5}$ \orcid{0000-0002-1940-4289},
Dominique Eckert$^{6}$,
Tony Mroczkowski$^{7}$ \orcid{0000-0003-3816-5372},
\newauthor
Fabio Gastaldello$^{8}$ \orcid{0000-0002-9112-0184},
Yen-Ting Lin$^{9}$,
Mark Devlin$^{5}$,
Charles E. Romero$^{5}$ \orcid{0000-0001-5725-0359},
\newauthor
Mark Birkinshaw$^{10}$,
Craig Sarazin$^{11}$, 
Cathy Horellou$^{12}$ \orcid{0000-0002-3533-8584},
Tetsu Kitayama$^{13}$ \orcid{0000-0002-9486-0356},
\newauthor
Keiichi Umetsu$^{9}$,
Mauro Sereno$^{14,15}$ \orcid{0000-0003-0302-0325},
Brian S.\ Mason$^{16}$ \orcid{0000-0002-8472-836X},
John A. ZuHone$^{17}$  \orcid{0000-0003-3175-2347},
\newauthor
Ayaka Honda$^{1}$,
Hiroki Akamatsu$^{18}$,
I-Non Chiu$^{9}$,
Kotaro Kohno$^{19,20}$,
\newauthor
Kai-Yang Lin$^{9}$,
Elinor Medezinski$^{21}$,
Satoshi Miyazaki$^{22}$,
Ikuyuki Mitsuishi$^{23}$,
\newauthor
Atsushi J. Nishizawa$^{24}$,
Masamune Oguri$^{25,19,26}$,
Naomi Ota$^{27}$,
Florian Pacaud$^{28}$,
\newauthor
Marguerite Pierre$^{29}$,
Jonathan Sievers$^{30}$,
Vernesa Smol\v{c}i\'c$^{31}$,
Sara Stanchfield$^{5}$, 
\newauthor
Keigo Tanaka$^{32}$,
Ryoichi Yamamoto$^{1}$,
Chong Yang$^{1}$, 
and
Atsushi Yoshida$^{23}$
\\
\\
$^{1}$Department of Physics, Hiroshima University, 1-3-1 Kagamiyama, Higashi-Hiroshima, Hiroshima 739-8526, Japan\\
$^{2}$Hiroshima Astrophysical Science Center, Hiroshima University, 1-3-1 Kagamiyama, Higashi-Hiroshima, Hiroshima 739-8526, Japan\\
$^{3}$Core Research for Energetic Universe, Hiroshima University, 1-3-1, Kagamiyama, Higashi-Hiroshima, Hiroshima 739-8526, Japan\\
$^{4}$Physics Program, Graduate School of Advanced Science and Engineering, Hiroshima University, 1-3-1 Kagamiyama, Higashi-Hiroshima, Hiroshima 739-8526, Japan\\
$^{5}$Department of Physics and Astronomy, University of Pennsylvania, 209 South 33rd Street, Philadelphia, PA, 19104, USA\\
$^{6}$Department of Astronomy, University of Geneva, ch. d’Ecogia 16,
1290 Versoix, Switzerland\\
$^{7}$ESO - European Southern Observatory, Karl-Schwarzschild-Str.\ 2, D-85748 Garching b.\ M\"unchen, Germany\\
$^{8}$INAF - IASF Milano, via Bassini 15, I-20133 Milano, Italy\\
$^{9}$Academia Sinica Institute of Astronomy and Astrophysics (ASIAA), No. 1, Section 4, Roosevelt Road, Taipei 10617, Taiwan\\
$^{10}$H.H. Wills Physics Laboratory, University of Bristol, Tyndall
Avenue, Bristol, BS8 1TL, UK\\
$^{11}$ Department of Astronomy, University of Virginia, 530 McCormick Road, Charlottesville, VA 22904-4325, USA \\
$^{12}$Chalmers University of Technology, Department of Space, Earth and Environment, Onsala Space Observatory, 439 92 Onsala,Sweden\\
$^{13}$Department of Physics, Toho University, 2-2-1 Miyama, Funabashi, Chiba 274-8510, Japan\\
$^{14}$INAF - Osservatorio di Astrofisica e Scienza dello Spazio di Bologna,
via Piero Gobetti 93/3, I-40129 Bologna, Italy\\
$^{15}$INFN, Sezione di Bologna, viale Berti Pichat 6/2, 40127 Bologna, Italy\\
$^{16}$National Radio Astronomy Observatory, 520 Edgemont Rd., Charlottesville VA 22903, USA\\
$^{17}$Harvard-Smithsonian Center for Astrophysics, 60 Garden St., Cambridge, MA 02138, USA\\
$^{18}$SRON Netherlands Institute for Space Research, Sorbonnelaan 2, 3584 CA Utrecht, The Netherlands\\
$^{19}$Institute of Astronomy, The University of Tokyo, 2-21-1 Osawa, Mitaka, Tokyo 181-0015, Japan\\
$^{20}$Research Center for the Early Universe, School of Science, The
University of Tokyo, 7-3-1 Hongo, Bunkyo, Tokyo 113-0033, Japan\\
$^{21}$Department of Astrophysical Sciences, Princeton University, Princeton, NJ 08544, USA\\
$^{22}$National Astronomical Observatory of Japan, Osawa 2-21-1, Mitaka,
Tokyo 181-8588, Japan\\
$^{23}$Department of Physics, Nagoya University, Aichi 464-8602, Japan\\
$^{24}$Institute for Advanced Research, Nagoya University Furocho,
Chikusa-ku, Nagoya, 464-8602 Japan\\
$^{25}$Department of Physics, University of Tokyo, Tokyo 113-0033, Japan\\
$^{26}$Kavli Institute for the Physics and Mathematics of the Universe (Kavli IPMU, WPI), University
of Tokyo, Chiba 277-8582, Japan\\
$^{27}$Department of Physics, Nara Women's University,
Kitauoyanishi-machi, Nara, Nara 630-8506, Japan\\
$^{28}$Argelander-Institut f\"ur Astronomie, University of Bonn, Auf dem H\"ugel 71, 53121 Bonn, Germany\\
$^{29}$AIM, CEA, CNRS, Universit\'e Paris-Saclay, Universit\'e Paris Diderot, Sorbonne Paris Cit\'e, F-91191 Gif-sur-Yvette, France \\
$^{30}$Department of Physics, McGill University, 3600 University Street
Montreal, QC H3A 2T8, Canada\\
$^{31}$Department of Physics, Faculty of Science, University of Zagreb, Bijenicka cesta 32, 10002 Zagreb, Croatia\\
$^{32}$Graduate School of Natural Science \& Technology, Kanazawa University, Kakuma-machi, Kanazawa, Ishikawa 920-1192, Japan\\
\\\\\\\\\\\\\\\\\\\\\\\\\\\\\\\\\\\\\\\\\\\\\\\\\\\\\\\\\\\\\\
}

\begin{document}

\date{\today}
\label{firstpage}
\pagerange{\pageref{firstpage}--\pageref{lastpage}} \pubyear{2020}

\maketitle

\newcommand{\simgt}{\lower.5ex\hbox{$\; \buildrel > \over \sim \;$}}
\newcommand{\simlt}{\lower.5ex\hbox{$\; \buildrel < \over \sim \;$}}
\newcommand{\tp}{\hspace{-1mm}+\hspace{-1mm}}
\newcommand{\tm}{\hspace{-1mm}-\hspace{-1mm}}
\newcommand{\gIII}{I\hspace{-.3mm}I\hspace{-.3mm}I}
\newcommand{\bmf}[1]{\mbox{\boldmath$#1$}}
\def\bbeta{\mbox{\boldmath $\beta$}}
\def\btheta{\mbox{\boldmath $\theta$}}
\def\bnabla{\mbox{\boldmath $\nabla$}}
\def\bk{\mbox{\boldmath $k$}}
\newcommand{\cA}{{\cal A}}
\newcommand{\cD}{{\cal D}}
\newcommand{\cF}{{\cal F}}
\newcommand{\cG}{{\cal G}}
\newcommand{\trQ}{{\rm tr}Q}
\newcommand{\Real}[1]{{\rm Re}\left[ #1 \right]}
\newcommand{\paren}[1]{\left( #1 \right)}
\newcommand{\red}{\textcolor{red}}

\newcommand{\blue}{\textcolor{blue}}
\newcommand{\norv}[1]{{\textcolor{blue}{#1}}}
\newcommand{\norvnew}[1]{{\textcolor{red}{#1}}}
\newcommand{\norvgreen}[1]{{\textcolor{green}{#1}}}
\newcommand{\commentc}[1]{{\bf \textcolor{purple}{CR: }\textcolor{DarkGreen}{#1}}}
\newcommand{\comments}[1]{{\bf \textcolor{DarkGreen}{SD: #1}}}
\newcommand{\commentt}[1]{{\bf \textcolor{Fuchsia}{TM: #1}}}
\newcommand{\change}[1]{{\bf \textcolor{BlueViolet}{#1}}}

\def\h70kpc{\mathrel{h_{70}^{-1}{\rm kpc}}}
\def\hkpc{\mathrel{h^{-1}{\rm kpc}}}
\def\hMpc{\mathrel{h^{-1}{\rm Mpc}}}
\def\Mvir{\mathrel{M_{\rm vir}}}
\def\cvir{\mathrel{c_{\rm vir}}}
\def\rvir{\mathrel{r_{\rm vir}}}
\def\Dvir{\mathrel{\Delta_{\rm vir}}}
\def\rsc{\mathrel{r_{\rm sc}}}
\def\rhoc{\mathrel{\rho_{\rm crit}}}
\def\Msol{\mathrel{M_\odot}}
\def\hMsol{\mathrel{h^{-1}M_\odot}}
\def\h70Msol{\mathrel{h_{70}^{-1}M_\odot}}

   \defcitealias{2016A&A...592A...1P}{XXL~Paper~I}
   \defcitealias{2016A&A...592A...2P}{XXL~Paper~II}
   \defcitealias{2016A&A...592A...3G}{XXL~Paper~III}
   \defcitealias{2016A&A...592A...4L}{XXL~Paper~IV}
   \defcitealias{2014ApJ...794..157M}{XXL~Paper~V}
   \defcitealias{2016A&A...592A...5F}{XXL~Paper~VI}
   \defcitealias{2016A&A...592A...6P}{XXL~Paper~VII}
   \defcitealias{2016A&A...592A...7A}{XXL~Paper~VIII}
   \defcitealias{2016A&A...592A...8B}{XXL~Paper~IX}
   \defcitealias{2016A&A...592A...9Z}{XXL~Paper~X}
   \defcitealias{2016A&A...592A..10S}{XXL~Paper~XI}
   \defcitealias{2016A&A...592A..11K}{XXL~Paper~XII}
   \defcitealias{2016A&A...592A..12E}{XXL~Paper~XIII}
   \defcitealias{2016PASA...33....1L}{XXL~Paper~XIV}
   \defcitealias{2016MNRAS.462.4141L}{XXL~Paper~XV}
   \defcitealias{2018A&A...620A...1M}{XXL~Paper~XVI}
   \defcitealias{2018A&A...620A...2M}{XXL~Paper~XVII}
   \defcitealias{2018A&A...620A...3B}{XXL~Paper~XVIII}
   \defcitealias{2018A&A...620A...4K}{XXL~Paper~XIX}
   \defcitealias{2018A&A...620A...5A}{XXL~Paper~XX}
   \defcitealias{2018A&A...620A...6M}{XXL~Paper~XXI}
   \defcitealias{2018A&A...620A...7G}{XXL~Paper~XXII}
   \defcitealias{2018A&A...620A...8F}{XXL~Paper~XXIII}
   \defcitealias{2018A&A...620A...9F}{XXL~Paper~XXIV}
   \defcitealias{2018A&A...620A..10P}{XXL~Paper~XXV}
   \defcitealias{2018A&A...620A..11C}{XXL~Paper~XXVI}
   \defcitealias{2018A&A...620A..12C}{XXL~Paper~XXVII}
   \defcitealias{2018A&A...620A..13R}{XXL~Paper~XXVIII}
   \defcitealias{2018A&A...620A..14S}{XXL~Paper~XXIX}
   \defcitealias{2018A&A...620A..15G}{XXL~Paper~XXX}
   \defcitealias{2018A&A...620A..16B}{XXL~Paper~XXXI}
   \defcitealias{2018A&A...620A..17P}{XXL~Paper~XXXII}
   \defcitealias{2018A&A...620A..18L}{XXL~Paper~XXXIII}
   \defcitealias{2018A&A...620A..19H}{XXL~Paper~XXXIV}
   \defcitealias{2018A&A...620A..20K}{XXL~Paper~XXXV}

\vspace{10em}
  \begin{abstract}
We present results from simultaneous modeling of high angular resolution GBT/MUSTANG-2 90 GHz Sunyaev-Zel’dovich effect (SZE) measurements and XMM-XXL X-ray images of three rich galaxy clusters selected from the HSC-SSP Survey. The combination of high angular resolution SZE and X-ray imaging enables a spatially resolved multi-component analysis, which is crucial to understand complex distributions of cluster gas properties. The targeted clusters have similar optical richnesses and redshifts, but exhibit different dynamical states in their member galaxy distributions: a single-peaked cluster, a double-peaked cluster, and a cluster belonging to a supercluster. A large-scale residual pattern in both regular Compton-parameter $y$ and X-ray surface brightness distributions is found in the single-peaked cluster, indicating a sloshing mode. The double-peaked cluster shows an X-ray remnant cool core between two SZE peaks associated with galaxy concentrations. The temperatures of the two peaks reach $\sim20-30$ keV in contrast to the cool core component of $\sim2$ keV, indicating a violent merger. The main SZE signal for the supercluster is elongated along a direction perpendicular to the major axis of the X-ray core, suggesting a minor merger before core passage. The $S_X$ and $y$ distributions are thus perturbed at some level, regardless of the optical properties. We find that the integrated Compton $y$ parameter and the temperature for the major merger are boosted from those expected by the weak-lensing mass and those for the other two clusters show no significant deviations, which is consistent with predictions of numerical simulations. 
\end{abstract}


\section{Introduction}

Galaxy clusters, whose compositions are dominated by dark matter, ionised gas and galaxies, 
are the largest gravitationally bound objects in the
Universe and sometimes aggregate in superclusters.
The abundance of galaxy clusters is sensitive to the growth
of matter density perturbations, and thus serves as a cosmological probe.
Thanks to recent technical progress,
galaxy clusters can be discovered by various observational methods:
optical \citep[e.g.][]{Rykoff14,Oguri14b,Rozo16,Oguri18}, X-ray
\citep[e.g.][]{Bohringer04,Piffaretti11,2018A&A...620A...5A}, thermal Sunyaev-Zel'dovich
effect (SZE) \citep[e.g.][]{PlanckSZ14,Bleem15,Sifon16} and weak-lensing
mass reconstruction \citep[e.g.][]{Miyazaki07,Miyazaki18}.
Optical techniques are unbiased against cluster mergers which non-linearly
change properties of the the intracluster medium (ICM),
but suffer from projection effects along the line of sight \citep{Okabe19}.
As X-ray emission from the ICM is proportional
to the square of the electron number density, projection effects are less important but this 
technique suffers from a cool core bias \citep{Eckert11,Rossetti17}. 
The surface brightness of the thermal SZE is proportional to the line of sight integral of the ICM electron pressure, and is independent of redshift \citep[see e.g.][]{Birkinshaw99,Mroczkowski19}.
When using weak-lensing shear to select clusters \citep{Miyazaki18,2020ApJ...891..139C}, the resulting sample does not rely on any baryonic physics, but may
potentially suffer from projection bias in the lensing signals.
While complementary, the redshift dependence and the tracer used (ICM, galaxies, or total mass) in different techniques can introduce different biases for each method.
It is therefore important for the upcoming era of cluster cosmological
studies to understand the selection function that arises in the construction of 
cluster catalogues from the true cluster population.
In particular, it is essential to understand the baryonic physics as
a function of dynamical state and the interplay between dark matter and baryons.

The Hyper Suprime-Cam Subaru Strategic Program
\citep[HSC-SSP;][]{HSC1stDR,HSC1styrOverview,Miyazaki18HSC,Komiyama18HSC,Kawanomoto18HSC,Furusawa18HSC,Bosch18HSC,Haung18HSC,Coupon18HSC,HSC2ndDR}
is an on-going wide-field optical imaging survey composed of three layers of different depths (Wide, Deep and UltraDeep). 
The Wide layer is designed to obtain five-band ($grizy$) imaging over $1400$~deg$^2$.
The HSC-SSP Survey achieves both excellent imaging quality ($\sim$0.7 arcsec
seeing in $i$-band) and deep observations ($r\simlt26$~AB~mag).
Therefore, the HSC survey currently has the best performance to search simultaneously for
galaxy clusters and to measure their weak-lensing masses \citep[for review;][]{Pratt19}.
\citet{Oguri18} constructed a cluster catalogue using the {\it Cluster finding Algorithm based
on Multi-band Identification of Red-sequence gAlaxies} \citep[CAMIRA;][]{Oguri14b}, which is a
red-sequence cluster finder that exploits stellar population synthesis model fitting.
The catalogue contains $\sim 1900 $ clusters at $0.1<z<1.1$ with richness
larger than $N=15$ in the $\sim 240$~deg$^2$ HSC-SSP S16A field.
The accuracy of photometric redshifts of the clusters is $\sigma_z/(1+z)\sim 0.01$ for the whole redshift range. Compared to shallower data from the Sloan Digital Sky Survey \citep[SDSS;][]{Rykoff14,Oguri14b} and the Dark Energy Survey \citep[DES;][]{Rykoff16}, many clusters beyond $z\sim0.8$ were discovered for the first time \citep{Oguri18}.
\citet{Okabe19} found $\sim 190$ major-merger candidates using a peak-finding method
of galaxy maps of the CAMIRA clusters and confirmed that the
mass ratio of the sub and main halo is higher than $0.1$ based on stacked weak-lensing analysis.
Our statistical approach uncovers
merger boosts in stacked {\it ROSAT} $L_X$ and {\it Planck} SZE scaling
relations for the CAMIRA clusters and equatorial-shock-heated gas in
cluster outskirts \citep{Ricker01,ZuHone11,Ha18} in both stacked X-ray and SZE images.
However, using a stacked analysis makes it difficult to discriminate
between the dynamical states of individual clusters, such as pre- and post- mergers.
In principal, the optically-selected CAMIRA clusters cover various dynamical states and stages
(relaxed, minor merger, major merger, pre-merger, and post-merger), 
and thus systematic multi-wavelength follow-up studies of individual clusters
are critically important to understand the relationship between gas
properties and dynamical states in more details.

In this paper, we carry out joint SZE and X-ray studies of three CAMIRA
clusters exhibiting different dynamical states to derive gas
distributions, and compare the gas properties with optical properties and weak-lensing masses.
The SZE data were taken using MUSTANG-2
\citep{Dicker14,Romero20} installed on the 100-meter Green Bank Telescope (GBT). 
MUSTANG-2 has an angular resolution of 9\arcsec\ full-width
half-maximum (FWHM) at 90 GHz and an instantaneous field of view of 4.25\arcmin, well matched to our resolution requirement and the angular size of our clusters.  
We use X-ray images from the XXL Survey \citep{2016A&A...592A...1P,2016A&A...592A...2P,2016A&A...592A...3G,2016A&A...592A...4L,2016A&A...592A...6P,2018A&A...620A...5A,2018A&A...620A...7G} that is the
largest observing program undertaken by {\it XMM-Newton}.
The XXL Survey covers two distinct sky areas for a total of 50 square degrees down to a sensitivity of
$6\times10^{-15}\,{\rm erg}\,{\rm cm}^{-2}\,{\rm s}^{-1}$ for point-like sources ([0.5-2] keV band).
The XXL survey provides us with the unique, complementary X-ray data set for the joint analysis. 
We use the HSC-SSP Survey data for optical and weak-lensing analyses.

This paper is organised as follows.
Sec. \ref{sec:target} describes our target properties.
Sec. \ref{sec:data} presents our observations, a method of joint SZE and
X-ray analysis, and our weak-lensing analysis.
Sec. \ref{sec:result} is devoted to the results and discussion, respectively.
We summarise our results in Sec. \ref{sec:con}.
Throughout this paper we use $\Omega_{m,0}=0.3$, $\Omega_{\Lambda,0}=0.7$ and $H_0=70h_{70}$\,km\,s$^{-1}$\,Mpc$^{-1}$.

\section{Targets} \label{sec:target}

We selected three clusters (Table \ref{tab:targets})
at redshifts of $z\sim0.4$ from the sample of the HSC-SSP CAMIRA clusters
\citep{Oguri18} to observe with MUSTANG-2.
As described in Sec~\ref{subsec:gbt}, 
recovery of faint signals on angular scales larger than an instrument's 
instantaneous field-of-view (FOV), can be problematic. 
Clusters with angular sizes 
comparable to MUSTANG-2's $\approx 4.25$ arcmin FOV are at medium ($z\sim0.4$) to high ($z>1.0$)
redshifts, making them well-suited for MUSTANG-2 follow-up.
However X-ray observations suffer from strong cosmological dimming and so for 
joint MUSTANG-2/X-ray analysis, the choice of $z\approx 0.4$ is close to optimal.
At these redshifts, the $9$ arcsec FWHM resolution of MUSTANG-2
enables us to resolve the pressure distribution with physical resolution of $\sim60$ kpc.
The point spread function (PSF) of {\it XMM-Newton} is
comparable to the angular resolution of MUSTANG-2,
and a joint analysis of MUSTANG-2 SZE and XXL X-ray observation enables measurements
of the two dimensional distributions of electron number density,
temperature, pressure, and entropy parameter, all with reasonably high angular resolution.

As pointed out by \citet{Okabe19}, optically-selected clusters are free from
bias against the ICM merger boost because the number of luminous
red galaxies is essentially conserved during cluster mergers, but X-ray and
SZE observables are affected by cluster merger phenomena.
This is simply caused by the collisionless nature of member galaxies and
collisional particles of the ICM.
Thus, the sample of optical clusters, composed of a wide
range of various dynamical states, is a very well-suited sample to investigate dynamical dependence of gas properties.

As our first observation, we selected three representative clusters of
different galaxy distributions (Table \ref{tab:targets}) from the
CAMIRA catalogue \citep{Oguri18} based on galaxy distributions \citep{Okabe19}. 
We first constructed Gaussian smoothed maps (${\rm FWHM} = 200 h_{70}^{-1}$ kpc) of number densities of red galaxies selected in the colour-magnitude plane.
We then identified peaks above a redshift-dependent threshold considering the contamination of extended galaxy distributions from nearby peaks caused by the smoothing procedure.
The threshold was empirically determined to be an average peak stacked over the CAMIRA clusters at each redshift slice. The multi-peaked clusters are likely to be major-merger candidates by stacked weak-lensing analysis.
The method cannot resolve substructures within the smoothing scale, less massive subhalos.
We cannot discriminate between pre- and post- mergers due to the collisionless feature of galaxies.

\begin{figure*}
 \begin{center}
 \includegraphics[trim=0 4mm 0 10mm,clip,width=\hsize]{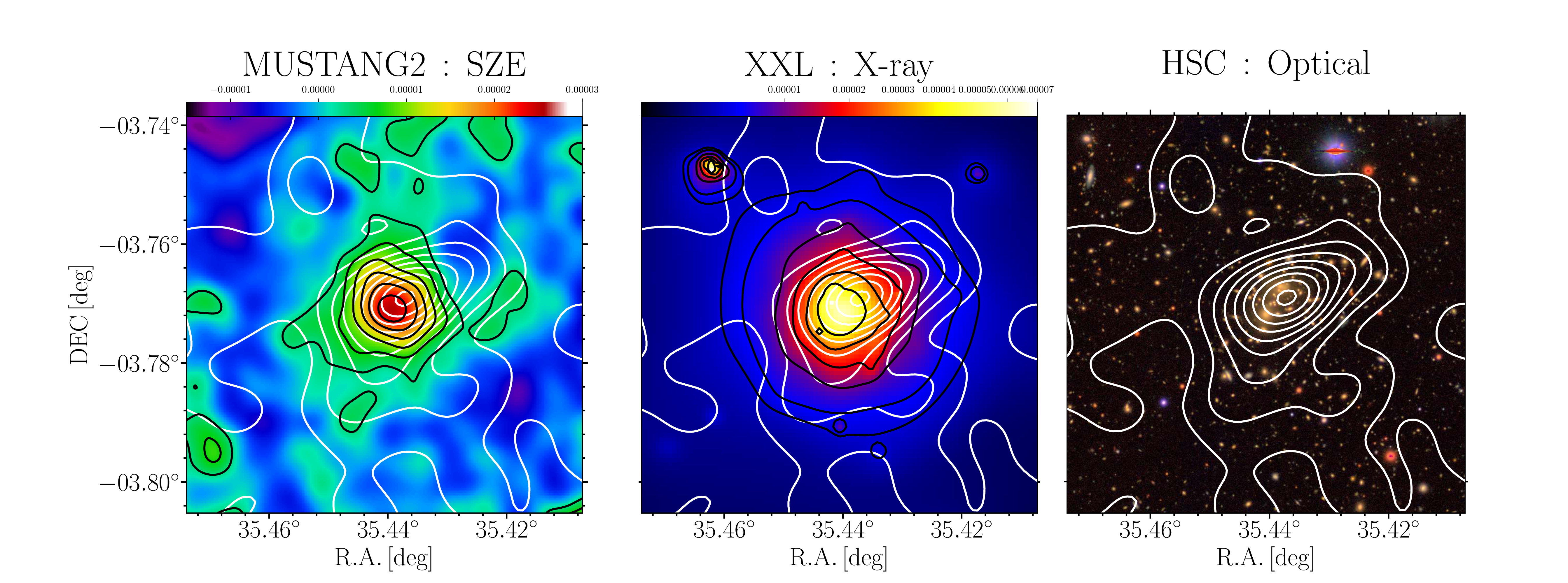}
 \includegraphics[trim=0 4mm 0 10mm,clip,width=\hsize]{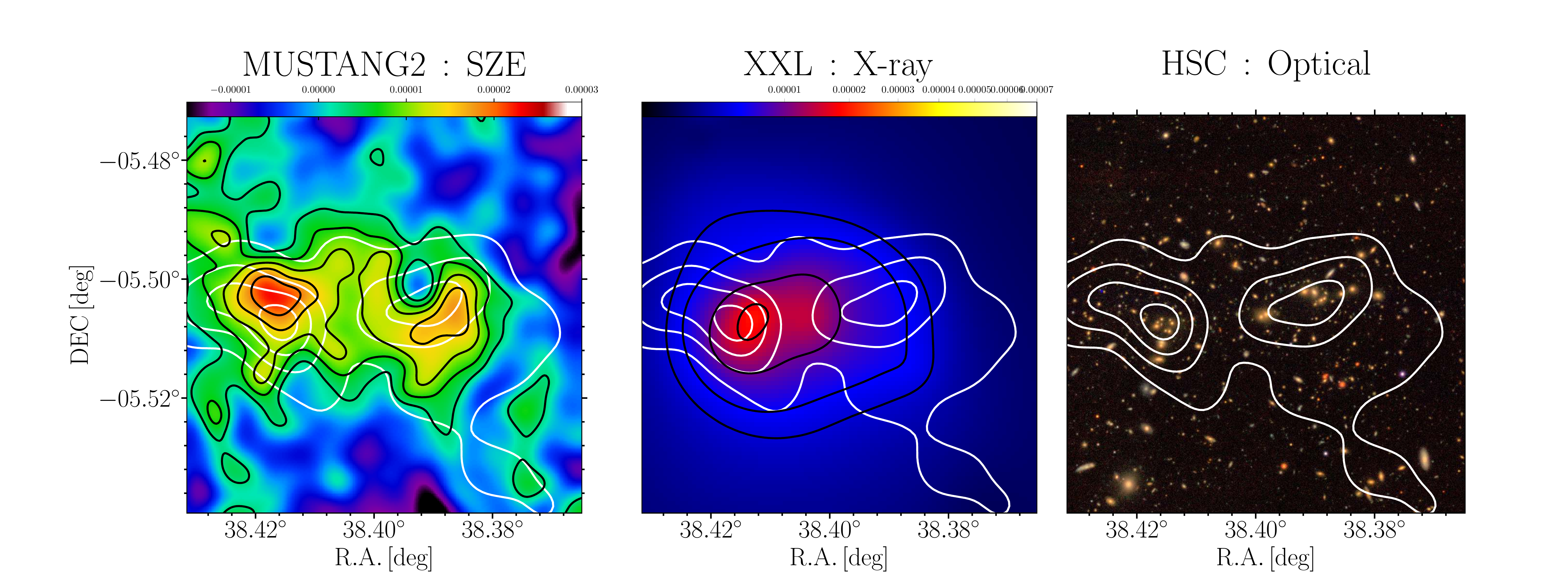}
 \includegraphics[trim=0 4mm 0 10mm,clip,width=\hsize]{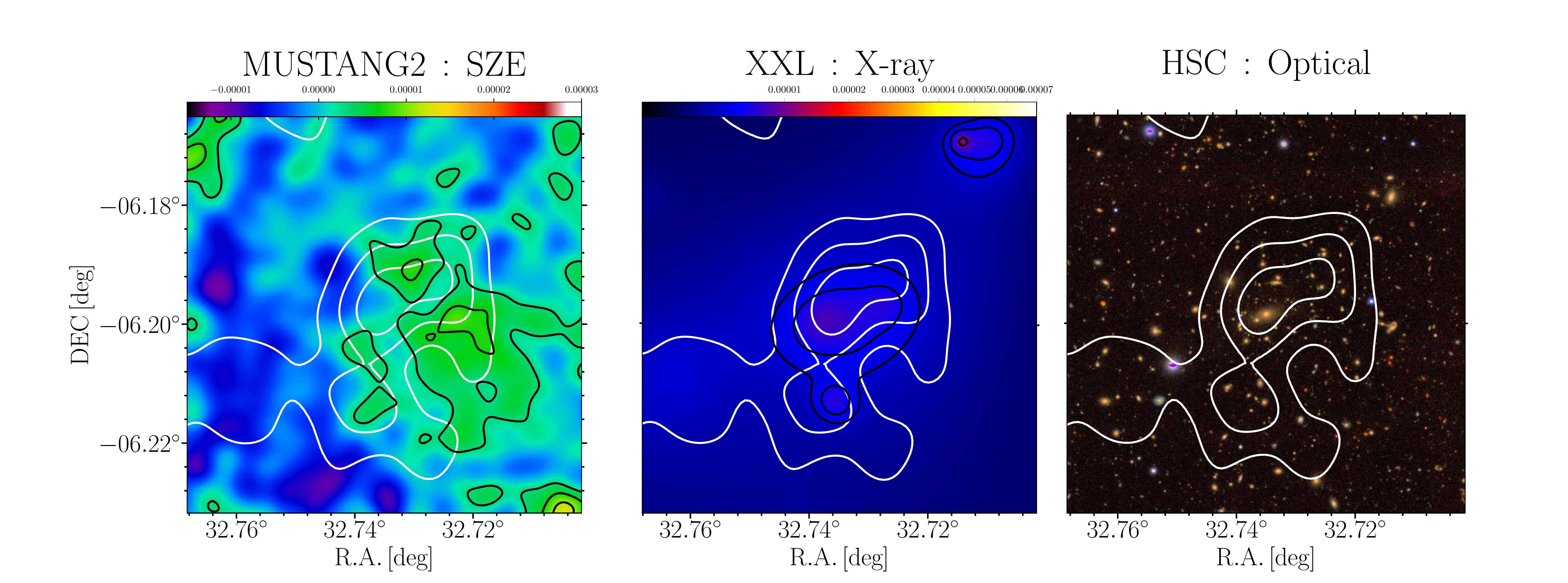}
 \end{center}
 \caption{SZE (left), X-ray (middle) and optical (right) imaging ($4\arcmin \times 4\arcmin$) for the three targeted
 clusters (from the top to the bottom; HSC J022146-034619, HSC J023336-053022, and HSC J021056-061154). 
 {\it Left}: GBT/MUSTANG-2 Compton $y-$map with Gaussian smoothing
 ($\sigma=8 \arcsec$) of raw images, yielding an effective resolution of 12.5\arcsec. The black contours are at $y=[3,7,11,15,19,23]\times 10^{-6}$. The RMS noises within 1\arcmin\ from the center of FOV of the smoothed map are $y=2, 6, \,{\rm and}\,3\times10^{-6}$ from the top to the bottom, respectively.
 The green and red colours show positive
 $y$ values and the blue and purple colours show negative
 values. Color scales are the same for all the clusters.
 White contours are galaxy distributions, taken from the right panel.
{\it Middle}: adaptively smoothed X-ray images in the soft band ($0.5-2.0$~keV) from the
 XXL survey. Black contours denote X-ray surface brightness
 ($[4,5.4,9.8,17,27,40]\times 10^{-6}\,{\rm ct\,s}^{-1}$). The white contours are the same as
 those in the right panel. Color scales are the same for all the
 clusters. {\it Right:} HSC-SSP optical $riz$-color image overlaid with
 galaxy contours (white) smoothed with a $\sigma=200\,{\rm kpc}$ Gaussian \citep{Okabe19}. Each contour is stepped by two additional luminous member galaxies, starting at a level of two luminous galaxies.}
 \label{fig:maps}
\end{figure*}

The first cluster, HSC J022146-034619, is classified as 
single-peaked in galaxy distribution within the projected radius of $2$ Mpc.
As shown in the top-right panel of Figure \ref{fig:maps}, the galaxy
distribution is concentrated around the cluster center.
The second cluster, HSC J023336-053022, exhibits two
galaxy peaks separated by about $520$ kpc (the middle-right
panel of Figure \ref{fig:maps}).
The third cluster,  HSC J021056-061154, shows an irregular galaxy distribution
(the bottom-right panel of Figure \ref{fig:maps}). 
At a $200$ kpc smoothing scale, the galaxy distribution has a single peak.
This cluster is a part of the supercluster at $z=0.43$ discovered by \citet{2016A&A...592A...6P}.
As a mass proxy, we adopt cluster richness, $N\simgt40$, as a selection function, which corresponds to $M_{500}\simgt2\times10^{14}h_{70}^{-1}M_\odot$ \citep{Okabe19}. The data used in our multi-wavelength analysis are summarized in Table \ref{tab:datalist}.

 \begin{table*}
  \caption{Properties of the clusters. 
  $^{a)}$ cluster richness from the CAMIRA catalogue \citep{Oguri18}. $^{b)}$ X-ray temperature within $300$ kpc
  \citep{2018A&A...620A...5A}. $^{c)}$: gas properties revealed by this paper.
   $^\dagger$: \citet{Okabe19}. $^\ddagger$: \citet{2016A&A...592A...6P}.}\label{tab:targets}
  \begin{center}
  \begin{tabular}{ccccccccc}
   \hline
   CAMIRA Name & Optical Morphology  & RA & DEC & $z$ & $N^{a)}$ & XXL name & $k_B T_{\rm 300kpc}^{b)}$ &  Dynamical State$^{c)}$\\
   &    &  [deg] & [deg] & & & & [keV] & \\
   \hline
   HSC J022146-034619 & single-peaked$^\dagger$
                     & 35.441
	             & -3.772
                     & 0.433
                     & 69
                     & XLSSC 006
                     & $4.2\pm0.5$
                     & sloshing \\
   HSC J023336-053022 & double-peaked$^\dagger$
                      & 38.398
	              &-5.506
	              & 0.436
		      & 47
		      & XLSSC 105
		      & $6.0\pm1.0$
	 	      & post-major merger \\
   HSC J021056-061154 & supercluster$^\ddagger$
                      & 32.735
	              & -6.198
	              & 0.429
		      & 41
		      & XLSSC 083
		      & $5.1\pm0.9$
		      & pre-minor merger \\
\hline
\end{tabular}
\end{center}
\end{table*}

\begin{table*}
\caption{Properties of the data in our multi-wavelength
 analysis. $^{a)}$: observing hours on source by MUSTANG-2 \citep{Dicker14}, the signal-to-noise ratio of the $\tilde{y}_d$ profiles (eq. \ref{eq:sn_y} and Figures, \ref{fig:HSC2_model}, \ref{fig:HSC3_model}, and \ref{fig:HSC1_model}), and the peak signal-to-noise ratio of the two dimensional $\tilde{y}_d$ maps. $^{b)}$: Obs Id for the pointing observation by the XXL Survey, and the signal-to-noise ratio of the $\tilde{S}_{X,d}$ profiles (eq. \ref{eq:sn_x} and Figures, \ref{fig:HSC2_model},\ref{fig:HSC3_model}, and \ref{fig:HSC1_model}). $^{c)}$: \citet{2016A&A...592A...1P}  $^{d)}$: \citet{HSC1stDR,HSCPhotoz17}  $^{e)}$: \citet{HSCWL1styr,Mandelbaum18}, $^{f)}$: the GMRT data of the XXL Survey \citep[610MHz;][]{Smolcic18}. $^{g)}$ FIRST archival data 
\citep[1.4GHz;][]{Helfand15}. $^{h)}$ TGSS archival data \citep[147.5 MHz;][]{Intema17TGSS}. $\dagger$: shape catalogue in the central region is not available. }
\label{tab:datalist}
    \centering
    \begin{tabular}{ccccccc}
\hline
CAMIRA name     & SZE & X-ray   & Optical & WL & Synchrotron  & \\
                & GBT MUSTANG-2$^{a)}$ & {\it XMM-Newton} & Subaru & Subaru &  GMRT & VLA/GMRT \\
\hline                         
HSC J022146-034619 & 6.1/36$\sigma$/$8.8\sigma$ &  0604280101(XXL)$^{b)}$/100$\sigma$ &  HSC-SSP$^{d)}$  & HSC-SSP$^{e)}$  & - & - \\
HSC J023336-053022 & 9.1/36$\sigma$/$4.5\sigma$ &  XXL$^{c)}$/24$\sigma$ & HSC-SSP   &  HSC-SSP  &  XXL$^{f)}$ & FIRST$^{g)}$/TGSS$^{h)}$ \\
HSC J021056-061154 & 4.4/5$\sigma$/$4.1\sigma$ &  XXL$^{c)}$/16$\sigma$ & HSC-SSP   &  HSC-SSP$^\dagger$  & - & - \\
\hline
\end{tabular}
\end{table*}

\section{Observation and Data analysis} \label{sec:data}

\subsection{GBT/MUSTANG-2 SZE analysis} \label{subsec:gbt}

MUSTANG-2 \citep{Dicker14} is a 223-feedhorn bolometer camera
installed on the 100-m GBT\footnote{The Green Bank Observatory is a major facility supported by the National Science Foundation and operated under cooperative agreement by Associated Universities, Inc.}. It has an angular resolution of $9\arcsec$ full-width
half-maximum (FWHM) and a 75--105~GHz bandpass. 
The instantaneous field of view is $4.25$ arcmin.
We observed each cluster with a $3$ arcmin radius daisy scan pattern similar to that
used for other clusters by MUSTANG-2 \citep{Romero20}, spending 6.5, 10, and 4.6 hours on-source for HSC J022146-034619, HSC J023336-053022, and HSC J021056-061154, respectively.
Every 20 minutes pointing and focus checks were carried out on 0217+0144 
allowing us to calibrate out drifts in detector gains or the atmosphere.  
Several times a night this source was tied to observations of Uranus for absolute calibration.  
Stacked observations of the calibrators allowed recovery of the {\it effective}
beamshape of the GBT.  This beam includes any filtering in the mapmaker, near sidelobes 
from focus drifts, and any remaining pointing errors.

Data were reduced using the MIDAS data pipeline.  Briefly, this pipeline first 
calibrates each detector with gains and beam volumes 
extrapolated between each observation of the point source 0217+0144.  The astronomical 
signal is mostly between $\sim$0.1~Hz (i.e. $\sim 10$ seconds, the time to scan across the map) and $\sim$10~Hz
(i.e. $\sim 0.1$ seconds, the time taken to scan across a point source).
At lower frequencies the signal is dominated by $1/f$ noise from the atmosphere and at 
higher frequencies there is noise from electrical pickup.
  A Fourier filter 
(0.08--30~Hz bandpass) is applied to each detector timestream to greatly improve the data quality.
After this, problematic detectors (for example ones with low gain or high noise) are flagged
along with portions of data showing glitches.  At this stage the timestreams are dominated by 
atmospheric emission.  This can be removed using a principal component analysis 
to produce cleaned timestreams which are made into the maps presented in this paper.  
More details of the MIDAS pipeline can be found in \citet{Romero20}.

\begin{figure}
 \begin{center}
 \includegraphics[width=\hsize]{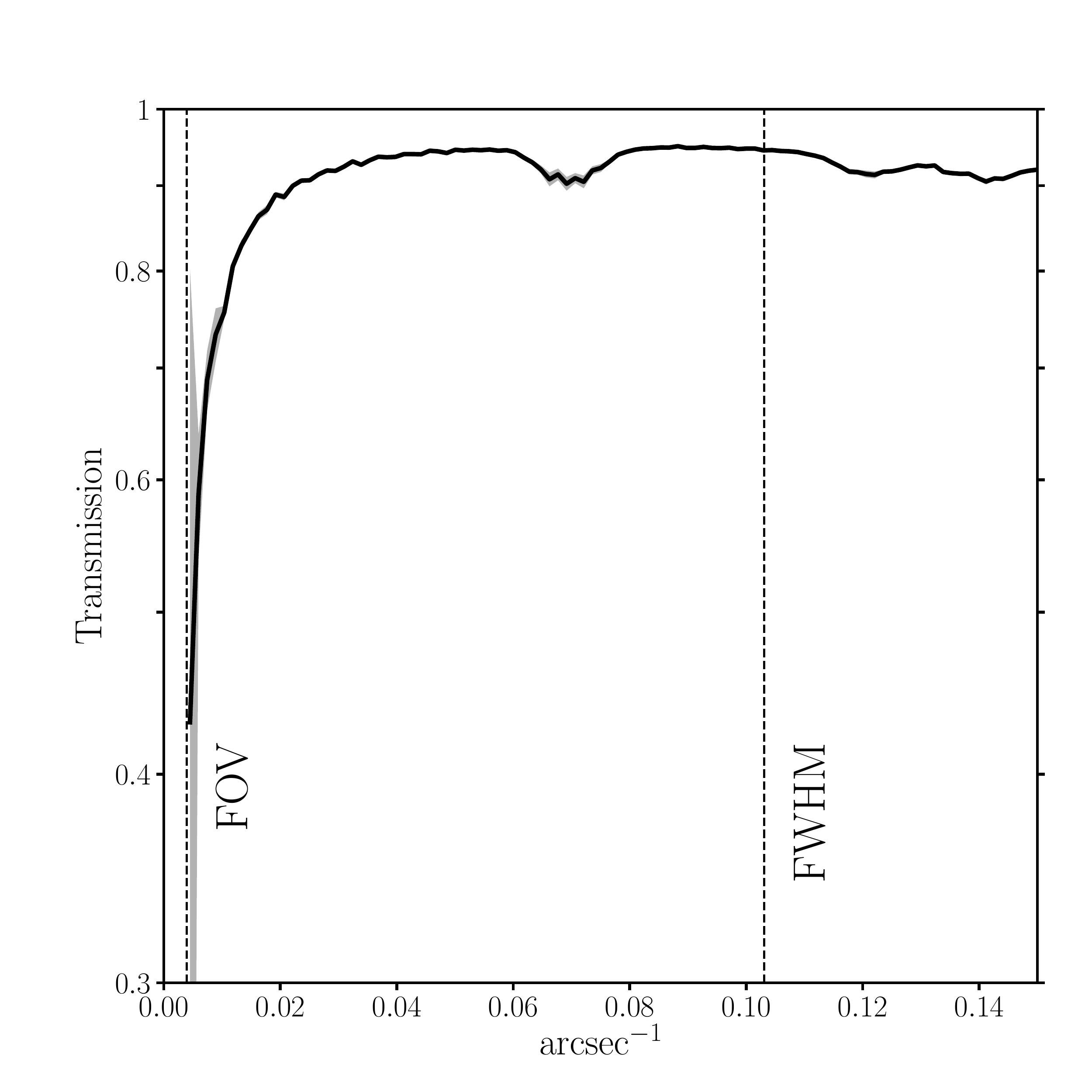}
 \end{center}
 \caption{ Effective average transfer function for our sample of clusters, $\hat{f}_{\rm TF}$, as a function 
 of angular wavenumber $k$ [arcsec$^{-1}$]. The gray transparent region at low $k$ shows that at low $k$ a few points have large error bars.
 The transfer functions of individual clusters are calculated as the square root of
the ratio of the one-dimensional power spectra of the observed fake sky and
input fake sky. Vertical dashed lines denote the relevant angular wavenumbers for the
FOV and FWHM.} 
 \label{fig:TF}
\end{figure}

Although the MIDAS pipeline reduces the RMS in the raw timestreams by several 
orders of magnitude, the maps it produces are not unbiased.  Structure on angular scales significantly larger  
than the size of the FOV are diminished in brightness.  This can be characterised by an angular 
transfer function ($f_{\rm TF}$; Figure \ref{fig:TF}). 
When quantitatively comparing observational data to model fits, it is
essential to correct for this transfer function (Sec \ref{subsec:fitting}).
The transfer function is calculated by passing randomised sky structure with equal power
on all spatial scales through simulated observing software which produces time ordered data
for a set of scans identical to those for each cluster.
Those time-streams are then processed with the same filtering as is done
on the real data and maps of the given instance of randomised sky are output. The transfer
function is defined as the ratio in Fourier space of the power spectrum
of the reconstructed image to the power spectrum of the input map
(Figure \ref{fig:TF}). 

The rest of the analysis of the MUSTANG-2 data presented in this paper 
is carried out in map space.  
We follow \citet{Romero15,Romero17,Romero20} for point source removal.
To calculate cluster profiles
we use radial averaging in segments (either 90 or 45 degrees) to bring
down the noise/get higher accuracy in the profiles without compromising
the ability to resolve the shape of the cluster.

As well as the transfer function, knowledge of the effective beam shape in the maps is critical. 
As described above, we made stacked beam maps for each cluster using 0217+0144. 
These beams were well described by the double Gaussian fit. 
A primary and secondary beam of an average of the three clusters have
FWHM of $9.7$ arcsec and $54$ arcsec, respectively. The peak ratio of the secondary
beam is $\sim5\times10^{-3}$ of the primary one. 
The secondary beam agrees with the expected near-sidelobes on the GBT
given the MUSTANG-2 illumination pattern and medium-scale aperture phase
errors not fully corrected by the out-of-focus (OOF) procedure.

\subsection{XXL X-ray analysis} \label{subsec:xxl}

We here briefly describe X-ray analysis of the XXL Survey \citep{2016A&A...592A...1P}. 
We processed the XXL data using the XMMSAS package and
calibration files v10.0.2 and the data reduction pipeline \citep{2016A&A...592A...2P} in order to obtain cleaned event files for each observation.
We extracted photon images in the [0.5−2.0] keV band for each EPIC instrument and created 
co-added EPIC images by summing the images obtained for each detector. 
In this paper, we use the co-added images in model fittings described in Sec. \ref{subsec:fitting}. 

We compare our results with X-ray complementary quantities from the literature, 
and briefly describe measurement methods.  
The measurement of the electron number density profile is described in detail by \citet{Eckert17}. 
The electron number density is measured by a deprojection method using surface brightness profiles 
that were extracted 
for each cluster using {\sc Proffit} \citep{Eckert11}. 
The X-ray temperature is measured and described in detail by \citet{2016A&A...592A...3G}. 
The X-ray spectra are extracted within a circular aperture of $300$ kpc centered on the X-ray positions.
The background is measured from an annulus centred on the cluster with the inner radius set to the detection radius and the outer radius as 400 arcsec. The resulting temperatures are summarised in Table \ref{tab:targets}.
Since there is deep X-ray pointing data (Obs Id:0604280101) for HSC J022146-034619 (XLSSC006), 
we also measure X-ray temperature profile following the XMM-Newton
cluster outskirts project \citep[XCOP;][]{Eckert17,Ghirardini18,Ettori19,Eckert19}.

\subsection{Gas modeling}\label{subsec:fitting}

We employ a Bayesian forward modeling method to measure gas properties of the ICM. 
In the modeling, we introduce a generalised Navarro, Frenk, and White
profile \citep[hereafter,
gNFW;][]{NFW97,Nagai07b,Mroczkowski09,Arnaud10,Planck11SZE,Okabe14b} of the electron number density and
the temperature of the ICM, 
\begin{eqnarray}
n_e(r)&=& 
n_0
\left(\frac{r}{r_s}\right)^{-\gamma_n}\left(1+\left(\frac{r}{r_{s,n}}\right)^{\alpha_n}\right)^{-\beta_n}, \label{eq:n} \\
 T_e(r)&=&
  T_0\left(\frac{r}{r_s}\right)^{-\gamma_T}\left(1+\left(\frac{r}{r_{s,T}}\right)^{\alpha_T}\right)^{-\beta_T}. \label{eq:T}
\end{eqnarray}
Here, $r$ is three-dimensional radius from cluster center. 
We note that the notations of slope parameters are different from definitions of \citet{Nagai07} 
in order to clarify parameter degeneracy during the analysis.
We assume a spherically symmetric model with $\alpha_n=\alpha_T=2$.
When the inner slope $\gamma$ is additionally assumed to be $\gamma=0$,
the model corresponds to a $\beta$ model \citep{Cavaliere76} which is well-used in X-ray analysis. 
The temperature scale radius, $r_{s,T}$, cannot be constrained well by a sharp $y$ distribution 
at $\theta>1$ arcmin because of the transfer function. We therefore adopt $r_{s,n}=r_{s,T}=r_s$. 
The electron pressure is directly calculated by $P_e=n_ek_BT_e$.

The SZE Compton-$y$ parameter and X-ray surface brightness are expressed
as a geometrical projection of the spherical profiles along the
line of sight, 
\begin{eqnarray}
 y (R) &=&\frac{\sigma_T}{m_e c^2}\int P_e(R,l) d l, \\
 S_X(R) &=& S_0 \int n_e(R,l)^2 d l + S_b 
\end{eqnarray}
where $r^2=R^2+l^2$, $R$ is the projected radius from the
cluster center, and $l$ is the distance along the line of sight.
Here, $\sigma_T$ is the Thomson cross-section for electron scattering, 
$m_e$ is the electron mass and $c$ is the light velocity.
Since the X-ray soft-band emissivity  ($0.5-2.0$ keV) is almost independent of gas
temperature \citep{Ettori13}, we ignore the temperature dependence.
$S_b$ and $S_0$ are the background components for X-ray data and the
conversion factor from the electron number density to the X-ray surface
brightness, respectively.

Given the model, we compute actual SZE and X-ray measurements on the sky
taking into account the instrument spatial responses, namely, the point
spread function (PSF) and the radial transfer function of the MUSTANG-2 (Sec \ref{subsec:gbt}). 
We pixelise the models onto a regular grid of angular position
$\bm{\theta}$ ($\theta=R/D_A$) and then convolve them
with the instrument response function using the two-dimensional Fourier
transform (${\mathcal FT}$),  
\begin{eqnarray}
 \tilde{y}_m(\bm{\theta})&=&(f_{\rm PSF}^{\rm SZ}f_{\rm TF})\otimes
  y(\bm{\theta}) \label{eq:model_y} \\
 \tilde{S}_{X,m}(\bm{\theta})&=&f_{\rm PSF}^{\rm X}\otimes S_X(\bm{\theta}) \label{eq:model_x}
\end{eqnarray}
where $f_{\rm TF}$ ($\hat{f}_{\rm TF}={\mathcal FT}(f_{\rm TF})$) and $f_{\rm PSF}^{\rm SZ}$ are the transfer function and the
PSF of the GBT/MUSTANG-2, respectively, and $f_{\rm PSF}^{\rm X}$ is the
PSF of the {\it XMM-Newton}. We use the transfer functions and PSFs of individual clusters.

Since the X-ray surface brightness depends on only the electron
number density and the $y$ parameter is specified by both the electron
number density and temperature, the constraints imposed by the SZE and
X-ray data enables us to resolve a degeneracy between the number density
and temperature in the $y$ parameter and then model the
three-dimensional profiles under the assumption of spherically
symmetric distributions.
We therefore simultaneously fit the SZE and X-ray data with the models
(eqs. \ref{eq:model_y}-\ref{eq:model_x}), in a similar manner to X-COP \citep{Eckert17,Ghirardini18,Ghirardini19,Ettori19,Eckert19} and other studies \cite[e.g.][]{2019arXiv191100560R}.
The joint log-likelihood is written as 
\begin{eqnarray}
 -2\ln {\mathcal L}= \sum_i \frac{(\tilde{y}_{d,i}-\tilde{y}_{m,i})^2}{\sigma_{y,i}^2} +
   \sum_j \frac{(\tilde{S}_{Xd,j}-\tilde{S}_{Xm,j})^2}{\sigma_{X,j}^2}+{\rm const},  \label{eq:L}
\end{eqnarray}
where $\tilde{y}_d$ and $\sigma_{y}$ are the MUSTANG-2 measurements of $y$ parameter and statistical errors and $\tilde{S}_{Xd}$ and $\sigma_{X}$ are the X-ray surface 
brightness distribution and statistical errors, respectively.
We do not include the X-ray temperature measurement in the joint
likelihood, because the spatial resolution of the spectroscopic
measurement is much worse than those of $y$ and $S_X$ distributions.
Since $n_0$, $T_0$ and $r_s$ are positive parameters, we treat them as
logarithmic quantities in our fitting procedures. All quantities are estimated using a central
biweight in order to down-weight outliers in skewed posterior distributions.

In the Bayesian modeling, we use radial profiles computed with 
logarithmic binning and linearly pixelised maps as the data array of $y$ and
$S_X$ of eq. \ref{eq:L}. The former and latter methods are called one-
and two- dimensional analyses, respectively. 
The former method is effective at reducing computational time and good at constraining 
the inner slopes of the electron number density and temperature profiles.
We convert from the PSF-convolved maps to radial profiles in computing
the likelihood. We choose XXL centers as central positions except for
the major-merger case.
The latter method is time-consuming but can consider multiple components of
the ICM and treat cluster centers as free parameters. 
We use $S_X$ and $y$ maps binned with pixel size of $10$ or $20$ arcsec to reduce
computational time, and thus the angular resolution of the central
distributions is worse than that in the one-dimensional analysis.
Therefore, the two analyses are complementary to each other. 
We exclude regions around $\sim0.3$ arcmin in radius centering radio point or X-ray point sources in
computing the log-likelihood.

We also estimates the signal-to-noise ratios of the $\tilde{y}_d$ and $\tilde{S}_X$ radial profiles, defined by 
\begin{eqnarray}
 (S/N)_y&=&\left(\sum_{\tilde{y}_{d,i}>0}\frac{\tilde{y}_{d,i}^2}{\sigma_{y,i}^2}\right)^{1/2}, \label{eq:sn_y}\\
 (S/N)_X&=&\left(\sum_i
	    \frac{(\tilde{S}_{X,i}-\tilde{S}_b)^2}{\sigma_{X,i}^2}\right)^{1/2},
 \label{eq:sn_x}
\end{eqnarray}
where the subscript $i$ denotes the $i$-th radial bin.

\subsection{Weak-lensing Mass Measurement} \label{subsec:wl}

We describe weak-lensing (WL) analyses of individual clusters.
Since the signal-to-noise ratio of the WL data is much lower than
those of the X-ray and SZE imaging, we do not include a weak-lensing
likelihood in the joint likelihood (eq. \ref{eq:L}) but independently measure individual cluster masses.
The independent analysis has the advantage that it does not impose the assumption of 
hydrostatic equilibrium (HE) in the modeling \citep[e.g.][]{Okabe08}.

For the shape measurement, we use the re-Gaussianization method
\citep{Hirata03} which is implemented in the HSC pipeline \citep[see
details in][]{HSCWL1styr}. 
Only galaxies satisfying the full-color and full-depth criteria from
the HSC galaxy catalogue were used in both our precise shape measurements 
and photometric redshift estimations. 
We select background galaxies behind each cluster using the color-color
selection following \cite{Medezinski18}.

The dimensional, reduced tangential shear $\Delta \Sigma_{+}$ in the
$k-$th radial bin can be 
computed by azimuthally averaging the measured tangential ellipticity,
$e_+=-(e_{1}\cos2\varphi+e_{2}\sin2\varphi)$;
\begin{eqnarray}
\Delta \Sigma_{+} (R_k) = 
\frac{\sum_{i} e_{+,i} w_{i} \langle \Sigma_{{\rm cr}}(z_{l}, z_{s,i})^{-1}\rangle^{-1}}{2 \mathcal{R}(R_k) (1+K(R_k)) \sum_{i} w_{i}}, \label{eq:g+}
\end{eqnarray}
\citep[e.g.][]{Miyaoka18,Medezinski18b,Okabe19,Miyatake19,Murata19}.
The inverse of the mean critical surface mass density for the $i$-th galaxy is computed by the probability function $P(z)$ 
from the machine learning method \citep[MLZ;][]{MLZ14} calibrated
with spectroscopic data \citep{HSCPhotoz17},
\begin{eqnarray}
 \langle \Sigma_{{\rm cr}}(z_{l},z_{s})^{-1}\rangle =
  \frac{\int^\infty_{z_{l}}\Sigma_{{\rm cr}}^{-1}(z_{l},z_{s})P(z_{s})dz_{s}}{\int^\infty_{0}P(z_{s})dz_{s}},
\end{eqnarray}
where $z_{l}$ and $z_{s}$ are the cluster and source redshift, respectively.
The critical surface mass density is expressed as $\Sigma_{{\rm cr}}=c^2D_{s}/4\pi G D_{l} D_{ls}$,
where $D_l$, $D_s$ and $D_{ls}$ are the angular diameter distances from
the observer to the cluster, to the sources and from the lens to the sources, respectively.  
The radius position, $R_k$, is defined by the weighted harmonic mean \citep{Okabe16b}.
We adopt the same central position as that of one-dimensional SZE and
X-ray analysis.
The dimensional weighting function is given by
\begin{eqnarray}
w=\frac{1}{e_{\rm rms}^2+\sigma_{e}^2}\langle \Sigma_{{\rm cr}}^{-1}\rangle^2\label{eq:weight}
\end{eqnarray}
where $e_{\rm rms}$ and $\sigma_e$ are the root mean square of
intrinsic ellipticity and the measurement error per component
($e_\alpha$; $\alpha=1$ or $2$), respectively.
The shear responsivity, $\mathcal{R}$, and the calibration factor, $K$, are
obtained by $\mathcal{R}=1-\sum_{ij} w_{i,j} e_{{\rm rms},i}^2/\sum_{ij} w_{i,j}$ and
$K=\sum_{ij} m_i w_{i,j}/\sum_{ij} w_{i,j}$ with the multiplicative
shear calibration factor $m$ \citep{HSCWL1styr,Mandelbaum18}, respectively.
We also conservatively subtract an additional, negligible offset term
for calibration.

We use the NFW profile \citep{NFW96} for individual cluster mass measurements. 
The generalised version of the NFW model (gNFW; eqs \ref{eq:n}-\ref{eq:T})
is too complicated for low signal-to-noise ratio lensing profiles of individual clusters,
and thus its slope parameters cannot be constrained. 
The gNFW model can be constrained by the stacked lensing profile measured with high signal-to-noise ratio \citep[e.g.][]{Okabe13,Okabe16b,Umetsu16}.
The three-dimensional mass density profile of the NFW profile is
expressed as, 
\begin{equation}
\rho_{\rm NFW}(r)=\frac{\rho_s}{(r/r_s)(1+r/r_s)^2},
\label{eq:rho_nfw}
\end{equation}
where $r_s$ is the scale radius and $\rho_s$ is the central density
parameter. The NFW model is also specified by the spherical mass,
$M_{\Delta}=4\pi \Delta \rho_{\rm cr} r_{\Delta}^3/3$, and the halo
concentration, $c_{\Delta}=r_{\Delta}/r_s$. Here, $r_\Delta$ is the overdensity radius.
We treat $M_\Delta$ and $c_\Delta$ as free parameters.
By integrating the mass density profile along the line of sight, we compute 
the model of the reduced tangential shear, $f_{\rm model}$, defined by 
\begin{eqnarray}
f_{\rm model}(R)= \frac{\bar{\Sigma}(<R)-\Sigma(R)}{1-\mathcal{L}_z \Sigma(R)},
\end{eqnarray}
where $\Sigma(R)$ is the local surface mass density at the projected radius $R$, 
$\bar{\Sigma}(<R)$ is the average surface mass density within the projected radius $R$, 
and $\mathcal{L}_z=\sum_{i} \langle \Sigma_{{\rm cr},i}^{-1}\rangle w_{i}/\sum_{i} w_{i}$.
Given the mass model, the log-likelihood of the weak-lensing analysis is described by
\begin{eqnarray}
&& -2\ln {\mathcal L}_{\rm WL}=\ln(\det(C_{km})) +  \label{eq:likelihood}
  \\
&& \sum_{k,m}(\Delta \Sigma_{+,k} - f_{{\rm model}}(R_k))C_{km}^{-1} (\Delta
 \Sigma_{+,m} - f_{{\rm model}}(R_m)), \nonumber
\end{eqnarray}
where $k$ and $m$ denote the $k-$th and $m-$th radial bins.
The covariance matrix, $C$, is composed of the uncorrelated
large-scale structure (LSS), $C_{\rm LSS}$, along the line of sight
\citep{Schneider98}, the shape noise $C_g$ and the errors of the source
redshifts, $C_s$ \citep[e.g.][]{Pratt19}. The elements of LSS lensing
covariance matrix are correlated with each other.

We also carry out the NFW model fitting with a free central position
using two-dimensional shear pattern \citep{Oguri10b}. The log-likelihood
is defined as 
\begin{eqnarray}
  -2\ln\mathcal{L}_{\rm WL} = \sum_{\alpha,
   \beta=1}^2\sum_{k,m}\left[\Delta \Sigma_{\alpha,k} - f_{{\rm model},\alpha}\left(\bm{R}_k\right) \right]\bm{C}^{-1}_{\alpha\beta, km}\nonumber \\
    \times\left[ \Delta \Sigma_{\beta,m} - f_{{\rm model},\beta}\left(\bm{R}_m\right) \right]+\ln(\det(C_{\alpha\beta,km})).
\end{eqnarray}
Here, the subscripts $\alpha$ and $\beta$ denote each shear component.  
The central positions are restricted to full-width boxes 
$2\,{\rm arcmin} \times2\,{\rm arcmin}$ centered on the brightest cluster galaxies (BCGs). 
The two-dimensional analysis is good at determining  the central
positions \citep{Oguri10b} and measuring masses of multi-components of
merging clusters \citep{Okabe11,Okabe15b,Medezinski16}.

In actual analyses, the maximum radial range to compute $\Delta \Sigma_\alpha$ is determined 
by excluding neighboring, massive CAMIRA clusters to avoid their contamination in lensing signals. 
We adopt an adaptive radial-bin choice \citep{Okabe16} for cluster mass estimation. 
The shape catalogue in the central region of HSC J021056-061154 is not provided because the region does not satisfy the full-colour and full-depth condition of the shape measurement. 
We thus measure WL masses without the central region of HSC J021056-061154.

\section{Results and Discussion}\label{sec:result}

\subsection{HSC J022146-034619}\label{subsec:hsc2}

\subsubsection{Joint analysis}

As shown in the top panel of Figure \ref{fig:maps},
the two-dimensional distributions of the member galaxies, $y$-parameter and
X-ray surface brightness have a single peak around the
BCG, coinciding within the PSF or smoothing scale of the observation. 
The $y$ and $S_X$ distributions exhibit regular morphology, while the
red member galaxy distribution is elongated to the northwest direction.

\begin{figure*}
 \begin{center}
 \includegraphics[width=\hsize]{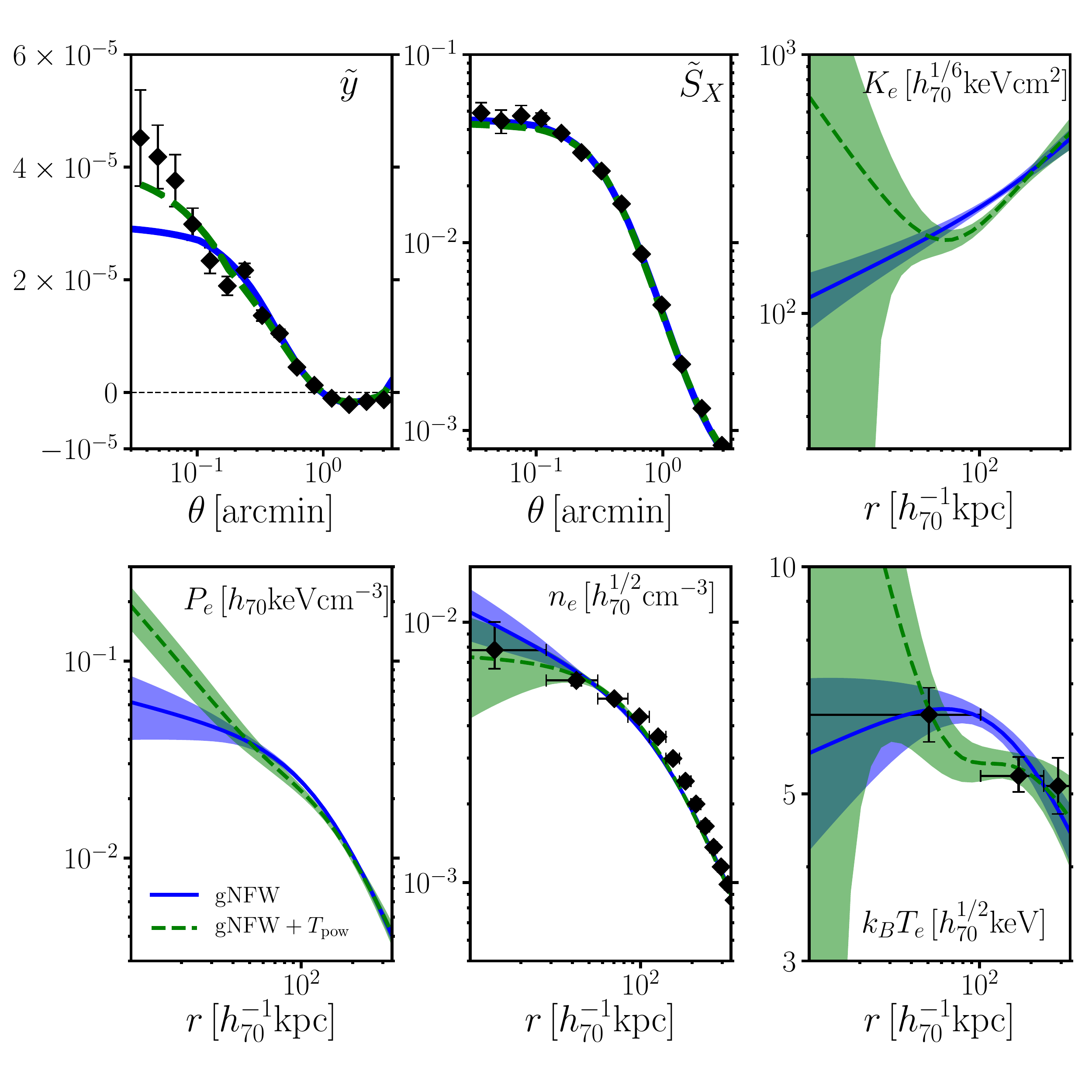}
 \end{center}
 \caption{{\it Top}: $\tilde{y}$ (left) and $\tilde{S}_X$ (middle) profiles of
   HSC J022146-034619.
   The signal-to-noise ratios computed by the observed $\tilde{y}_d$ and $\tilde{S}_{Xd}$ profiles are
   $\sim36\sigma$ and $\sim100\sigma$, respectively. They are computed
 from eqs \ref{eq:sn_y} and \ref{eq:sn_x}. The blue solid and green dashed lines are the best-fit gNFW and gNFW+$T_{\rm pow}$ models  
   derived from the joint SZE and X-ray analysis, respectively.
   The right panel shows the entropy index profile ($K_e$) computed from the bottom panels.
    The blue and green transparent regions are their $1\sigma$ errors. 
   {\it Bottom}: Three-dimensional
   profiles of the best-fit pressure ($P_e$, left), 
   density ($n_e$, middle) and temperature ($k_B T_e$, right).
   The black diamonds in the middle panel denote 
   the model-independent, deprojected profile of the electron number
   density. The black diamonds in the right panel are X-ray temperature
   measured by deep X-ray data.}
 \label{fig:HSC2_model}
\end{figure*}

We measure the $y$ and $S_X$ radial profiles,
and fit them with the gNFW model (eqs.\ref{eq:n}-\ref{eq:T}) using uniform priors. 
In order to discriminate between the actual observations and the non-convolved models, 
we represent the observations by $\tilde{y}_d$ and $\tilde{S}_{Xd}$ (eq. \ref{eq:L}).
The best-fit parameters are summarised in Table \ref{tab:gas}. 
The best-fit $y$ and $S_X$ profiles are shown by the blue solid line in the top-left and
top-middle panels of Figure \ref{fig:HSC2_model}.
Due to the transfer function (see Figure \ref{fig:TF}) of the MUSTANG2 observation, 
the observed and best-fit $y$ profiles sharply decrease at $\theta>1$ arcmin.
In a very central region of $\theta\simlt0.07$ arcmin $\sim4$ arcsec ($R\simlt20$ kpc) comparable to BCG scale, 
we find a $3\sigma$ discrepancy between the observed $\tilde{y}_d$ profile and the best-fit $\tilde{y}$ profile. Although we dropped the assumptions of $\alpha$ and $r_T$ in the gNFW profiles, the excess of the $y$-parameter cannot be explained by the gNFW model alone. 
In order to solve the discrepancy, 
we add the power-law model for the temperature profile (eq. \ref{eq:Tpow}) to the gNFW temperature model (eq. \ref{eq:T}); $T_e=T_{\rm gNFW}+T_{\rm pow}$.
The power-law model is specified by 
\begin{eqnarray}
 T_e=T_{p0} \left(\frac{r}{r_0}\right)^{-p} \label{eq:Tpow}
\end{eqnarray}
where $T_{p0}$ is the normalization, $r_0=1$ kpc is a pivot radius and $p$ is a slope. 
We refer to it as a gNFW+$T_{\rm pow}$ model. 
The best-fit parameters are summarised in the middle panel of Table \ref{tab:gas} and shown by the green dashed lines in Figure \ref{fig:TF}. 
The best-fit profiles for the gNFW and gNFW+$T_{\rm pow}$ models are in
good agreements at $\theta>0.07$ arcmin.

The bottom panels of Figure \ref{fig:HSC2_model} are, from left to right, the three-dimensional
profiles of the pressure, the electron number density and the electron temperature.  
The errors shown by blue and green transparent regions are calculated by the error covariance matrix. 
The electron pressure profiles for the gNFW model and the gNFW+$T_{\rm pow}$ model 
have a flat core and a cuspy structure, respectively. 
The electron number density profiles for the two models are similar to each other.
We compare with the deprojected electron number density and find a good agreement. 
The electron temperature profile for the gNFW model has a flat core, 
while the gNFW+$T_{\rm pow}$ model has a steep profile. 
The two profiles at $r>200$ kpc agree well with each other. 
We note that the temperature uncertainty of the gNFW$+T_{\rm pow}$ model is larger than the number density or pressure uncertainties, because the number density and pressure are directly linked to the $y$ and $S_X$ profiles in the likelihood. The uncertainty introduced by the additional power-law temperature distribution is anti-correlated with that of the number density in order to reproduce the $y$ and $S_X$ profiles. The temperature uncertainty of the $g_{\rm NFW}$ model is smaller than that of the $g_{\rm NFW}+T_{\rm pow}$ model, because the gNFW temperature profile is sensitive to the $y$ and $S_X$ profiles with small measurement errors at large radii.

We also measured X-ray temperature using deep X-ray
data based on the X-COP method \citep{Eckert17}. 
In X-ray temperature measurements of the 1st and 2nd inner bins, 
we consider a mixture of incoming photons at each annulus using {\sc Cross-arf} in {\sc Xspec}. 
The X-ray temperature profile agrees well with the SZE temperature profiles. 
However, the emission-weighted temperature using the XXL temperature ($4.2^{+0.5}_{-0.7}$ keV;  Table \ref{tab:targets}) within $300$ kpc is slightly lower than the X-COP measurement using deep X-ray data, but the discrepancy is only $2.3\sigma$ level.

\begin{figure*}
 \begin{center}
 \includegraphics[width=\hsize]{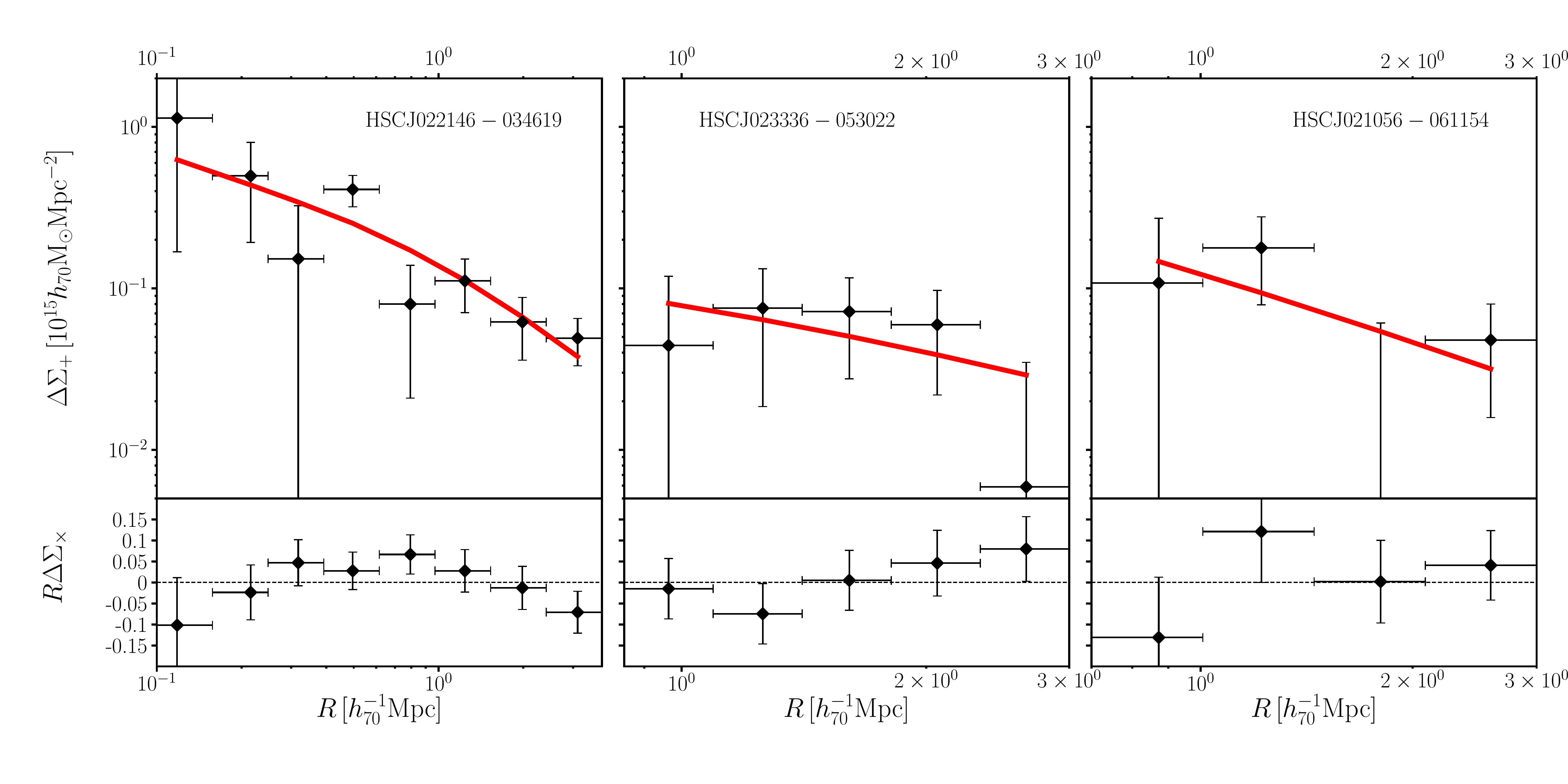}
 \end{center}
 \caption{{\it Top}: Tangential shear profiles for the
 three subsamples of HSC J022146-034619, HSC J023336-053022, and HSC
  J021056-061154 (from left to right). {\it Bottom} : The product ($10^{15} h_{70}M_\odot{\rm Mpc^{-1}}$) of the $45$
 degree rotated component, $\Delta \Sigma_\times$, with the projected distance, $R$, as a function of $R$.}
 \label{fig:WL}
 \end{figure*}

We measure weak-lensing masses using a tangential shear profile (the
left panel of Figure \ref{fig:WL} and Table \ref{tab:wlmass}). 
A comparison of weak-lensing and HE masses is discussed in Sec \ref{sub:mass_com}.

\begin{table*}
 \caption{Best-fit gas model parameters. The upper and lower panels are the
 best-fits by the one-dimensional analysis and the multi-component two-dimensional analysis. 
 The middle panel is the best-fits including the power-law model (pow) of the temperature for HSC J022146.  }\label{tab:gas}
  \begin{center}
  \begin{tabular}{lccccccc}
   \hline
   Name & $n_0$ & $k_B T_0$ & $r^{\rm gas}_s$ & $\beta_n$ & $\gamma_n$ &   $\beta_T$ & $\gamma_T$\\
               & $[h_{70}^{1/2}{\rm cm}^{-3}]$ & [$h_{70}^{1/2}{\rm
	   keV}$] & [$h_{70}^{-1}{\rm kpc}$] \\
   \hline
   HSC J022146 & $4.66_{-0.70}^{+1.30}\times10^{-3}$
               & $7.47_{-0.48}^{+0.72}$
  	       & $137.48_{-28.84}^{+22.46}$
	              &$0.684_{-0.033}^{+0.058}$
		      &$0.328_{-0.162}^{+0.076}$
		      &$0.315_{-0.043}^{+0.115}$
	              &$-0.106_{-0.151}^{+0.101}$ \\
    HSC J023336        &  $1.34_{-0.24}^{+0.43}\times10^{-3}$
                               &  $7.00_{-0.86}^{+1.31}$
                               &  $523.64_{-187.37}^{+100.39}$
	                       &  $1.308_{-0.274}^{+0.272}$
		               &  $0.428_{-0.134}^{+0.067}$
		               &  $0.622_{-0.232}^{+0.389}$ 
			       &  $-0.186_{-0.170}^{+0.109}$\\
      HSC J021056    &   $5.39_{-1.98}^{+6.35}\times10^{-4}$
                      &$15.41_{-5.59}^{+9.22}$
	              &$525.30_{-276.15}^{+629.61}$ 
	              &$0.511_{-0.158}^{+0.581}$ 
		      &$0.579_{-0.148}^{+0.052}$ 
		      &$0.727_{-0.246}^{+0.296}$  
	              &$-0.692_{-0.685}^{+0.243}$ \\ 
	              \hline
	   HSC J022146  (gNFW) & $6.96_{-1.74}^{+4.33}\times10^{-3}$
               & $6.96_{-1.04}^{+1.01}$
  	       & $99.03_{-26.67}^{+30.00}$
	              &$0.795_{-0.092}^{+0.304}$
		      &$0.263_{-0.675}^{+0.228}$
		      &$1.382_{-0.733}^{+0.648}$
	              &$-2.359_{-1.090}^{+1.386}$ \\
	     HSC J022146 (pow)   & - 
	              & $0.86_{-0.50}^{+1.20}$
	              & - 
	              & -
	             & -
	             & -
	              & $0.973_{-0.287}^{+0.521}$ \\
   \hline
   HSC J023336 (center) &  $9.66_{-2.82}^{+5.01}\times10^{-4}$
                               &  $11.64_{-2.44}^{+2.14}$
                               &  $522.28_{-130.00}^{+182.34}$
	                       &  $1.229_{-0.316}^{+0.285}$
		               &  $0.622_{-0.190}^{+0.114}$
		               &  $0.914_{-0.344}^{+0.345}$
			   &  $-1.370_{-0.888}^{+0.720}$ \\
   HSC J023336 (east) & $5.49_{-0.95}^{+1.12}\times10^{-4}$
                      & $33.91_{-6.54}^{+3.82}$
	              & $322.40_{-47.54}^{+28.87}$ 
	              & $1$ (fixed)
		      & $0$ (fixed)
		      & $1$ (fixed)
	              & $0$  (fixed)\\
   HSC J023336 (west) & $3.77_{-0.82}^{+1.72}\times10^{-4}$
                             & $54.91_{-14.57}^{+14.75}$
                             & $322.40$ (linked)
	              & $1$ (fixed)
		      & $0$  (fixed)
		      & $1$  (fixed)
			   & $0$  (fixed) \\
    HSC J021056 (east)   &   $5.87_{-2.20}^{+3.38}\times10^{-4}$
                      &$16.37_{-12.03}^{+10.03}$
	              &$485.27_{-169.00}^{+203.28}$ 
	              &$0.920_{-0.413}^{+0.514}$ 
		      &$0.566_{-0.288}^{+0.334}$ 
		      &$1.458_{-0.419}^{+0.353}$  
	              &$-1.128_{-0.739}^{+0.556}$ \\ 
   HSC J021056 (west)   &   $5.10_{-1.16}^{+0.58}\times10^{-4}$
                      &$14.62_{-1.43}^{+0.50}$
	              &$551.15_{-70.26}^{+46.30}$ 
                      & $1$ (fixed)
		      & $0$  (fixed)
		      & $1$  (fixed)
			   & $0$  (fixed) \\
      \hline
\end{tabular}
\end{center}
\end{table*}

 \begin{table}
  \caption{WL masses. $^\dagger$: mass computed from 2D WL analysis. $^\ddagger$: shape
 catalogue in the central region is not available.}\label{tab:wlmass}
  \begin{center}
  \begin{tabular}{lcc}
   \hline
   Name  & $M_{200}$ & $M_{500}$ \\
               & [$10^{14}\h70Msol$] & [$10^{14}\h70Msol$] \\
   \hline
   HSC J022146-034619 & $8.34_{-2.49}^{+3.22}$
                      & $5.69_{-1.43}^{+1.65}$ \\
   HSC J023336-053022$^\dagger$ & $2.46_{-1.08}^{+1.27}$
                      & $1.51_{-0.71}^{+0.86}$ \\
   HSC J021056-061154$^\ddagger$ & $5.42_{-3.84}^{+3.84}$
                    & $4.30_{-3.25}^{+3.01}$ \\
   \hline
\end{tabular}
\end{center}
       \end{table}

\subsubsection{Sloshing Feature} \label{subsec:hsc2_dis}

Both the $S_X$ and $y$ distributions for HSC J022146-034619 exhibit regular morphology in the sky plane
in contrast to the other two clusters. 
However, in the gNFW profile alone it is difficult to explain the excess in the observed $\tilde{y}_d$ profile at very central region of $\theta \simlt4\arcsec$, corresponding to $R\simlt20$ kpc, as shown in Figure \ref{fig:HSC2_model}. The radial size is comparable to the beam radius.
The excess requires the additional hot component. 
Since the three-dimensional profiles computed with and without the hot component agree well with each other on large scale, the global gas structure does not change and only local modification occurs. 
We also assume an elliptical gNFW temperature model elongated along the line of sight for fitting, but found it difficult to explain the excess by the reasonable parameter choices. 
Therefore, the feature implies that the gas is locally interacting, heated or perturbed.

\begin{figure*}
 \begin{center}
 \includegraphics[width=\hsize]{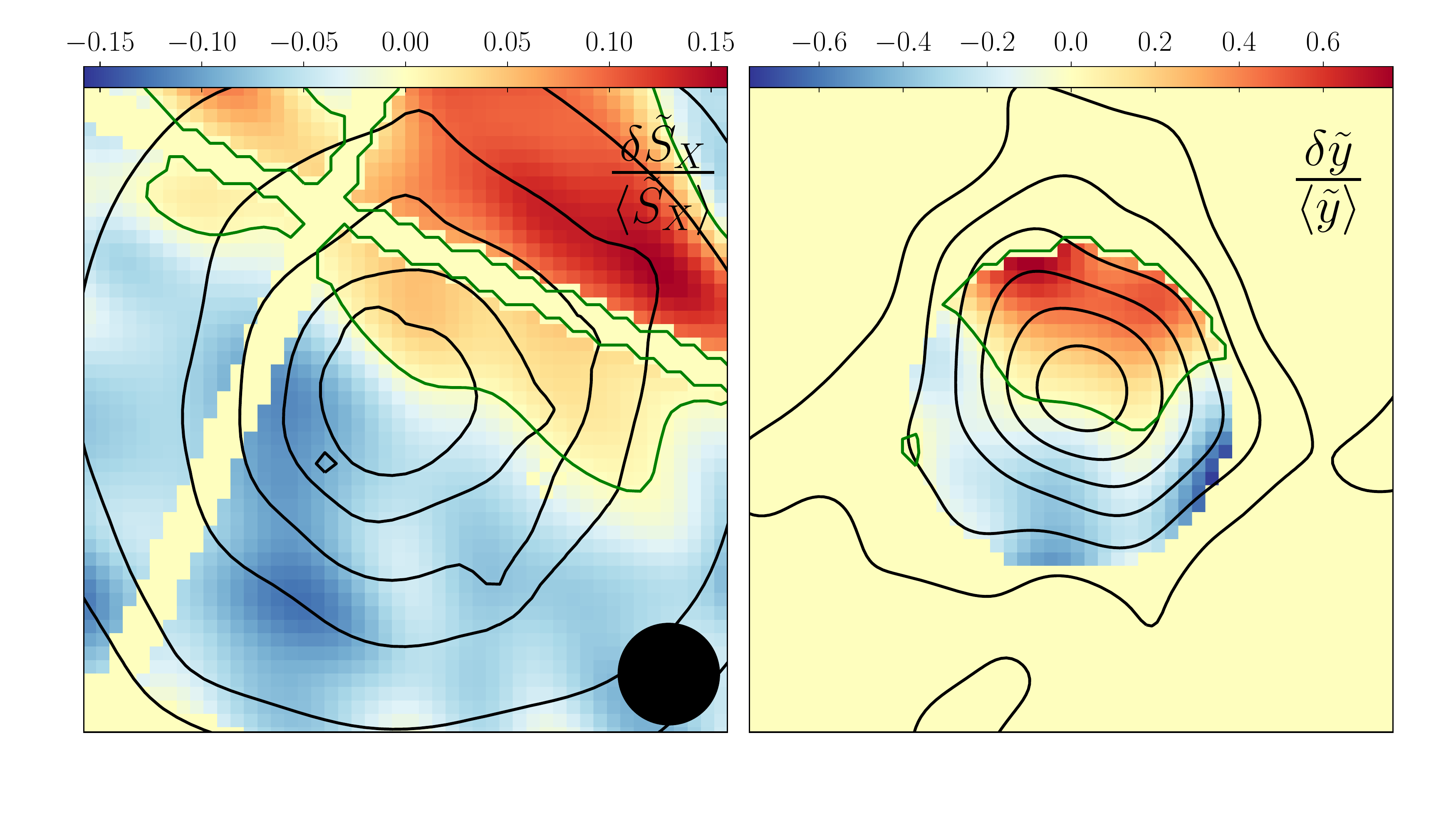}
 \end{center}
   \caption{Fractional residual maps of the X-ray surface brightness (left) and the $y$ parameter (right) of the box of $2$ arcmin $\times$ $2$ arcmin of HSC J022146-034619, centering with the XXL center. 
   The mean X-ray surface brightness and $y$ are interpolated from the azimuthally-averaged profiles.
   {\it Left}: the residual $S_X$ map as computed from MOS1 and MOS2.
   The CCD gaps are evident as a cross pattern. 
   Green contours are $\delta \tilde{S}_X=0$. The black circle at the lower right corner shows the FWHM circle of the smoothing scale ($\sigma=8$ arcsec) in order to reduce noisy feature. Black contours is $\tilde{S}_X$ distribution which is the same as in the top-middle panel of Figure \ref{fig:maps}. 
   {\it Right}: the residual $y$ map on the XMM grid of $2.5$-arcsec pixels. The map is masked outside $\theta=0.5$ arcmin
   where $\tilde{y}$ is of low signal/noise.
   The smoothing is applied as in the left panel. Green contours are $\delta \tilde{y}=0$. Black contours are the $\tilde{y}$ distribution which is as in the top-left panel of Figure \ref{fig:maps}. 
   }
 \label{fig:HSC2_res}
\end{figure*}

\begin{figure}
\begin{center}
 \includegraphics[width=\hsize]{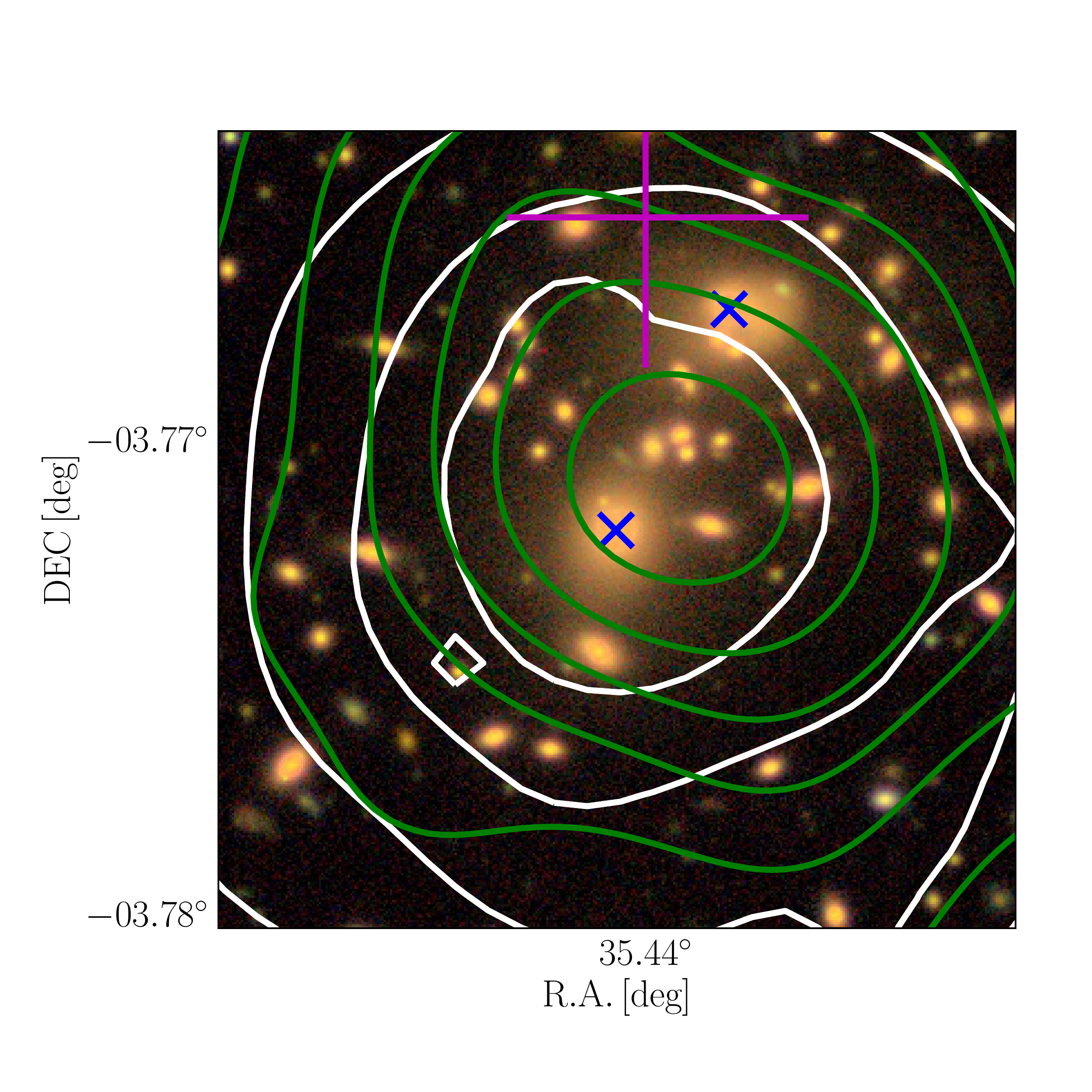}
 \end{center}
 \caption{Optical $riz$-color image around the BCG of HSC J022146-034619 ($1$ arcmin $\times$ $1$ arcmin). 
 The white and green contours denote $y$ and $S_x$ distributions (Figure \ref{fig:maps}).
 The blue crosses represent the central BCG and the second brightest galaxy at the northwest. The large magenta symbol marks the WL-determined center and its error.}
 \label{fig:HSC2_BCG}
\end{figure}

A clearer display of this feature is shown in Figure \ref{fig:HSC2_res}, where we compute fractional residual $S_X$ and $y$ maps 
between the observed images and average images.
The residual images ( $\delta \tilde{S}_X$ and $\delta \tilde{y}$ ) are derived by subtracting the averaged images.  
The average images are computed by azimuthally-averaged profiles through interpolation and thus free from any assumptions of analytical models. 
We then normalise them by the averaged images and obtain $\delta \tilde{S}_X/\langle \tilde{S}_X \rangle$ and $\delta \tilde{y}/\langle \tilde{y} \rangle$ to consider the radial dependence of $\langle \tilde{S}_X \rangle$ and $\langle \tilde{y} \rangle$. For {\it XMM-Newton}, there is a large CCD gap in the PN detector around the central region, and the residual map is computed from the MOS1 and MOS2 detectors. 
Since there are small CCD gaps even in MOS1 and MOS2, we independently compute $\delta \tilde{S}_X/\langle \tilde{S}_X \rangle$ excluding CCD gaps and then combined the two residual maps. 
For MUSTANG-2, the $\tilde{y}$ profile is negative at $\theta\simgt 1$ arcmin, and we thus mask the region to $\theta > 0.8$ arcmin. Since the XMM pixel size (2.5 arccec) is different from the MUSTANG-2 pixel size (1 arcsec), we computed the residual $y$ map using the XMM pixel size. 
Since the two residual maps are still noisy, we adopt Gaussian smoothing with $\sigma=8$ arcsec which is the same as Figure \ref{fig:maps}. To avoid the masked region in the residual $y$ map, we conservatively limit the region at $\theta<0.5$ arcmin at which $\tilde{y}$ is almost constant (Figure \ref{fig:HSC2_model}). The resulting residual maps are shown in Figure \ref {fig:HSC2_res}. Positive and negative excesses appear in the northern and southern areas in the two maps. 
The sums of the signal-to-noise ratios of the deviations ($\delta \tilde{y}$ and $\delta \tilde{S}_X$) in every pixel within $0.5$ arcmin from the center are $3.5\sigma$ in $\delta \tilde{y}$, $5.0\sigma$ in $\delta \tilde{S}_X$, and $6.1\sigma$ in total.
The residual patterns are coherently distributed, which indicates a presence of gas sloshing.
The gas disturbances could also trigger local heating \citep[e.g.][]{Ascasibar06,ZuHone10,Vazza12}.
Even when we use the BCG position as the central position (Figure \ref{fig:HSC2_BCG}), 
these coherent residual pattern does not disappear. 
We quantify the pixel-to-pixel cross-correlation between the residual patterns with different centers and find the correlation coefficients of $0.86$ in $\delta \tilde{y}/\langle \tilde{y} \rangle$ and of $0.97$ in $\delta \tilde{S}_X/\langle \tilde{S}_X \rangle$.

It is difficult to search by photometric information for a subhalo which triggered the sloshing mode.  
A large number of spectroscopic redshifts will be crucial to identity the subhalo, 
but only nine redshifts are available to date.  
Since the red galaxy distribution \citep{Okabe19} could not offer clues on the subhalos location within $R_{200}$, 
two subhalo candidates can be expected.
The first candidate is a subhalo within the smoothing scale of $200$ kpc from the BCG and the second candidate is a less massive subhalo. 

There is a second luminous galaxy at $(\alpha,\delta)=(35.4382,-3.7673)$ 
which is $106$ kpc northwest from the BCG. Its stellar mass is about one-fourth that of the BCG.  
The presence of the second brightest galaxy ( the second BCG ) is not rare in optical clusters.
We perform the two-dimensional WL analysis using a single NFW model and its free central location. The resulting center is close to the second BCG rather than the BCG (Table \ref{tab:center} and Figure \ref{fig:HSC2_BCG}).
Although the two galaxies are too close to resolve their mass structures by WL analysis, the position would be explained by the superposition of the mass associated with the second bright galaxy and the main halo.  
If the difference in their redshifts gives their relative peculiar velocity along the line of sight,
$v=c \delta z / (1+z) \sim450\,{\rm km\,s^{-1}}$ is likely to be subsonic motion. Therefore, the scenario that the moving second BCG triggers the hot component and sloshing pattern does not qualitatively contradict the observational results.

\begin{figure*}
 \begin{center}
 \includegraphics[width=\hsize]{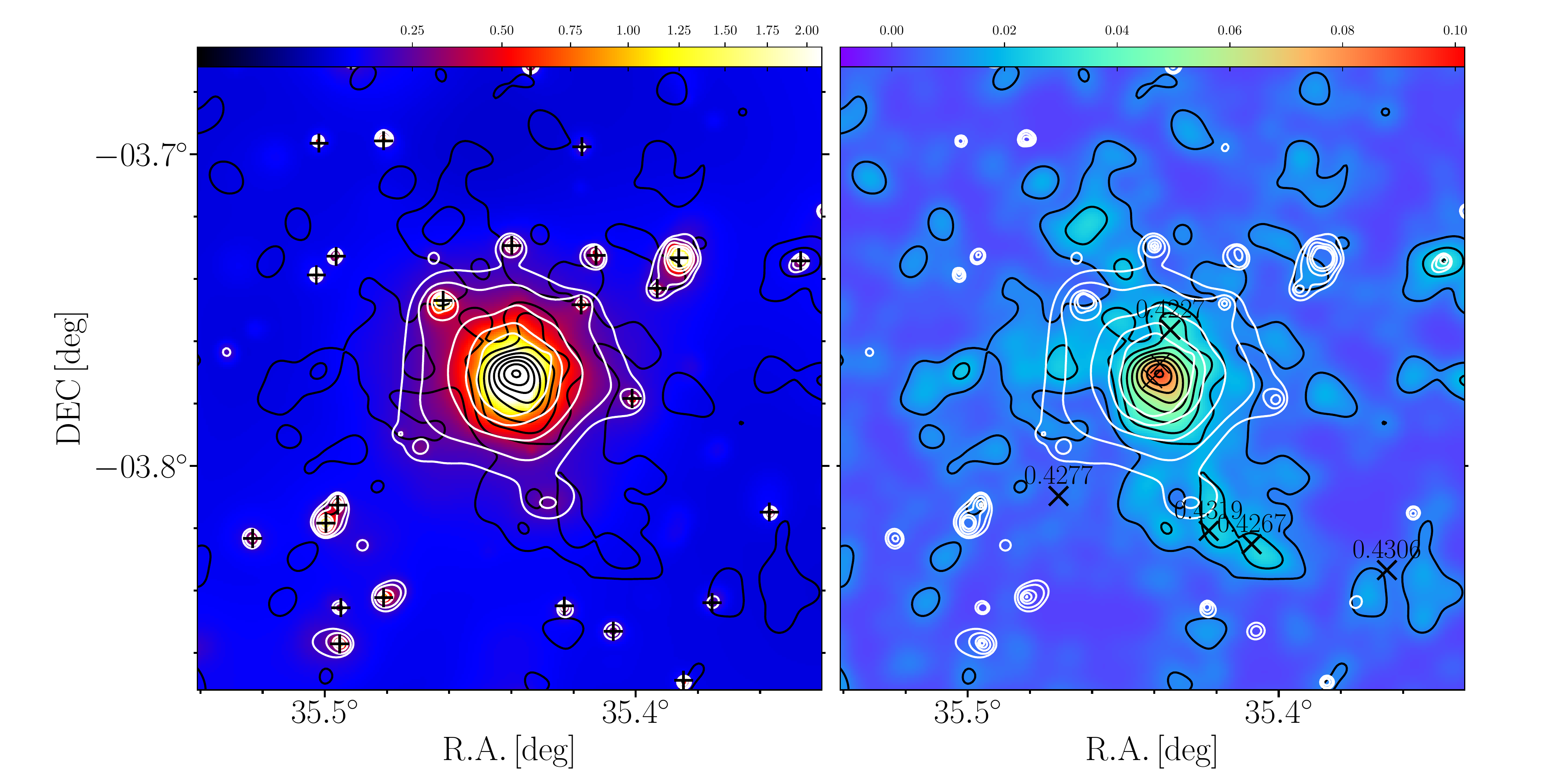}
 \end{center}
   \caption{Adaptively smoothed X-ray map (left) and galaxy distribution (right) sliced by photometric redshift probability function $P(z)$ for HSC J022146-034619. The full-width size is $12\arcmin$. 
   {\it Left}: the X-ray tailed feature from the cluster center to the southern region is marginally found.
   Black pluses denote X-ray point sources or foregrounds identified by cross-matching optical and X-ray images. 
   White and black contours are X-ray and galaxy distributions, respectively. 
   {\it Right}: the galaxy number density distribution smoothed with FWHM=$200$ kpc. Black X points are spectroscopically identified galaxies of which redshifts are shown in black texts, excluding the BCG and second bright galaxy. The galaxy distribution is elongated along the north and south direction on large scale. X-ray faint galaxy substructure is found around (35.4087,-3.82514). If the galaxy substructure were passing from the cluster center to the south, it could trigger the sloshing mode in the central region.
   }
 \label{fig:HSC2_map_large}
\end{figure*}

To search for a second halo candidate, we next made a galaxy number map of red and blue galaxies selected by full probability function of the photometric redshift \citep{HSCPhotoz17}.
Following \citet{Ichikawa13} to map the surrounding galaxy distribution on the large scale, 
we compute a probability of each galaxy located at a slice of $\pm \Delta z$, defined by
\begin{eqnarray}
 p_{\rm gal}=\int_{z_l-\Delta z}^{z_l+\Delta z} P(z)dz/ \int_0^\infty P(z)dz,
\end{eqnarray}
where $P(z)$ is the full probability function, $z_l$ is a cluster redshift and 
$\Delta z$ is the redshift slice.
Taking into account photometric redshift errors, we adopt $\Delta z=0.1$ in a similar way to \citet{Eckert17}.
The galaxy number distribution is shown in the right panel of Figure \ref{fig:HSC2_map_large}. 
The galaxy distribution is elongated along the north-south direction. The spectroscopically identified galaxies \citep{2018A&A...620A...7G} are shown by red pluses. A galaxy group is found around $(\alpha,\delta)=(35.4087, -3.8252)$ which is at $\sim1.3$ Mpc south of the BCG.
The group is mainly composed of red luminous galaxies. 
Diffuse X-ray emission is also found around $(\alpha,\delta)=(35.4269, -3.8115)$ which is $\sim 0.8$ Mpc south of the BCG and at $\sim0.4$ Mpc north of the southern galaxy group. 
The X-ray contours along the line connected between the X-ray main peak and the southern X-ray emission is slightly curved outwards.
One of possible interpretations is that the subhalo passed through the cluster center and gas was stripped away by ram-pressure.
Since the southern galaxy group is compact and less massive, we measure its projected mass following a subhalo mass measurement \citep{Okabe14a}. The projected mass, $M_{\zeta_c}=3.5\pm0.8\times10^{13}M_\odot$, is only about $1/25$ of the main halo.

Since having a less massive halo fall into a cluster occurs fairly often
we cannot make a conclusive statement about which subhalo triggered the sloshing mode. Systematic future spectroscopic observations will reveal the details.

\subsubsection{Comparison with numerical simulations}

\begin{figure}
 \begin{center}
 \includegraphics[width=\hsize]{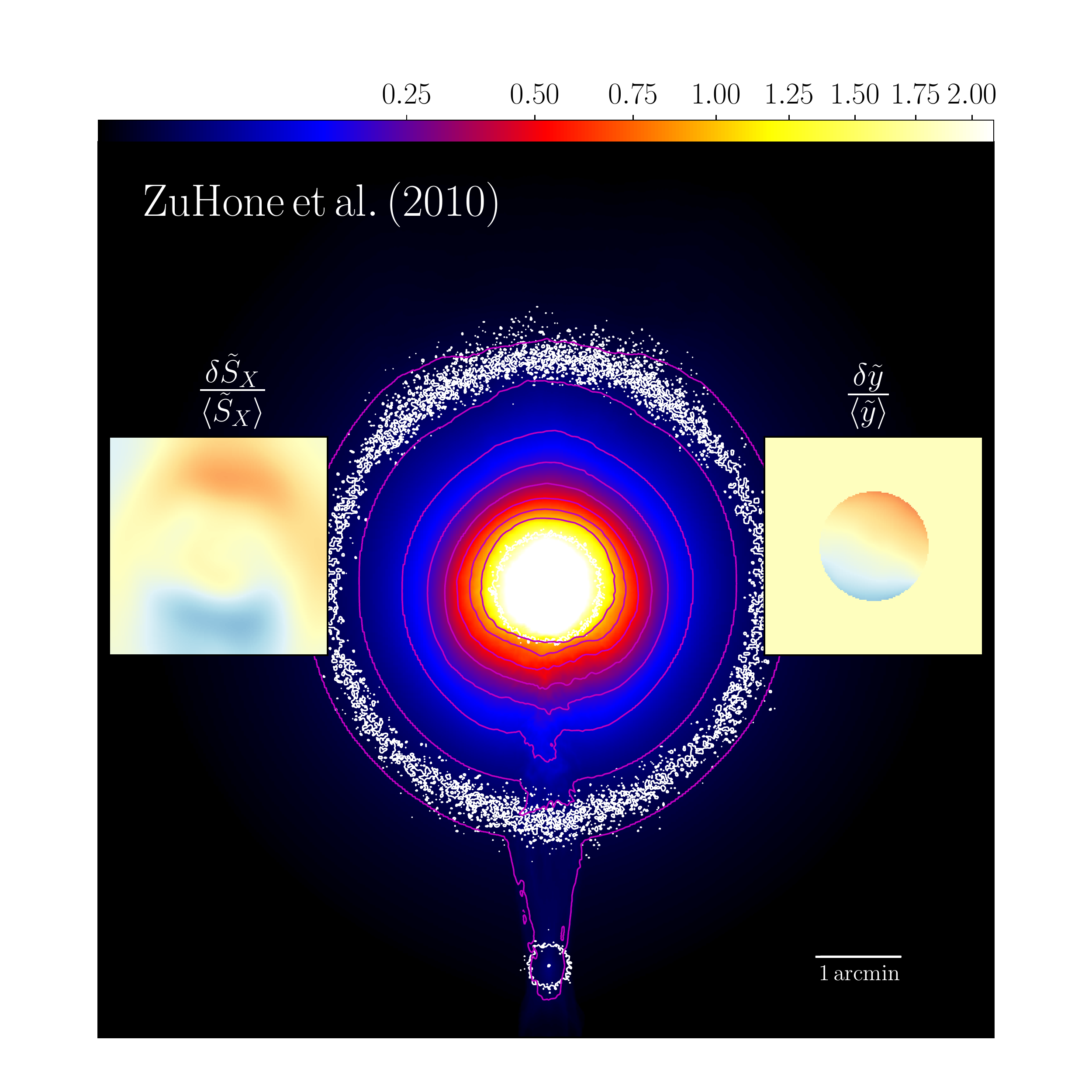}
 \end{center}
   \caption{One of examples of simulated sloshing images retrieved from \citet{ZuHone10,ZuHone18}. White and magenta contours are projected mass and X-ray surface brightness distributions, respectively. Inset panels are fractional residual $\tilde{S}_X/\langle \tilde{S}_X \rangle$ (left) and $\tilde{y}/\langle \tilde{y} \rangle$ maps (right) taking account of the PSFs and the transfer function of the box of $2$ arcmin $\times$ $2$ arcmin. The appearance of the residual images resembles our residual images (Figure \ref{fig:HSC2_res}). The configurations of the X-ray and mass clumps are similar to those in Figure \ref{fig:HSC2_map_large}.
   }
 \label{fig:ZuHone_sloshing}
\end{figure}

\citet{ZuHone10} studied sloshing features in the gas core using  $N$-body/hydrodynamic numerical simulations for which data is publicly available \citep{ZuHone18}. In order to visually understand the observed feature, we compute mock residual maps of simulated clusters at the cluster redshift considering the PSFs and the transfer function. We use the data-set of $M_{200}=10^{15}M_\odot$ for the main cluster and a subhalo with mass ratio $1:20$ and impact parameter of $200$ kpc. 
The sloshing modes appeared in all phases after the first impact. The direction of residual emission and pattern depend on the viewing angle. We pick a phase at $2.5$ Gyr after the closest encounter as a typical example. At this phase, the subhalo is on its way to the second impact after turn-around. 
Figure \ref{fig:ZuHone_sloshing} is an edge-on view from the merger plane for the simulated $S_X$ and mass distributions. The subhalo is located at the south and a tailed gas feature is found.
Inset figures are the fractional residual maps convolved with the PSFs and the transfer function excluding any observational noise. The coherent residual patterns and the location of the subhalo are similar to our observation (Figures \ref{fig:HSC2_res} and \ref{fig:HSC2_map_large}).

\subsection{HSC J023336-053022} \label{subsec:hsc3}

\subsubsection{Joint analysis}

The $y$ map shows a clear double-peaked structure associated with galaxy
concentrations (the middle horizontal panel of Figure \ref{fig:maps}).
An X-ray core with a round shape is found at the intermediate position
between the two high $y$ components. 
The X-ray core can be visually decomposed into two substructures. 
The X-ray peak coincides with the positions of the eastern $y$ and galaxy structures. 
The secondary X-ray peak is located close to 
a surface-brightness weighted centroid.
The $y$ parameter around the X-ray morphological center is lower than those of
the double-peaked $y$ components. The high density and low pressure suggests that the X-ray core is likely to be a cool core.
A possible scenario explaining the observed feature is that
two clusters with cool cores are colliding with each other and two
shock-heated regions are triggered ahead of the gas cores.
The cluster is likely to be at a phase just after core passage.

\begin{figure*}
\begin{center}
 \includegraphics[width=\hsize]{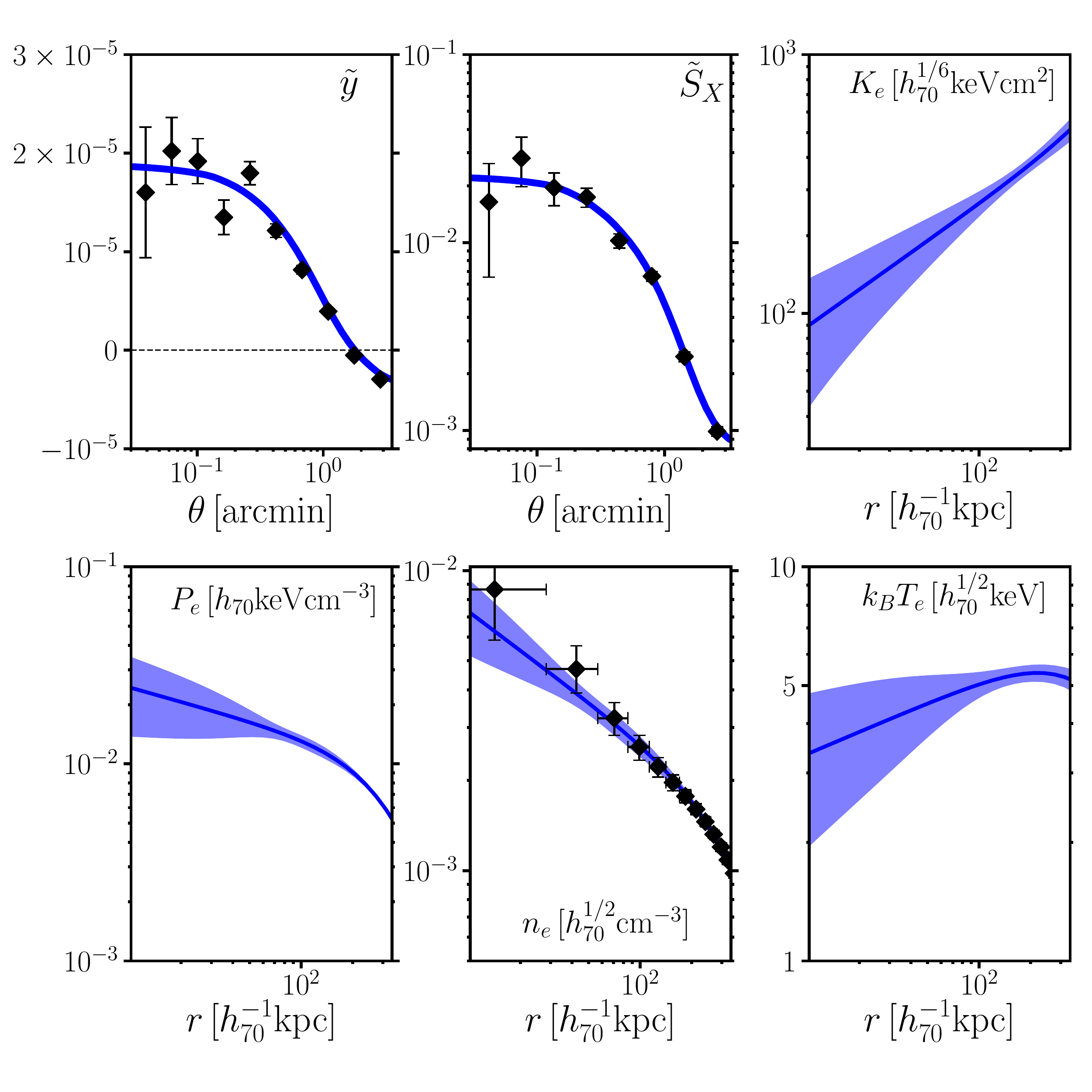}
\end{center}
\caption{The same figure as Figure \ref{fig:HSC2_model}, but for HSC J023336-053022. 
   The signal-to-noise ratios of the $\tilde{y}_d$ and $\tilde{S}_{Xd}$ profiles are $\sim36\sigma$ and $\sim24\sigma$, respectively.}
\label{fig:HSC3_model}
\end{figure*}

As a first attempt at modeling, we assume a spherically symmetric gas distribution in a
similar way as the other clusters (Secs
\ref{subsec:hsc2} and \ref{subsec:hsc1}), 
though the $y$ and $S_X$ distributions are composed of the three gas components.  
The azimuthally-averaged $y$ profile (the top-left panel of Figure
\ref{fig:HSC3_model}) is almost constant at small radii and steeping
at $\theta\sim1$ arcmin because of the angular transfer function.
For this cluster, we choose an X-ray count-rate weighted center within 100 kpc
of the X-ray peak, because we find the $S_X$
and $y$ profile centers are misaligned with the XXL center. 
As expected from the observed $y$ profile, the modeled pressure profile has a shallow slope.
The electron number density shows a steeply decreasing function and agrees
with the model-independent, deprojected electron number density.
The temperature profile slightly increases as the radius increases to 
compensate for the steep function of the electron number density. 
The X-ray-like emission-weighted, projected temperature \norv{($k_BT_{\rm SZ+X}$)} within $300$ kpc is $5.2_{-0.3}^{ +0.3}$ keV.

We next perform a multi-component analysis to model the double-peaked $y$ structure and single $S_X$ structure. 
The peak signal-to-noise ratios of the western and eastern $y$ structures are $3.4$ and $4.5\sigma$, respectively. 
The signal-to-noise ratios (eq \ref{eq:sn_y}) of the $\tilde{y}_d$ profiles within an $0.5$ arcmin circle centered on the two peaks are $21.7\,\sigma$ and $27.9\,\sigma$, respectively. The signal-to-noise ratio of the $\tilde{S}_X$ profile within $0.5$ arcmin centering the X-ray peak is $15.9\,\sigma$. We therefore introduce three gas
components as a function of three-dimensional space, ${\bm x}$,
\begin{eqnarray}
 g({\bm x})=g_{\rm gNFW}^{\rm C}({\bm x}_c^{\rm C};{\bm x})+g_{\beta}^{\rm W}({\bm
  x}_c^W;{\bm x})+g_{\beta}^{\rm E}({\bm x}_c^E;{\bm x}),
\end{eqnarray}
where $g=n_e$ or $g=T_e$ and ${\bm x}_c$ is the central position.
The subscript and superscript denote the type of models and central positions,
respectively.
We adopt the gNFW model for the central gas component in order to
express the presence of the cool core. 
In order to describe the western and eastern hot-thin regions, we adopt the $\beta$ model for simplicity.
Since the outer slopes of the number density and temperature profiles of the
two hot regions cannot be constrained well, we fixed the slope $\beta_n=\beta_T=1$. 
The slope is much steeper than the typical value of clusters derived by X-ray surface brightness profile \citep[e.g.][$\beta\sim0.67$]{2016A&A...592A...2P} in order to describe the gas locally heated by cluster merger shocks.
We also assume that the scale radii of the western and eastern components are the same.
In other words, the differences between the western and eastern gas properties are described by the central temperature and density values.

When we compute the projected $y$ and $S_X$ distributions,
we assume that the two gas components do not interact with each other. 
We treat central positions as free parameters, restricted to boxes
of $1\,{\rm arcmin}\times1\,{\rm arcmin}$ centered on the brightest cluster galaxies of the western and
eastern components and the cool core for the $W$, $E$ and $C$ components, respectively. 
We use two-dimensional images binned by ten pixels for the $y$ map and four pixels for
the $S_X$ map, respectively. The pixel sizes of the two binned images correspond to $10\arcsec$.

\begin{table}
  \caption{Centers determined by two-dimensional fitting analyses.}\label{tab:center}
  \begin{center}
  \begin{tabular}{lcc}
   \hline
 Component  & $\alpha$ & $\delta$ \\
            & [deg]   & [deg] \\
   \hline
HSC J022146-034619 & \\
Gas  &  $35.4389_{-0.0001}^{+0.0001}$ &  $-3.7705_{-0.0001}^{+0.0001}$ \\
Mass &  $35.4400_{-0.0034}^{+0.0029}$ &  $-3.7653_{-0.0032}^{+0.0041}$ \\
\hline
HSC J023336-053022 & \\
Gas C  & $38.4081_{-0.0010}^{+0.0009}$ & $-5.5052_{-0.0010}^{+0.0014}$ \\
Gas W  & $38.3889_{-0.0010}^{+0.0007}$ & $-5.5053_{-0.0004}^{+0.0004}$ \\ 
Gas E  & $38.4153_{-0.0007}^{+0.0020}$ & $-5.5016_{-0.0005}^{+0.0006}$ \\
Mass W  & $38.3895_{-0.0072}^{+0.0058}$ & $-5.5030_{-0.0068}^{+0.0051}$ \\ 
Mass E  & $38.4178_{-0.0103}^{+0.0067}$ & $-5.5069_{-0.0070}^{+0.0062}$ \\
\hline 
HSC J021056-061154 & \\
Gas E  &  $32.7338_{-0.0035}^{+0.0010}$ & $-6.1975_{-0.0009}^{+0.0010}$ \\
Gas W  &  $32.7208_{-0.0035}^{+0.0021}$ & $-6.1998_{-0.0027}^{+0.0018}$ \\
Ell Gas E &   $32.7358_{-0.0015}^{+0.0022}$ & $-6.1978_{-0.0008}^{+0.0007}$ \\
Ell Gas W & $32.7228_{-0.0011}^{+0.0009}$ & $-6.2003_{-0.0015}^{+0.0013}$ \\
   \hline
\end{tabular}
\end{center}
\end{table}

\begin{figure*}
 \begin{center}
 \includegraphics[width=\hsize]{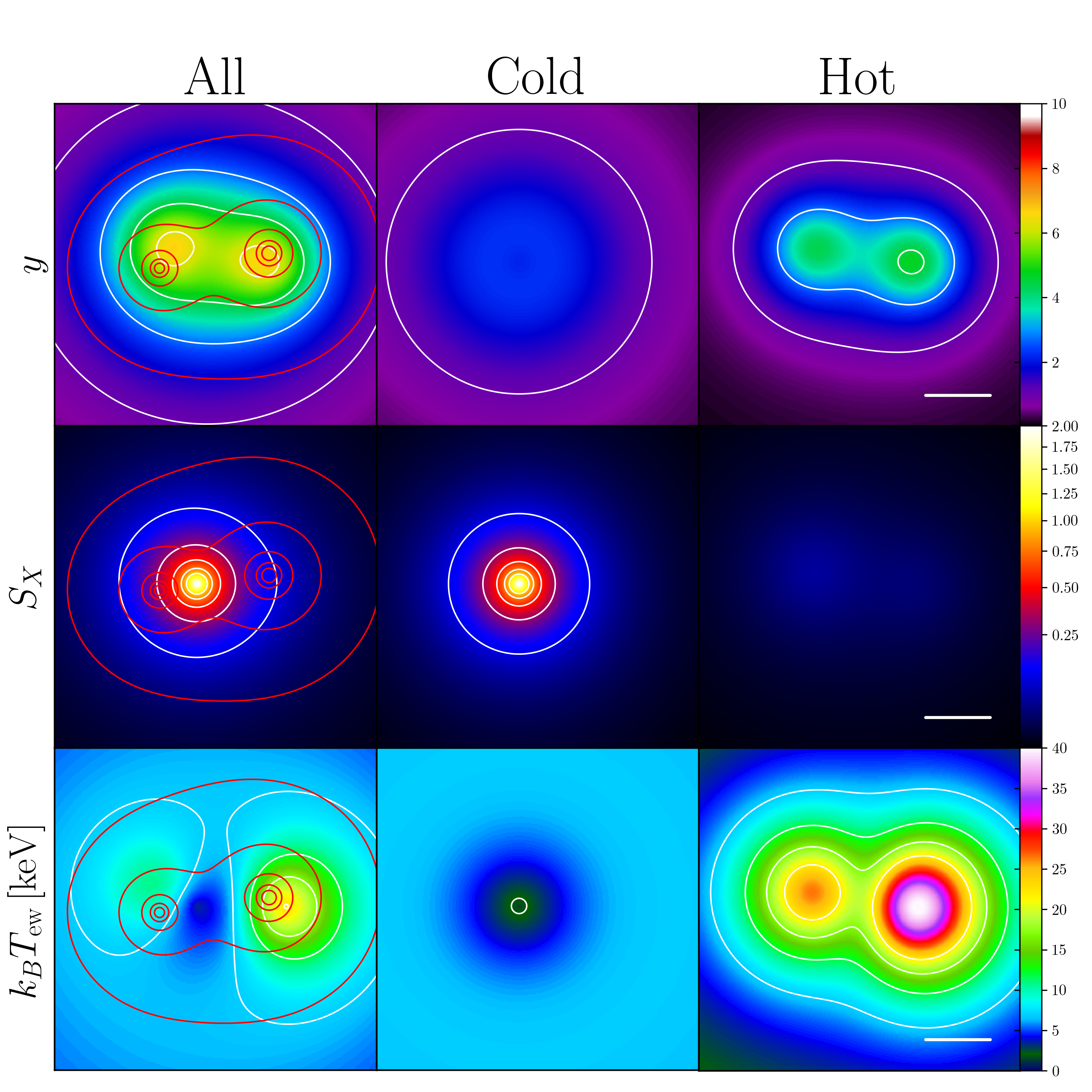}
 \end{center}
   \caption{Model maps for HSC J023336-053022: $y$ map [$10^{-5}$] ({\it Top}), $S_X$
   map in arbitrary units ({\it Middle}), and X-ray-like emission-weighted
 temperature, $k_B T$ [keV] ({\it Bottom}). The full-width size is 5arcmin. Neither the transfer
 function nor the PSF has been applied. 
  From left to right, the panels
   show the total gas components, the cold gas component, and the hot
   gas component, respectively. The red contours denote the projected
   mass contours derived by two-dimensional WL analysis ($[0.1,0.3,1.1,2.3,4]\times 10^{14}M_\odot$ Mpc$^{-2}$, stepped by square root). The white lines shown in the right panels denote $1$ arcmin.}
 \label{fig:HSC3_maps}
\end{figure*}

The resulting model maps and parameters are shown in Figure
\ref{fig:HSC3_maps} and Table \ref{tab:gas}, respectively.
The best-fit centers are shown in Table \ref{tab:center}.
The model maps do not take into account the transfer function and the PSFs to understand distributions of physical properties.
The models succeed in reproducing the double-peaked distribution of the
$y$ distribution (top-left panel) and the single cool core (middle and bottom panels).
The X-ray distribution is completely different from the WL mass distribution, as usual in on-going mergers \citep{Okabe08}.
The X-ray-like emission-weighted, projected temperature ($k_BT_{\rm SZ+X}$) of the central cluster component changes from $\sim2$
keV at the cool core to $\sim7$ keV at an intermediate radius of $2.5$ arcmin. 
The temperature within $300$ kpc from the best-fit centre agrees with the expectation from the WL mass (Sec \ref{sub:scaling}).
Although the temperature of the intermediate radius is slightly higher than expected from the WL mass, 
$\tilde{y}$ at the radius is negative due to the transfer function.  
The X-ray-like emission-weighted projected temperatures within the projected radius $300$ kpc from the same center of the 1D analysis is $k_B T_{\rm SZ+X}=6.6_{-1.3}^{+1.4}$ keV which agrees with that of the 1D analysis within the $\sim 1.1\sigma$ of the 2D analysis. We here take into account the full error covariance matrix of the gas parameters. 
The X-ray-like emission-weighted projected temperatures within
the projected radius $300$ kpc from the best-fit centres in the western
and eastern hot components \norv{(see details in Sections \ref{sec:Tcomp} and  \ref{sub:scaling}; $k_BT_{\rm SZ+X}$)} are $28.4_{-6.0}^{+5.9}$ keV and $20.2_{-3.4}^{+3.5}$ keV, respectively. 
We note that the projected temperatures depend not only on the normalization $T_0$ but also the overall temperature and density distributions. The uncertainties in the projected temperatures fully take into account the error covariance matrix of the gas parameters.

When we include the cool component, the projected temperature in the western and eastern regions are $11.3_{-1.6}^{+ 1.6}$ keV and $6.8_{-1.3}^{+1.4}$ keV, respectively. 
We note that a relativistic correction is small in the observing frequency \citep[e.g.][]{Hughes98,Mroczkowski19}.
Their temperatures are three or four times higher than that of the cool component in the same regions.
However, the electron number density of the two hot regions is lower than that of the cool component. 
We compute the integral of the electron density over a cylindrical volume within
a projected radius of $300$ kpc from the best-fit central position of the
western/eastern component.
The ratios of the electron number of the western/eastern component to
the total component are 0.45/0.38, respectively.
It thus indicates that a small fraction of the ICM is locally heated by
the cluster merger.

If we assume that the outer slope for the hot component is $\beta=2$, 
the normalized temperatures, $T_0$, in the western and eastern hot components become
lower by $\sim20$ per cent and by $\sim6$ per cent, respectively. 
Since the other parameters are also accordingly changed, the cylindrical temperatures of all the components, the western hot component and the eastern hot component within $300$ kpc, change by only $-2,-10$ and $-2$ per cent, respectively, less than the $1\sigma$ uncertainty. 

We compare the Akaike information criterion (AIC) and Bayesian information criterion (BIC) of the multi-component analysis with those of the single gNFW model ($g_{\rm gNFW}^{C}$) analysis using the two dimensional $y$ and $S_X$ images. 
The differences  
are $\Delta{\rm AIC}={\rm AIC}_{\rm multi}-{\rm AIC}_{\rm gFNW}=-709<0$ and $\Delta{\rm BIC}={\rm BIC}_{\rm multi}-{\rm BIC}_{\rm gFNW}=-656<0$, respectively. Therefore, the additional two components based on visual inspections improve the modelling.

A future joint analysis of the high angular resolution, MUSTANG-2 data with the small FOV and the mid angular resolution SZE data  (e.g. AdvACT;$\sim1^\prime$) covering larger area would be helpful to constrain well the outer slopes of the gas temperature and density distributions and improve the parameter degeneracy caused by the transfer function.
Although this study assumes the gNFW model (or $\beta$ model), 
the shape of cluster merger shock surface could be a paraboloid-like feature and the asymmetric gas distribution model  
would be powerful in a future analysis. The geometrical assumption is also important for the deprojection and the volume-filling factor of each component in the three dimensional space.

To measure cluster richnesses of the western and eastern components,
we split into two galaxy components by right ascension of a bright
galaxy ($\alpha=38.3991$) around the cool core. 
Based on the S16A catalogue \citep{HSC1stDR,Oguri18}, the richness of the western
and eastern galaxies are 25 and 14, respectively. The total stellar masses
are $M_*^{\rm W}=3\times10^{12}M_\odot$ and $M_*^{\rm
E}=10^{12}M_\odot$.  
When we use the S18A catalogue \citep{HSC2ndDR},
the result does not significantly change; $M_*^{\rm
W}=3\times10^{12}M_\odot$ and $M_*^{\rm E}=2\times10^{12}M_\odot$.
Since cluster richness and halo mass are positively correlated
\citep{Okabe19}, the western galaxy component is likely to be the main cluster.

We measure weak-lensing mass using a tangential shear profile (Figure
\ref{fig:WL}).  
and multi-component masses by a two-dimensional weak-lensing analysis 
using a two-dimensional shear pattern \citep{Okabe11,Okabe15b,Medezinski16}. 
Since the concentration parameters of the two halos and the mass of the
subcluster are ill-constrained (because of a small number of background
galaxies), we assume the mass-concentration relation in \citet{Diemer15}. 
The central positions of the two halos are treated as free parameters in
a similar way as the two-dimensional SZE and X-ray analysis.

We first fit with uniform priors and obtain
$M_{200}^{W}=1.28_{-0.65}^{+1.17}\times10^{14}h_{70}^{-1}M_\odot$
and $M_{200}^{E}<0.75\times10^{14}h_{70}^{-1}M_\odot$.
This indicates that the western component is the main cluster, consistent
with its higher richness. 
The central positions determined by WL analysis, shown in Table \ref{tab:center}, 
coincide with BCG positions of the two galaxy components. 
Since we give an upper limit of the mass of the eastern component, 
we repeat fitting with fixed centers at best-fit positions.
The individual halo masses are $M_{200}^{W}=1.54_{-0.66}^{+1.29}\times10^{14}h_{70}^{-1}M_\odot$
and $M_{200}^{E}=0.90_{-0.40}^{+0.99}\times10^{14}h_{70}^{-1}M_\odot$.
The best-fit mass ratio between the subcluster and the total cluster is roughly $2:3$.
Considering the error matrix, the mass ratio is $0.54_{-0.28}^{+0.93}$.
The resulting matter distribution is shown by the red lines in Figure \ref{fig:HSC3_maps}. 
Even when we fit the entire shear pattern with a single NFW model, 
the best-fit center is consistent with the western BCG position rather than the eastern BCG position, 
indicating that the western component is the main cluster.

\subsubsection{Merger dynamics} \label{subsec:hsc3_dis}

The offsets in the $y$ and $S_X$ distributions (Figure \ref{fig:maps} and \ref{fig:HSC3_maps}) 
show violent merger phenomena, indicating that the merger is at the phase just after core-passage.
Even so, we first discuss whether the possibility of a pre-merger phase can be ruled out or not.
The expected masses given the measured temperatures of $>20$ keV associated with the eastern and western mass structures are $M_{500}^{E}=4.5_{-1.2}^{+1.4}\times10^{15}h_{70}^{-1}M_\odot$ and
$M_{500}^{W}=7.9_{-2.5}^{+2.9}\times10^{15}h_{70}^{-1}M_\odot$, respectively. 
We assume that the clusters follow the mass-temperature scaling relation \citep[][; see also Sec. \ref{sub:scaling}]{2016A&A...592A...4L}. 
The numbers of clusters with masses higher than the estimated masses 
are expected to be $N_{\rm clu}^{\rm E}<\mathcal{O}(10^{-5})$ and $N_{\rm clu}^{\rm W}<\mathcal{O}(10^{-7})$ in the overlapped footprint of the HSC-SSP and XXL surveys of $24.1$ deg$^2$ and the redshift slice of $|\Delta z|<0.1$ around the cluster redshift. The upper limits are constrained by the lower uncertainty bounds of the temperatures.
We here assume the mass function of \citet{Tinker08} with {\it Planck} cosmology \citep{Planck18}. 
Therefore, the hypothesis that the cluster is at the pre-merger is unlikely from the point of view of the concordance cosmology. When we include the cool component, $N_{\rm clu}^{\rm W}\simlt 0.04$ and $N_{\rm clu}^{\rm E}\simlt 0.7$ and thus the result does not change.

We estimate the merger time-scale after core passage. 
Since dark matter is likely composed of collisionless particles, 
the WL-determined central position can constrain the merger timescale \citep{Okabe11,Okabe15b}.
From the two-dimensional WL analysis (Sec \ref{subsec:hsc3}), assuming a point mass approximation, 
the projected distances between the two clusters and between the subhalo
and the center-of-mass are estimated to be $\sim580$ kpc and $\sim370$ kpc, respectively.
They are lower than the three-dimensional overdensity radii, $r_{200}=1212.3_{-227.1}^{+219.2}\,{\rm kpc}$ and $r_{500}=782.3_{-191.0}^{+149.9}\,{\rm kpc}$ which are derived by 2D WL analysis (Table \ref{tab:wlmass} and Sec \ref{subsec:wl}).
\citet{Okabe19} found a segregation in a probability density function (PDF) of collision velocities
of the optically-defined merging clusters and well-known merging clusters with diffuse synchrotron radio emissions.
The peak velocities of the optically-defined merging clusters and  merging clusters with
diffuse radio emissions are $\sim1000\,{\rm km\,s^{-1}}$ and $\sim2500\,{\rm km\,s^{-1}}$, respectively.
The former and latter cases give the merger timescales after core 
passage of $\sim0.36$ Gyr and $\sim0.14$ Gyr, respectively.

We also estimate the merger velocity from the deviation from a mass-temperature scaling relation \citep{2016A&A...592A...4L}. 
Assuming that the gas temperature before the merger follows the mass-temperature scaling-relation (see details in Sec. \ref{sub:scaling}),
we infer a pre-merger temperature from the eastern WL mass, and derive a Mach number, ${\mathcal M}^{E}=6.2_{-1.8}^{+1.9}$, from the ratio between the eastern hot and pre-merger temperatures.
The resulting collision velocity is $v^{E}=3940_{-457}^{+386}\,{\rm km\,s^{-1}}$ with the sound velocity of $c_s=636_{-99}^{+155}\,{\rm km\,s^{-1}}$ expected by the WL mass. Thus, 
the merger time-scale is $0.16_{-0.05}^{+0.07}$ Gyr,assuming a one-dimensional velocity of $v^{E}/\sqrt{3}$.
The difference between spectroscopic redshifts \citep{2018A&A...620A...7G} of two luminous galaxy associated with the western and eastern structure gives a relative velocity of $\delta v_{\rm l.o.s}\sim1700\,{\rm km\,s^{-1}}$. 
If the subhalo is not moving along the Dec direction, the time-scale is 
$\sim0.1$ Gyr with $((v^E)^2-\delta v_{\rm l.o.s}^2)^{1/2}\sim3500\,{\rm km\,s^{-1}}$.
Even when we use another mass-temperature scaling relation \citep{Umetsu19},  
the estimated time-scale does not significantly change. 
The estimated collision velocity is acceptable in cosmological simulations \citep{Bouillot15} and observations \citep{Okabe19}, 
but is at high end of the PDF of collision velocities for the optically-defined merging clusters 
and similar to those of merging clusters with diffuse radio emission.

\subsubsection{Comparison with numerical simulations} \label{subsubsec:HSC3sim}

\begin{figure*}
 \begin{center}
 \includegraphics[width=\hsize]{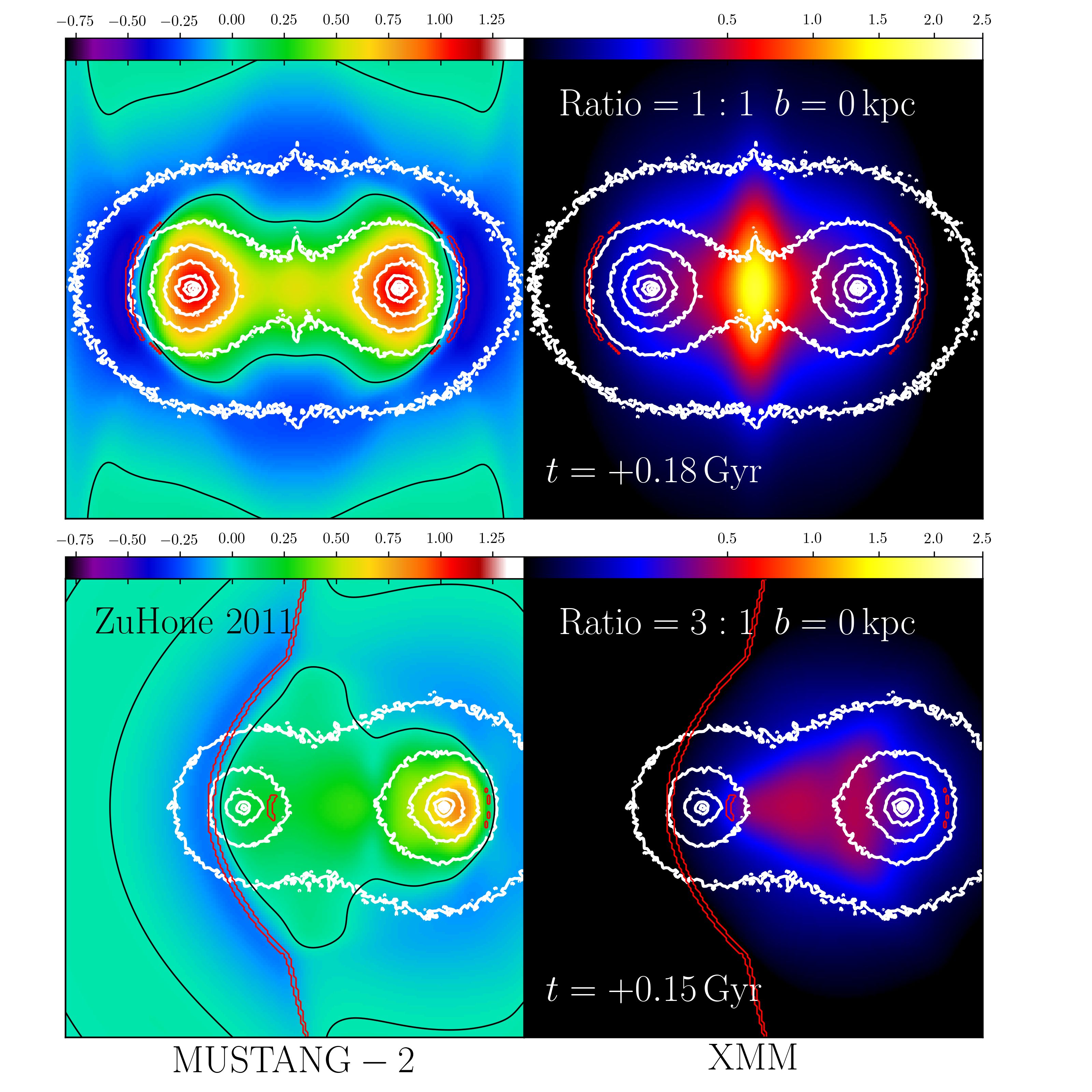} 
 \end{center}
 \caption{Simulated MUSTANG-2 (left) and {\it XMM-Newton} (right) images ($4$ arcmin $\times$ $4$ arcmin) using numerical simulations of \citet{ZuHone11} publicly available through \citet{ZuHone18}, taking into account the PSFs and the transfer function. The top panels are the edge-on view of a binary merger with equal-mass ratio and impact parameter $b=0$ kpc at $0.18$ Gyr after first core-passage, which is similar to the case of the major merger, HSC J023336-053022. The white contours denote projected mass density. The black contours in the left panels denote the lines for $\tilde{y}=0$. The red lines are shock features appearing in the projected sky. The bottom panels show the edge-on view with a mass ratio of $3:1$ for visual comparison to the X-ray morphology. 
 }
 \label{fig:ZuHone}
 \end{figure*}

In order to visually interpret merger phenomena, we made simulated MUSTANG-2 and {\it XMM-Newton} images using the numerical simulations of \citet{ZuHone11} that are made publicly available through \citet{ZuHone18}. \citet{ZuHone11} computed $N$-body/hydrodynamic simulations of binary mergers with mass ratios of $1:1$, $1:3$ and $1:10$ with impact parameters of $0$, $500$, and $1000$ kpc, respectively. The main cluster mass is $6\times10^{14}M_\odot$, which is slightly higher than those of our samples.
We set simulated clusters at the observed redshifts of the merging cluster and then pick out a simulated cluster so that their peak separations between $y$ and $S_X$ maps measured from $90$ deg from the merger plane  
are similar to our observations. The dynamical time and mass ratio are close to our cases. 
The simulated images, convolved with the PSFs and the transfer function, are shown in Figure \ref{fig:ZuHone}. 
The X-ray merger positions are rotated to be along the $x$-axis. Figure \ref{fig:ZuHone} shows the $\tilde{y}$ and $\tilde{S}_X$ maps of an equal-mass merger with zero impact parameter. The simulated images of the $y$ parameter clearly show a double-peaked structure. Bow-shock fronts (red lines) are located at outer-edges of the two hot components. These bow shocks appear as weak changes in the $y$ map. We find that it is difficult to constrain the Mach number from the $y$ distribution.  The $y$ peak regions are heated by input cluster merger shocks ($\mathcal{M}\sim4.7$). The X-ray peak of the simulated image is in-between
the double peak $y$ structure and its morphology is highly elongated and perpendicular to the merger axis. This feature is slightly different from the best-fit results and the observed features in that the X-ray peak is closer to the eastern component and the X-ray morphology is not elongated perpendicular to the axis between the two hot components. 
The difference could be caused by differences in mass ratio, viewing angle or both. The weak-lensing analysis indicates that the western component is the main cluster, and thus the X-ray peak could be shifted to the east because the X-ray core is moving from the west to the east. For the visual purpose, we plot the simulated images of the merger with the mass ratio of $3:1$ in the bottom panel of Figure \ref{fig:ZuHone}, in the similar way as the top panels. The $y$ peak is associated with the main cluster and the stripping X-ray core is elongated along the merger axis, which is is similar to the observed feature. 
Since the current data cannot constrain the line-of-sight structure of the merger, there remains the uncertainty in the viewing angle.
Although it is difficult to find a perfectly matched simulation, the visual comparison helps us understand the plausible 
configuration that created the observed double-peak $y$, the single X-ray peak, and the mass distribution. Future theoretical studies using the observational parameters would be better for understanding the details.

In the simulation of the equal-mass merger, the temperature in these regions reaches $\sim25$ keV from the initial value $\sim5$ keV, 
supporting the presence of the hot-component in the major merger. As for the mass ratio of $3:1$ case, the temperature of the main cluster becomes $\sim18$ keV from $\sim5$ keV and the temperature around the observed $y$ peak of the subcluster increases from $\sim2.5$ keV to $\sim15$ keV. We note that a prominent shock with $\sim20$ keV appears around the negative $y$ value at the east region from the subcluster. The asymmetric temperature distribution is similar to the joint analysis results. 
A quantitative discussion of the two-dimensional comparison is very difficult because the observation cannot constrain the line-of-sight information and the public library of the stimulated images does not cover all the orientation angle. Therefore, the quantitative discussion of the merger boost in Sec. \ref{sub:scaling} uses the integrated quantities through mass observable scaling relations.

We stress that the joint SZE and X-ray analysis using high-angular resolution data provides a powerful means to extract the
multiple gas structures and uncover hot components with $k_BT\sim20-30$ keV. Such spatially-resolved high temperature measurements are difficult with current X-ray satellites.
Future {\it Chandra} observations will test our interpretation by detecting the X-ray surface brightness jumps as shown in red curves in the upper panel of Figure \ref{fig:ZuHone}, and will independently estimate the Mach number.

\subsubsection{Absence of diffuse radio emission} \label{subsec:radio_stat}

The major-merger candidates defined by \citet{Okabe19} evenly cover a new parameter region of mass, mass ratio, and dynamical stages. One main difference from the well-known mergers with diffuse synchrotron radio emissions, the so-called radio halos and relics, is that the PDF of the merger velocity of the optically defined mergers is shifted to lower speeds. \citet{Okabe19} found no diffuse radio emission in $\sim190$ merging clusters by both visual inspection and spectral index maps using the NVSS \citep[1.4 GHz;][]{Condon98NVSS} and TGSS \citep[147.5 MHz;][]{Intema17TGSS} archival data because 
it is very difficult to discriminate between radio lobes and halos/relics. 
Combined with a presence of hot regions revealed by the joint SZE and X-ray analysis, it is a good opportunity to search again for radio halos/relics in the major-merging cluster.

\begin{figure*}
 \begin{center}
 \includegraphics[width=\hsize]{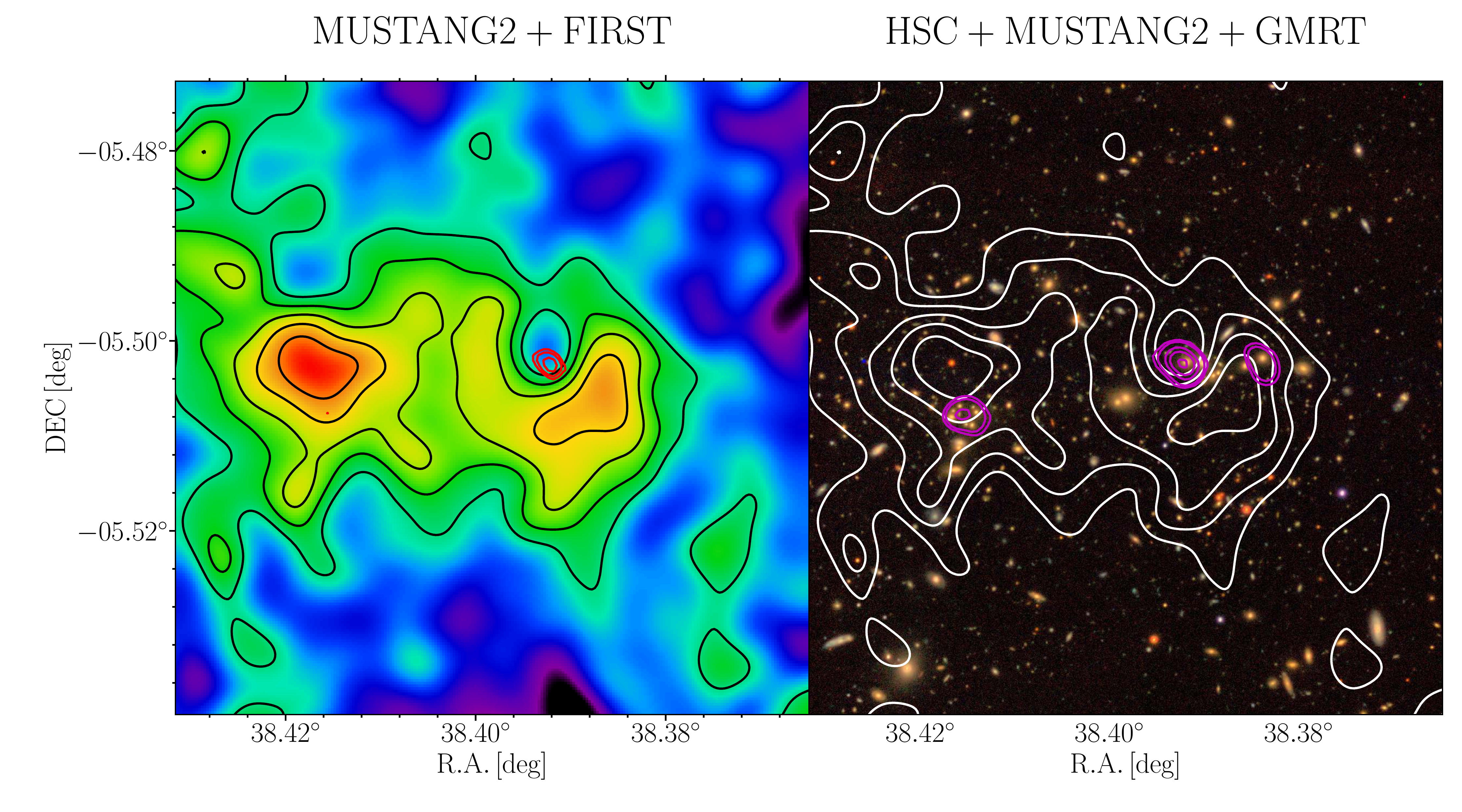}
 \end{center}
 \caption{{\it Left:} the $y$ map for HSC J023336-053022 overlaid with the FIRST (red) contours ($1.4$ GHz). The contour levels are $4, 8, 16,{\rm and}\,32\times\sigma_{\rm rms}$, where $\sigma_{\rm rms}^{\rm FIRST}=1.5\times10^{-4}\,{\rm Jy\,beam^{-1}}$.
 {\it Right:} HSC-SSP optical $riz$-color image overlaid with the $y$ map contours (Figure \ref{fig:maps}) and the GMRT contours (magenta). The contour levels are $4, 8, 32,64,128,{\rm and}\,512\times\sigma_{\rm rms}$, where $\sigma_{\rm rms}^{\rm GMRT}=6\times10^{-5}\,{\rm Jy\,beam^{-1}}$. The three low-frequency radio sources are associated with member galaxies. 
 The black contours in the left panel and the white contours in the right panel show the distribution of the Compton $y$ parameter (same levels as in Figure\ref{fig:maps}, left panel, second row).
 }
 \label{fig:HSC3_NVSS}
\end{figure*}

We here use the FIRST \citep[1.4GHz;][]{Helfand15} data and the GMRT data of the XXL Survey \citep[610MHz;][]{Smolcic18}. The beam sizes for the FIRST and GMRT data are $5\arcsec$ and $6.5\arcsec$, respectively.
In the FIRST image we find a point source (FIRST J023334.1-053008)
associated with the BCG at spectroscopic redshift $z = 0.4319$ in the western component of the cluster (Figure \ref{fig:HSC3_NVSS}).
The SZE flux around the radio source is suppressed by compact radio source contamination.
The radio contamination depends on the spectral index at high frequency \citep[e.g.][]{2009ApJ...694..992L,Gralla14}.
We note that the radio source region is excluded in the modeling fitting.
At 610 MHz \citep{Smolcic18} there are three sources (XXL-GMRT J023339.6-053028,
XXL-GMRT J023334.1-053008, and XXL-GMRT J023332.1-053008).
One of them coincides with the source listed in the FIRST catalogue;
the three radio sources are associated with cluster members (right panel of Figure \ref{fig:HSC3_NVSS}).
The radio AGN activity might be recently triggered by the cluster merger ($\sim0.15$ Gyr) because the typical timescale of AGN activities is short $0.01-0.1$ Gyr \citep{Soker16}.

We do not find evidence of diffuse radio sources, 
though the SZE and X-ray data exhibit the presence of hot gas components triggered by the violent merger.
This conflicts with the other cases of CIZA J2242.8+5301 \citep{vanWeeren10} hosting a prominent filamentary radio relic, Abell 2146 \citep{Hlavacek-Larrondo18} with double relics, and the Bullet cluster \citep{Markevitch02} with a radio halo.
The estimated merger timescale is comparable to $\sim 0.1-0.2$ Gyr for the Bullet cluster \citep{Markevitch02}, and $0.2-0.3$ Gyr for CIZA J2242.8+5301 and Abell 2146. One of differences is cluster mass. CIZA J2242.8+5301 \citep{Okabe15b}, Abell 2146 \citep{King16}, and the Bullet cluster \citep{Bardac06} are all very massive  
($M_{200}\sim10^{15}h_{70}^{-1}M_\odot$), while the mass of this cluster is only one-tenth that ($M_{200}\sim10^{14}h_{70}^{-1}M_\odot$). Thus, the release of gravitational energy differs by more than one order of magnitude. 
\citet{Cassano13} have shown that the $k$-corrected radio power at $1.4$ GHz for diffuse radio emissions have a strong mass dependence; $P_{1.4\rm GHz}\propto M_{500}^{p}$ with $p=3.77\pm0.57$. The upper limit of $P_{\rm FIRST}\sim3\times10^{23}\,{\rm W\,Hz^{-1}}$ 
with $3\sigma$ level is higher than $0.5\times10^{23}\,{\rm W\,Hz^{-1}}$ expected by the relation of \citet{Cassano13} using the WL mass, and thus we do not rule out a possibility to detect diffuse radio emission by future observations with $1.4$ GHz.
Assuming the spectral index $\alpha=1.3$ \citep{Cassano13}, the expected radio powers at the GMRT and TGSS frequencies are $\sim2\times10^{23}\,{\rm W\,Hz^{-1}}$ and $\sim10^{24}\,{\rm W\,Hz^{-1}}$, respectively. The $k$-corrected powers of the $3\sigma$ rms noise levels of the GMRT and TGSS observations are $\sim10^{23}\,{\rm W\,Hz^{-1}}$ and $\sim4\times10^{24}\,{\rm W\,Hz^{-1}}$, respectively. Therefore, expected diffuse radio sources \citep{Cassano13} are not detected by the GMRT observation \citep{Smolcic18}.

\begin{figure*}
 \begin{center}
 \includegraphics[width=\hsize]{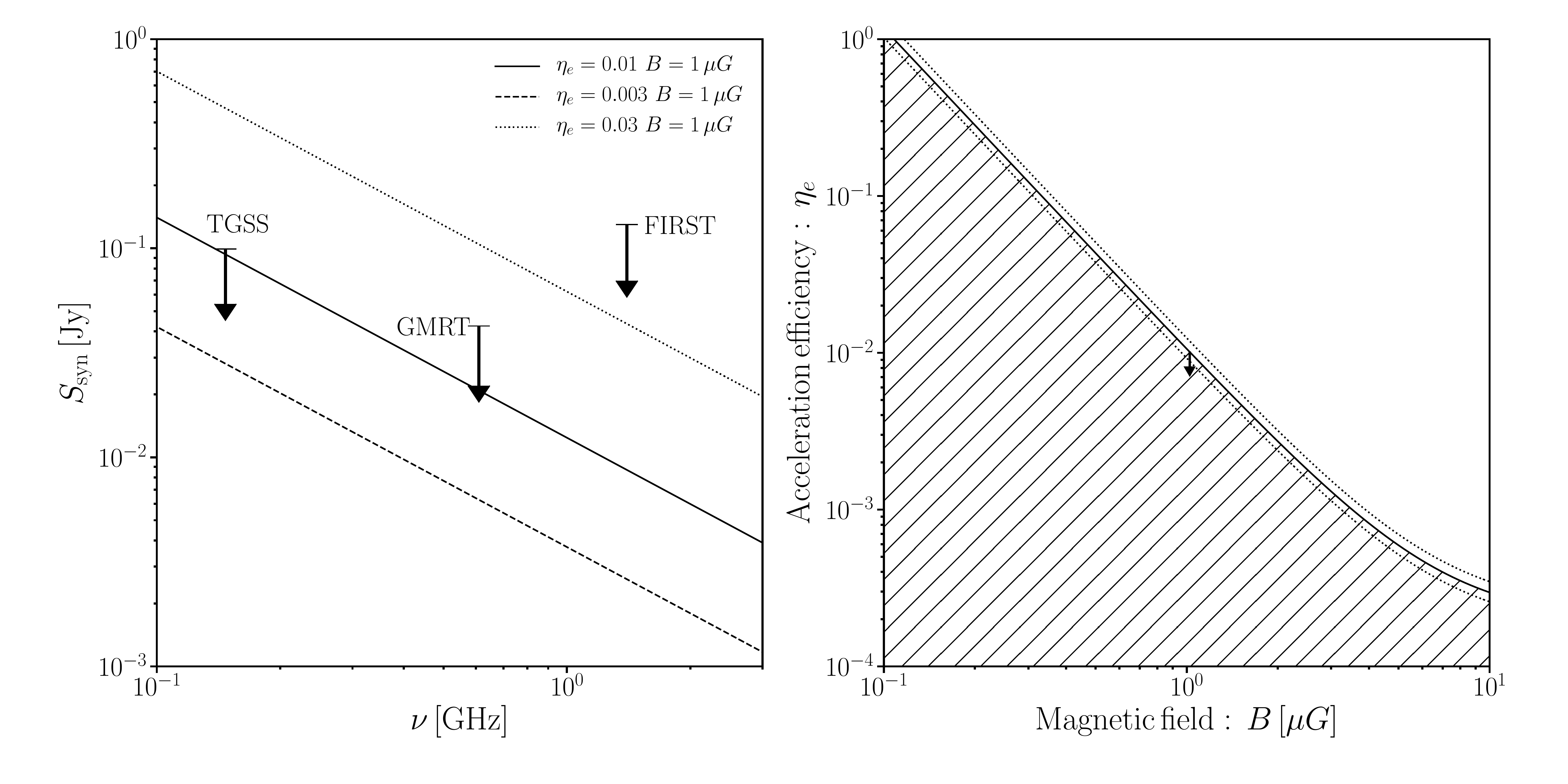}
 \end{center}
 \caption{{\it Left}: synchrotron radio flux density. Radio upper limits are given by $3\sigma_{\rm rms}$ levels. The solid, dashed, and dotted lines are the radio flux density expected by the multi-component joint SZE and X-ray analysis with different acceleration efficiency, $\eta_e$. {\it Right}: the upper limits of the acceleration efficiency as a function of magnetic strength. The dotted lines denote the errors computed from the uncertainty of $P_e$.
 Non-detection of diffuse radio emission in the hot gas component gives the low acceleration efficiency of $<10^{-2}$ at $B>1\,\mu G$. }
 \label{fig:DSA}
\end{figure*}

Thanks to the multi-component gas analysis and high-angular resolution synchrotron radio data, 
we can constrain the upper limit of the particle acceleration efficiency $\eta_e$ via the first-order Fermi acceleration.   
We assume that a merger shock created the hot gas component and simultaneously injected cosmic-ray electrons. 
An injection spectra at downstream of the shock is defined by $Q_e=Q_{e,0}\gamma^{-p}$, where $p$ is the particle index, $\gamma$ is the Lorentz factor, and  $Q_{e,0}$ is the injection normalization. Assuming a diffusive shock acceleration  \citep[DSA;][]{Drury83}, the particle index is given by $p=2(\mathcal{M}^2+1)/(\mathcal{M}^2-1)$. 
Cluster merger shocks convert kinetic energy to thermal energy. The total thermal energy of electrons heated by the merger shock can be computed from the pressure of the hot gas component measured by the multi-component analysis, $E_{{\rm th},e}=\int (3/2) P_e dV$.  
When we constrain the upper limit of the acceleration efficiency, we assume that all the kinetic energy is converted to the thermal energy. 
The injection energy is described by $\int_{\gamma_{\rm min}} Q_{e,0} \gamma^{-p} E d\gamma = \eta_e E_{\rm th,e}/\Delta t$ using a constant efficiency factor $\eta_e$, where $\Delta t$ is the merger time scale and $E$ is the energy per cosmic ray electron.
We assume $\gamma_{\rm min}=2$. Given the injection spectra and the inverse-Compton and synchrotron energy losses, the steady state spectra has a form of $N(\gamma) \propto \gamma^{-(p+1)}$ \citep{Sarazin99}. Then, the synchrotron emissivity is $dL_{\rm syn}/d\nu \propto B^{(p+2)/2}\nu^{-p/2}$, where $B$ is the magnetic field strength and $\nu$ is a frequency. 
A cooling time scale for cosmic-ray electrons emitting at the GMRT frequency is $0.04$ Gyr assuming a typical magnetic field $B=1\,\mu G$ and shorter than the merger time scale $\sim0.15$ Gyr. We assume Mach number ${\mathcal M}=6.2$ and $\Delta t=0.16$ Gyr (Sec. \ref{subsec:hsc3_dis}). Even when we change by $\mathcal{M}=\pm2$ or $\Delta t=\pm0.05$, the results do not significantly change. 
We consider the cylindrical volume of the eastern hot component within the central $1$ arcmin.
The left panel of Figure \ref{fig:DSA} shows a comparison of the $1\sigma$ upper limits of the radio observations and the synchrotron flux density expected by $0.3,1$ and $3$ percent acceleration efficiency at $B=1\,\mu G$. The upper limits of the radio observations are estimated by the $3\sigma$ level. The acceleration efficiency should be less than $\mathcal{O}(10^{-2})$, which is lower than typical values used in theoretical models of galaxy clusters \citep[e.g.][]{Kang13,Vazza16}
 and strong shocks in supernova remnants \citep[e.g.][]{Miniati01}. The right panel of Figure \ref{fig:DSA} shows the upper limit on $\eta_e$ as a function of $B$. Since magnetic strengths of order of ${\mathcal O}(1)\,\mu G$ \citep[e.g.][]{Vogt03} or higher in radio relics \citep{Nakazawa09,vanWeeren11b}, the acceleration efficiency is less than sub percent and thus is very inefficient.  Similar results are recently reported in well-known radio-relic clusters \citep{Botteon19}.

Another possibility for particle acceleration is re-acceleration of supra-thermal electrons which are possibly ejected from radio AGN sources. If the AGNs were not triggered before merger shocks was sweeping, the scenario would not conflict with non-detection of diffuse radio sources, given their short time cycles of $0.01-0.1$ Gyr \citep{Soker16}.

\subsection{HSC J021056-061154}\label{subsec:hsc1}

\subsubsection{Joint analysis}

HSC J021056-061154 exhibits a complex distribution of member galaxies, comprising 
a central component elongated along the west-east direction and
southern substructures (the bottom-right panel of Figure \ref{fig:maps}).
The central X-ray surface distribution is elongated along the
east-west direction, similar to the central galaxy distribution.
The X-ray peak position coincides with the BCG. 
The X-ray flux is much lower than that of the other clusters.
The bottom-left panel of Figure \ref{fig:maps} shows an anisotropic distribution of the Compton
$y$ parameter. The eastern and western regions 
show negative and positive $y$ values, respectively.
The high $y$ region is elongated along the direction perpendicular to the major axis of the X-ray core. 
Taking into account the angular transfer function, this feature indicates
that the peak position of the $y$ parameter, that is, the hot region, is offset $45$ arcsec to the west of the BCG.

\begin{figure*}
\begin{center}
 \includegraphics[width=\hsize]{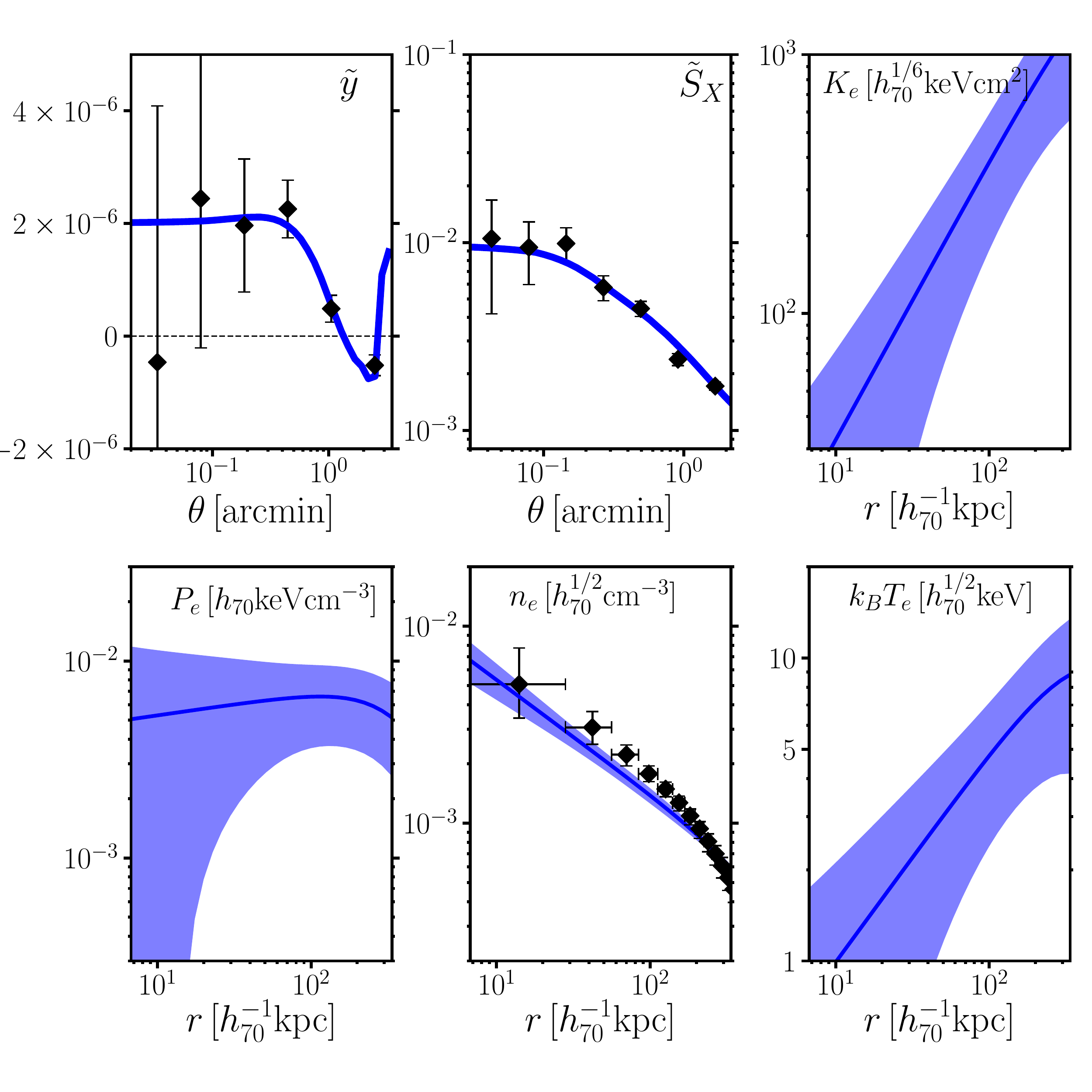}
\end{center}
\caption{The same figure as Figure \ref{fig:HSC2_model}, but for
 HSC J021056-061154.
   The signal-to-noise ratios of the $y$ and $S_X$ profiles are
   $\sim5\sigma$ and $\sim16\sigma$, respectively.
The increase in the $\tilde{y}_m$ profile at large radii is caused by the transfer function.} 
\label{fig:HSC1_model}
\end{figure*}

Figure \ref{fig:HSC1_model} shows the $y$ and $S_X$ radial profiles.
The signal-to-noise ratio of the $y$ profile is much lower than
those of the other clusters. Since it is difficult to constrain the
outer-slope, we adopt a Gaussian prior $\beta_T=0.67\pm0.30$ for the modeling.
The temperature is an increasing function out to $\sim500$ kpc. 
When we refit with a model with constant temperature, the model is disfavored to describe the $\tilde{y}$ profile. 
The measurement uncertainty of the temperature is larger than that of
the electron number density because of the low signal-to-noise ratio of the $y$ profile.

\begin{figure*}
 \begin{center}
 \includegraphics[width=\hsize]{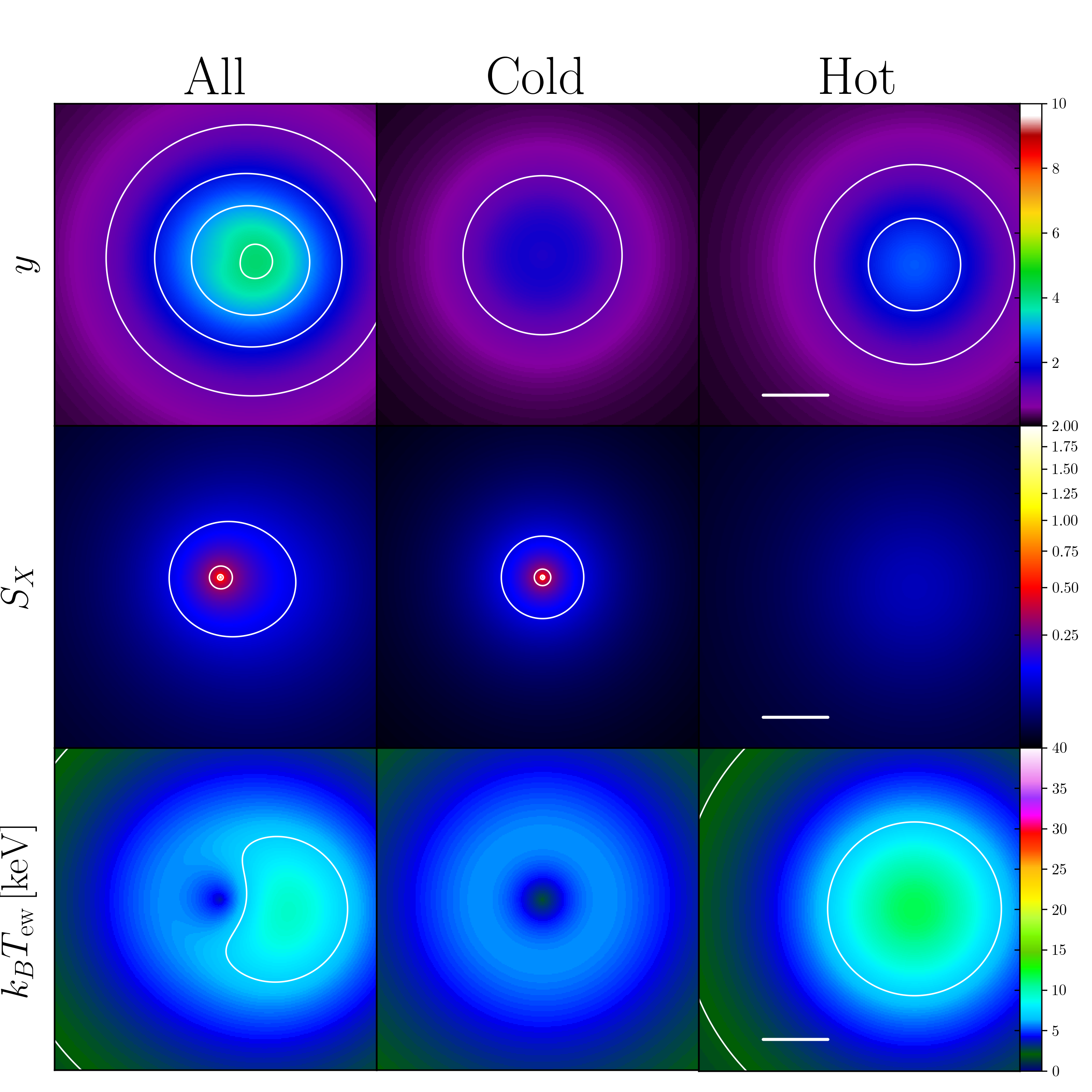}
 \end{center}
   \caption{Model maps for HSC J021056-061154. Colors, contours and lines are
 the same as those in Figure \ref{fig:HSC3_maps}.
   }
 \label{fig:HSC1_maps}
\end{figure*}

We next perform the two-dimensional analysis in a similar way to Sec \ref{subsec:hsc3}.
Since the signal-to-noise ratio is low, we use the binned images with
pixel size of $20$ arcsec.
In the modeling, we consider the eastern and western components to describe the main cluster and
the offset hot component, respectively. We adopt the spherically symmetric gNFW model for the
eastern main component and the spherically symmetric $\beta$ model for the western hot
component. We fixed $\beta_n^{\rm W}=\beta_T^{\rm W}=1$ to describe the localised hot component
in a similar way to Sec \ref{subsec:hsc3}. 
As shown in the lower-middle panel of Figure \ref{fig:HSC1_maps},
the main cluster has a cool core in which the X-ray-like emission-weighted temperature
changes from $\sim 3$ keV to $\sim 6$ keV as the radius increases. 
The X-ray-like emission-weighted temperature for the main component within $300$
kpc from the XXL center is $\sim5$ keV.
The features of the main component 
are not significantly different from  the results of the one-dimensional analysis. 
The western temperature reaches $\sim11$ keV.
The best-fit centers are shown in Table \ref{tab:center}.
The superposed temperature map (the lower-left panel of Figure
\ref{fig:HSC1_maps}) exhibits an anisotropic distribution. 
A small gas fraction of the western region has a high temperature, while
the eastern region is not affected by the high temperature component.
The feature indicates that the cluster is likely to be a binary cluster prior to merger.
The superposed X-ray surface brightness distribution is elongated along the east-west direction.

Since the cluster is located at the edge of the HSC-SSP survey field, the shape
catalogue does not fully cover the entire region of the
cluster because of the full-colour and full-depth requirement (Sec. \ref{subsec:wl}). 
Therefore, the measurement uncertainty of the WL mass is very
large (the right panel of Figure \ref{fig:WL} and Table \ref{tab:wlmass}). 
We cannot carry out a multi-component WL analysis because of this limitation.
A comparison of weak-lensing and HE-derived masses is discussed in Sec \ref{sub:mass_com}.

\subsubsection{Pre-minor merger physics} \label{subsec:hsc1_dis}

The $y$ map of this cluster (Figure \ref{fig:maps}) is chaotic in structure, with a peak offset to the west of X-ray peak.  
In particular, the major axis of the $y$ map is elongated perpendicular to that of the X-ray core.
We adopt an elliptical model to describe the gas structure.

\begin{figure*}
 \begin{center}
 \includegraphics[width=\hsize]{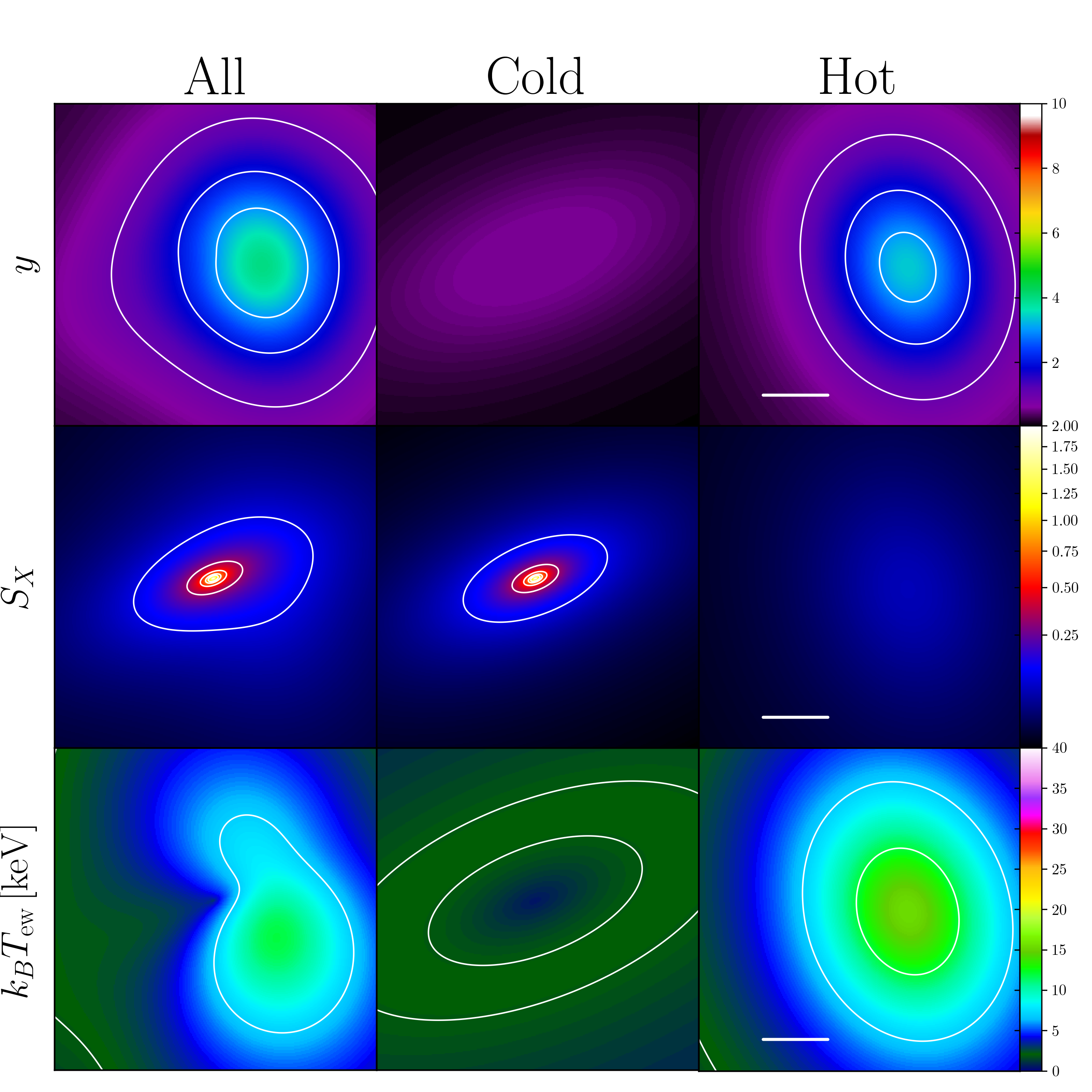}
 \end{center}
   \caption{Elliptical model maps for HSC J021056-061154. Colors and contours are
 the same as those in Figure \ref{fig:HSC3_maps}.
   }
 \label{fig:HSC1_maps_ell}
\end{figure*}

The ellipticity of the the projected distribution of the electron number density and temperature 
is defined by $e=1-b/a$, where $a$ and $b$ are the major and minor axes of gas properties following \citet{Oguri10b}.   
We introduce an orientation angle of the major axis, $\phi_e$, measured
from the north to the east. 
The distance of an iso-contour from the centre is given by 
\begin{eqnarray}
r&=&(x'^2/(1-e)+(1-e)y'^2)^{1/2}, \\
x'&=&x \cos \phi_e+y\sin \phi_e, \nonumber \\
y'&=&-x \sin \phi_e+y\cos \phi_e, \nonumber 
\end{eqnarray}
where $x'$ and $y'$ are the rotated coordinates.
The best-fit orientation angles of the eastern and western components
are $\phi_e^{\rm E}=112_{-8}^{+7}$ deg and $\phi_e^{\rm
W}=17_{-10}^{+9}$ deg, respectively.
The elongation of the hot gas in the projected sky is almost
perpendicular to that of the cold gas. This suggests that the hot region
has been heated by a merger shock triggered by the infalling dense gas.
The ellipticities are $e^{\rm E}=0.53_{-0.13}^{+0.10}$ and $e^{\rm W}=0.29_{-0.18}^{+0.16}$. 
The best-fit model map is shown in Figure \ref{fig:HSC1_maps_ell}. 
Since the eastern region is not yet heated, the cluster is interpreted to be in a pre-merger phase.

We did not consider the elongation along the line of sight. Another possibility is that an ellipticity of the temperature distribution differs from that of the density distribution. If the temperature distribution were more elongated on the sky plane than the gas distribution, the $y$ parameter would be lower than expectations from the spherical model. A full 3D reconstruction of gas properties is left for future studies to address the geometrical assumption.

A group of several member galaxies in the west end of the central overdensity region is associated with the western hot region. The stellar mass of this group, $M_*^{\rm W}=3\times10^{11}M_\odot$, is lower than the total stellar mass of $M_*=4\times10^{12}M_\odot$ found using the S16A catalogue. When we use the S18A catalogue, our estimates become 
$M_*^{\rm W}=7\times10^{11}M_\odot$ and $M_*=4\times10^{12}M_\odot$. 
Therefore, the substructure mass triggering the hot region is one-tenth or less of the mass of the main cluster.
The configuration and mass suggest that this system is likely to be a minor merger before core passage. Assuming a collision velocity of $1000\,{\rm km\,s^{-1}}$ \citep{Okabe19}, the spatial separation of the two galaxy concentrations implies that the cluster is $\sim0.3\,{\rm Gyr} (\sec \theta/1)$  before core passage.  Here, $\theta$ is an inclination angle between the merger axis and the sky plane. 
The ratio between the projected temperatures of the hot and cold components gives the Mach number ${\mathcal M}\sim4$ and the one-dimensional collision velocity $v/\sqrt{3}\sim1500\,{\rm km\,s^{-1}}$ and the estimated time-scale, $\sim 0.2\,{\rm Gyr}$, is not significantly changed. 

\subsubsection{Comparison with numerical simulations}

\begin{figure*}
 \begin{center}
 \includegraphics[width=\hsize]{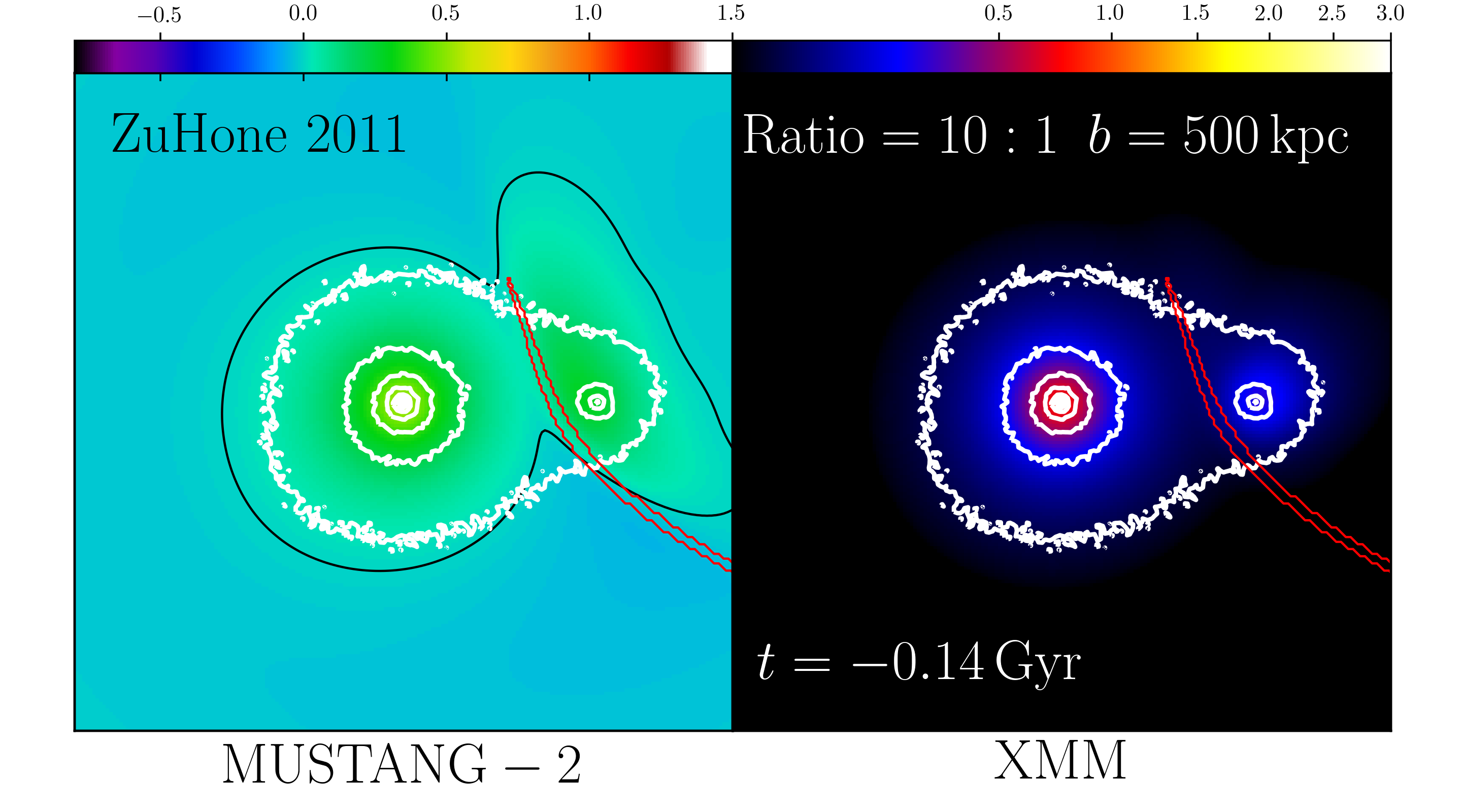} 
 \end{center}
 \caption{Simulated MUSTANG-2 (left) and {\it XMM-Newton} (right) images
 ($4$ arcmin $\times$ $4$ arcmin) using numerical simulations of
 \citet{ZuHone11} publicly available through \citet{ZuHone18}, taking
 into account the PSFs and the transfer function. The panels are the edge-on view of
 the merger with $1:10$ mass ratio and $b=500$ kpc at $0.14$ Gyr before
 first core-passage, which is a similar case to the minor merger, HSC
 J021056-061154. Contours are the same as those in Figure \ref{fig:ZuHone}.
 }
 \label{fig:ZuHone_HSC1}
 \end{figure*}

For visual purposes, we compute mock images using numerical simulations of \citet{ZuHone11} in a similar way of Sec. \ref{subsubsec:HSC3sim}. 
Figure \ref{fig:ZuHone_HSC1} shows an edge-on view of a simulated merger with a $1:10$ mass ratio and $b=500$ kpc at $t=0.14$ Gyr before first core-passage. 
The features such as the double-peaked X-ray surface brightness distribution and the elongated $y$ map associated with the western subcluster are found. The simulated $y$ map after processing with the transfer function, has negative values at $\theta\simgt 1$ arcmin. 
This morphology resembles the observation. 
The western component is heated by a merger shock (red lines). The simulated $y$ image shows a high value around the main cluster, while our observation does not detect such a feature, perhaps caused by differences in the assumptions of the cool core between simulations and observations, and/or a different main cluster mass. 
Although the details of the simulated image are not identical to our image, 
the characteristic properties of the precursor phase of a minor merger well represents the observation.  

\subsection{Comparison with SZE and X-ray images in previous observations}

We compare the SZE and X-ray imaging for the three clusters with those reported from previous observations.
As aforementioned, the high resolution mapping and their multi-wavelength analysis reveal that the gas properties are more or less disturbed regardless of the number of galaxy density peaks.
Thus, even minor mergers along with galaxies or small subhalos which are not significantly detected by the peak-finding method of the major-merger finder \citep{Okabe19} trigger perturbations in distributions of gas temperature and density. The observed properties are summarised as follows.
The single-peaked cluster shows the sloshing pattern in both the $y$ and $S_X$ distributions (Sec. \ref{subsec:hsc2_dis}). 
The double-peaked cluster with the mass ratio of $\sim2:3$ has the single X-ray core between the two $y$ peaks (Sec. \ref{subsec:hsc3}).
The part of the supercluster exhibits the elongated hot region perpendicular to the major axis of the cool gas core (Sec. \ref{subsec:hsc1_dis}).
Our multi-wavelength results of SZE, X-ray, WL and optical measurements agree with predictions of the imaging patterns in numerical simulations \citep{ZuHone10,ZuHone11,ZuHone18}; the single-peaked cluster, the double-peaked cluster, and the part of the supercluster are likely to be at sloshing, post-major, and pre-minor merging phases, respectively.

Sloshing features are found in many clusters through X-ray observations \cite[e.g.][]{Lagana10}.
Atacama Large Millimeter/submillimeter Array (ALMA) observation with $5$ arcsec resolution
found in RX J1347.5-1145 that $y$ and $S_X$ maps have a single peak, but there is a significant offset between the two peaks \citep{Kitayama16}. The MUSTANG observation \citep{Mason10,Korngut11} shows a reduced $y$ parameter around the X-ray peak due to strong contamination from the central, radio-loud AGN.
A significant residual $y$ is found in RX J1347.5-1145 when fitting and subtracting a profile with the SZ centroid fixed to be the same position as the X-ray centroid \citep{Plagge13,Ueda18,DiMascolo19b}.
However, \citet{DiMascolo19b} found no significant residuals, on scales from 5\arcsec\ to 10\arcmin, when jointly fitting an ellipsoidal pressure profile model with a floating centroid fit to the ALMA, ACA, Bolocam, and {\it Planck} data.

\citet{Kitayama04} have measured the temperature of a subcluster in a merging cluster, RX J1347.5-1145 combining the SZE data from Sub-millimeter Common User Bolometer Array (SCUBA) and  Nobeyama Bolometer Array (NOBA) and {\it Chandra} X-ray data and found a hot component with $\sim20$ keV, for the first time. Their finding of the hot temperature in the merging cluster is similar to those in HSC J023336-053022. 

The {\it Chandra} Observation of the Bullet cluster \citep{Markevitch02} shows a bow-shock ahead of a stripping cool core and X-ray emission from a main cluster of which the core is elongated along the direction perpendicular to the merger axis. \citet{Halverson09} detected a SZE signal associated with the main cluster using APEX-SZ with $1$ arcmin resolution. \citet{DiMascolo19} estimate a Mach number $\mathcal{M}=2.08\pm0.12$ of the bow-shock with ALMA and Atacama Compact (Morita) Array (ACA) assuming an instantaneous equilibration of the electron and ion temperatures. However, no high-resolution SZE observation covers the entire shock region to date.

The Arcminute Microkelvin Imager \citep[AMI;][]{Rumsey17} discovered in CIZA J2242.8+5301 an equatorial-shock-heated gas with elongation is perpendicular to both its merger axis and the major axis of the cool core. The relationship of the morphology of $y$ and $S_X$ maps is the same as found in the pre-minor merger, HSC J021056-061154.

\citet{Menanteau12} found that X-ray and $y$ distribution in the El Gordo cluster is offset similar to the case of RX J1347.5-1145. An ALMA observation relying on X-ray data for priors \citep{Bas16} constrained the Mach number $\mathcal{M}=2.4_{-0.6}^{+1.3}$ at the edge of X-ray surface brightness associated with radio relic \citep{Botteon16}.

Both stacked {\it Planck} SZE and {\it RASS} X-ray maps for low-redshift and massive merging clusters \citep{Okabe19} found that the $y$ and $S_X$ distributions at cluster outskirts are elongated along the direction perpendicular to the merger axis, though the X-ray main core is elongated along the merger axis.

An offset between X-ray and Compton $y$ parameter distributions seems common in merging clusters including literature and our sample. A perpendicular orientation angle between the major axes of high $y$ and X-ray core distributions is also found in some clusters. However, the double-peaked $y$ distribution has not yet been reported by observational studies.

\subsection{Temperature Comparison} \label{sec:Tcomp}

We compare the temperatures derived from the joint SZE and X-ray analyses
with X-ray temperatures from spectroscopic measurements \citep{2016A&A...592A...3G,2016A&A...592A...4L,2018A&A...620A...5A}. 
The XXL survey measured the X-ray temperature within $300h_{70}^{-1}$
kpc from their X-ray centers. We also use X-ray temperature derived from the deep
on-target observation for HSC J022146-034619 from the X-COP measurement (Sec \ref{subsec:hsc2}).

\begin{figure}
 \begin{center}
 \includegraphics[width=\hsize]{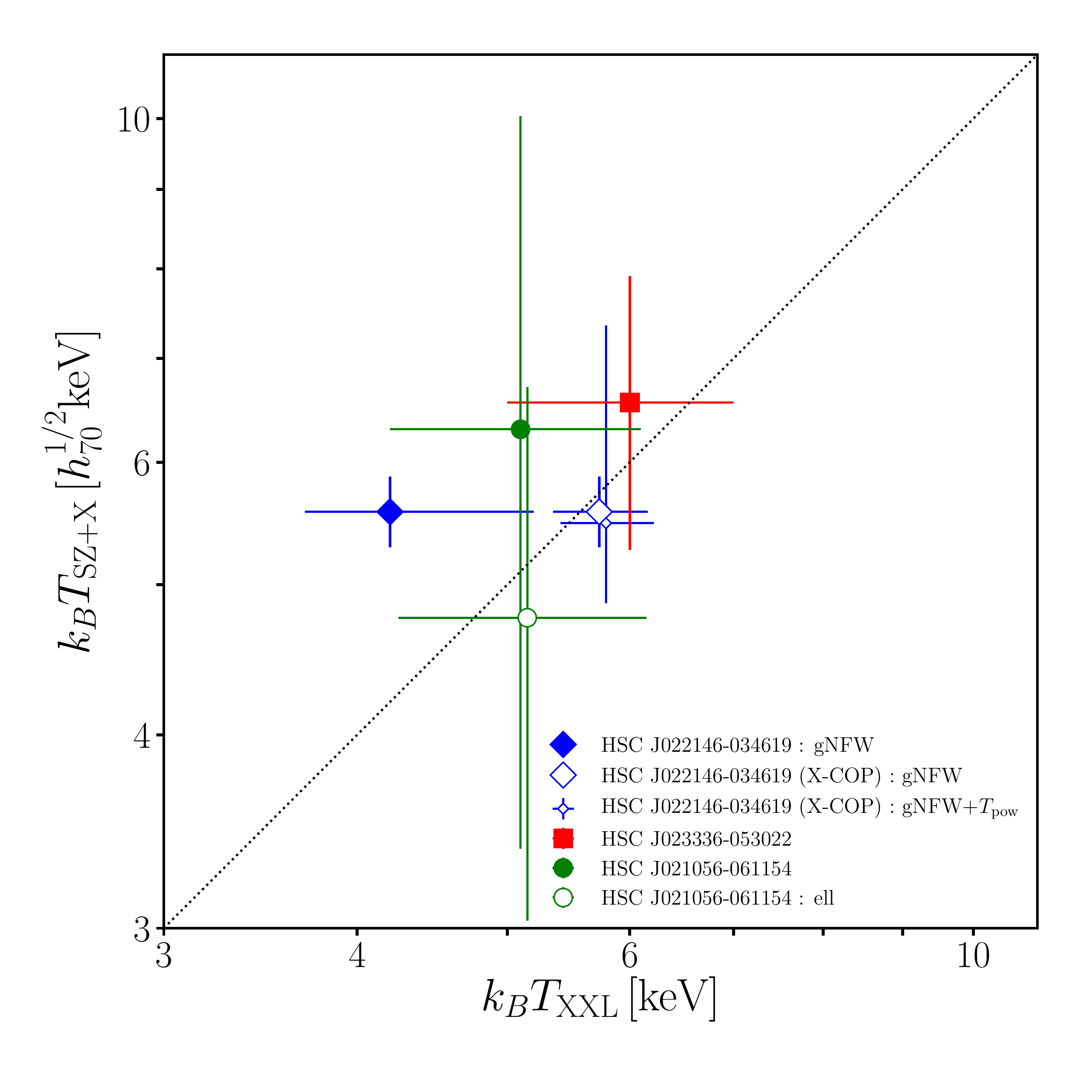}
 \end{center}
 \caption{Comparison of X-ray and SZE temperatures. Blue diamond, red square, and green circle are HSC J022146-034619, HSC J023336-053022 and HSC J021056-061154, respectively. The open blue diamonds for HSC J022146-034619 denote the X-COP temperature measurement using deep on-source data. The small open diamond is the SZE temperature based on the gNFW$+T_{\rm pow}$ model.
  The X-ray temperatures of the elliptical model for HSC J021056-061154 (open green
  circle) and the gNFW+$T_{\rm pow}$ model for HSC J022146-034619 (small open diamond) are shifted by $1.01$ for display purposes.}
 \label{fig:T}
 \end{figure}

We compute the cylindrical emission-weighted temperatures from our best fits
within $300h_{70}^{-1}$ kpc from the XXL centers.
Since HSC J023336-053022 and HSC J021056-061154 have complicated $y$ distributions, we use the
results of two-dimensional analyses. Figure \ref{fig:T} shows a
temperature comparison between the joint SZE and X-ray
analysis, $k_B T_\textsc{sz+x}$, and the X-ray spectroscopic measurement,
$k_B T_\textsc{xxl}$.
We find in HSC J022146-034619 that the temperature using the XXL survey data 
and that from X-COP using the deep pointing observation differ by $\sim 2$ keV, which is a $2.3\sigma$ difference;
 our result agrees with the latter one. 
The temperature of the gNFW+$T_{\rm pow}$ model for HSC J022146-034619 
does not significantly change from that of the gNFW model. 
For the three clusters, the central projected temperatures derived the joint SZE and X-ray analyses agree well with the X-ray temperatures.

\citet{Mroczkowski09} have found using Sunyaev–Zel'dovich Array (SZA) data that one-dimensional, radial temperature profiles determined from joint SZE and X-ray analysis are in reasonably good agreement with those obtained from an independent X-ray spectroscopic analysis.
\citet{Romero17} have compared SZE temperatures with X-ray temperatures for 14 clusters and found a good agreement $\langle T_\textsc{sz+x}/T_\textsc{x}\rangle =1.06\pm0.23$ in gas mass weighted temperatures. 
Although the previous studies did not carry out a multiple component analysis as demonstrated in this paper, 
our results agree with their results.

\subsection{Deviations from Scaling Relations} \label{sub:scaling}

In this section we study whether the dynamical states of the three
clusters affect their positions relative to the scaling scaling relations between total mass
and temperature ($k_BT_\textsc{sz+x}$), between total mass and the integrated Compton
parameter ($Y_{\rm cyl}$) and between the total and gas mass ($M_{\rm gas}$). 
We compute gas quantities from the best-fit temperature and
density profiles to avoid the PSF smearing effect and the radio transfer
function. We use the results of multi-component analyses for HSC J023336-053022
and HSC J021056-061154.

\subsubsection*{$M-T$ relation}

\begin{figure}
 \begin{center}
 \includegraphics[width=\hsize]{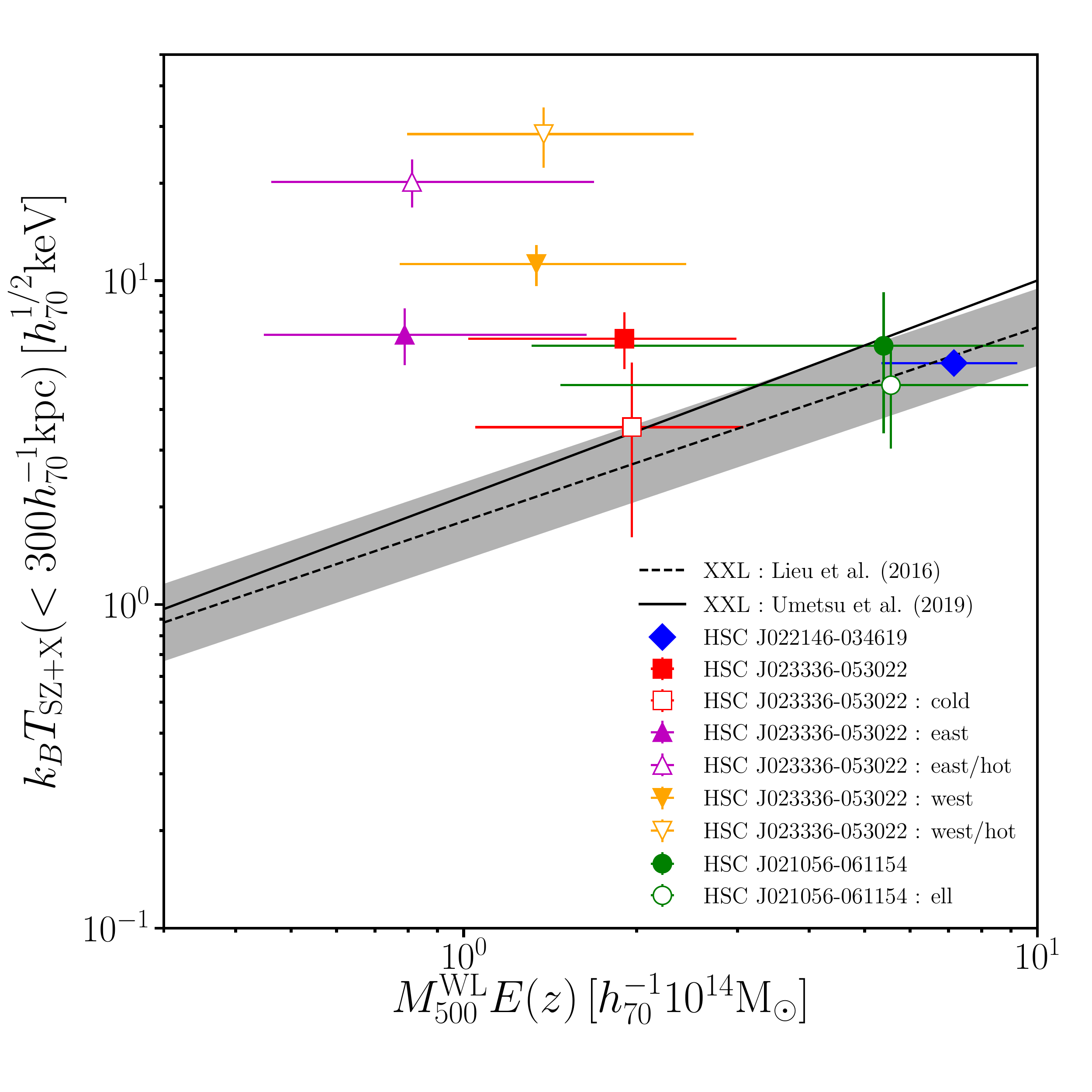}
 \end{center}
 \caption{Mass-temperature scaling relation at $\Delta=500$. The
 emission-weighted temperatures are computed within $300h_{70}^{-1}$ kpc
 from cluster centers. The solid and dashed lines denote the best-fit scaling relations for the XXL clusters
 of \citet{Umetsu19} and \citet{2016A&A...592A...4L}, respectively.
 The gray region is a combination of the $1\sigma$ uncertainty of the
 normalization and intrinsic scatter. Blue diamond, red square, orange down-triangle, magenta up-triangle and green circle are HSC J022146-034619, 
 the total, western and eastern component for HSC J023336-053022 and HSC J021056-061154, respectively.
 The open symbols denote the hot and cold components for the
 major merger or the elliptical model for the minor merger, respectively. 
 The masses of open symbols are shifted by a factor $1.03$ for display
 purposes. The temperatures of the major
 merger, HSC J023336-053022, are much higher than those of the scaling
 relation. 
 }
 \label{fig:M-T}
 \end{figure}

We first compute emission-weighted temperature, $k_B T_\textsc{sz+x}$, within projected
radius $R=300h_{70}^{-1}$ kpc from cluster centers following the XXL papers \citep{2016A&A...592A...4L}.
The emission-weighted temperature is computed by the best-fit parameters of the joint SZE and X-ray analysis; the X-ray emission weight and the uncertainty are calculated from the error covariance matrix. 
The dashed line in Figure \ref{fig:M-T} represents the mass--temperature
scaling relations compiled from
the XXL \citep{2016A&A...592A...1P}, COSMOS \citep{Kettula13} and CCCP
\citep{Hoekstra15} surveys.
\citet{Umetsu19} have carried out a weak-lensing analysis of XXL clusters using the
HSC-SSP 16A shape catalogue and found a slightly lower mass scale than that  of
\citet{2016A&A...592A...4L} (the solid line), though they are consistent within $\sim1\sigma$ errors. 
The temperature of the single-peaked cluster, HSC J022146-034619, agrees with \citet{2016A&A...592A...4L}.
We cannot find significant deviations for the minor merger, HSC
J021056-061154, from the best-fit scaling relations, regardless of the models.
The temperature for the major merger, HSC J023336-053022, is
two or three times higher than implied by the two scaling relations.
The significance level of the deviation compared to scatter, $[k_B T-f_T(M)]/\sigma_T$, is $4.7_{-3.0}^{+5.8}$, 
where $f_T(M)$ is the best-fit scaling relation \citep{2016A&A...592A...4L} with $M=M_{500}^{\rm WL}E(z)$ 
and $\sigma_T$ is a combination of the normalization uncertainty and intrinsic scatter of the scaling relation. 
We consider both the WL mass and temperature uncertainties in the error calculation.
The temperature of the central cool component still follows the scaling
relations. 
The temperatures of the eastern and western hot components are at $37.5_{-18.8}^{+25.9}$
and $38.5_{-18.8}^{+28.5}$ $\sigma-$levels higher than those expected
based on their WL masses, respectively. The main source of the errors is the uncertainties of weak lensing masses.
Although the intrinsic scatter extracted from the large sample \citep{2016A&A...592A...4L} would statistically include minor and major merger effects, the instant major merger boosts the temperatures from the baseline.
In previous studies \citep[e.g.][]{Ricker01,Poole07}, 
numerical simulations for major mergers with a mass ratio of $1:3$ 
have shown that the temperature increases by a factor of two just
after first pericentre passage. Our result is in good agreement with numerical simulations (see Figure \ref{fig:time}).

\subsubsection*{$M-Y_{\rm cyl}$ relation}

\begin{figure}
 \begin{center}
 \includegraphics[width=\hsize]{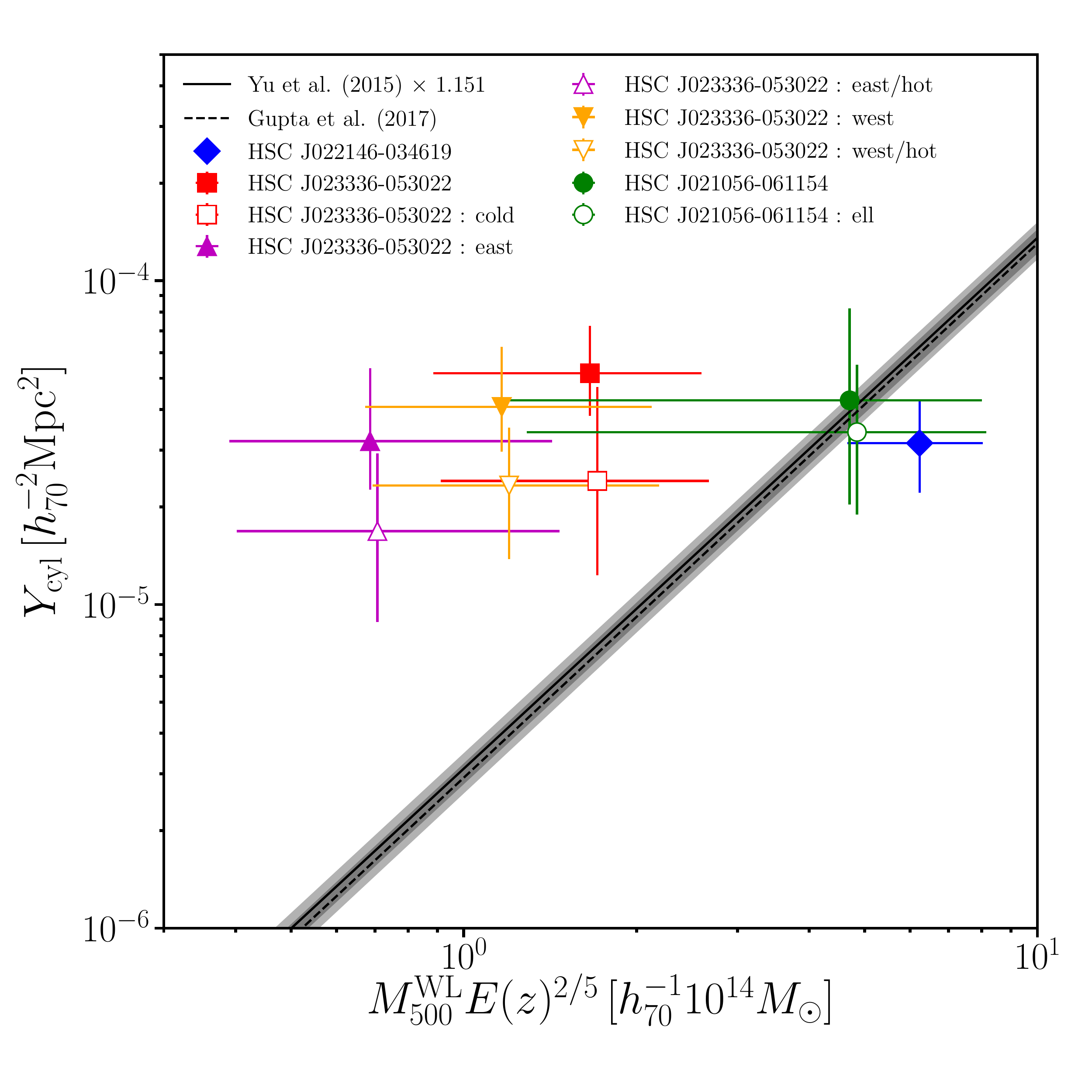}
 \end{center}
 \caption{Scaling relation of the WL mass and cylindrical $Y$ parameter at
  $\Delta=500$.  The solid and dashed lines denote the best-fit scaling
  relations from the numerical simulations of \citet{Yu15} and \citet{Gupta17}, respectively.
  Colour symbols and gray region are the same as Figure \ref{fig:M-T}.
The $Y$ parameters of the major merger significantly depart
 from the scaling relation.}
 \label{fig:Ysz}
 \end{figure}

We compute the cylindrical Compton parameter, $Y_{\rm cyl}$ (in units of Mpc$^2$), as follows:
\begin{eqnarray}
Y_{\rm cyl}=2\pi D_A^2\int_0^{\theta_{500}} y(\theta') \theta' d\theta'
\label{eg:Ycyl}
\end{eqnarray}
where $D_A$ is the angular-diameter distance from the observer to the cluster in Mpc,
and the enclosed radius, $R_{500} = \theta_{500} D_A$, 
is determined in the WL analysis\footnote{We note that arcmin$^2$ and steradians are other common units for $Y$ used in the literature, in which case the $D_A^2$ factor should be omitted from Equation\eqref{eg:Ycyl}
and corresponding units of angles or solid angles should be used in the integral.}.
We also propagate errors of the WL-determined $r_{500}$ into the integrated $Y$
parameters. The radii, $r_{500}$, for the western and eastern components in HSC J023336-053022 are computed from the $M_{500}$ derived from the 2D multi-component WL analysis (Sec. \ref{subsec:hsc3}). The projected distance between the two peaks in the $\tilde{y}_d$ map is lower than $r_{500}$ (Sec. \ref{subsec:hsc3_dis}).
\citet{Gupta17} found that a conversion factor from a spherical
integrated $Y$ parameter to a cylindrical Y parameter is $1.151$. 
We convert from a spherical $Y$ parameter using a numerical simulation \citep{Yu15} to the cylindrical
$Y$ parameter which is shown by the solid line in Figure \ref{fig:Ysz}.
The normalizations of two numerical simulations
\citep{Yu15,Gupta17} agree well with each other.
The $Y$ parameter of the single-peaked cluster, HSC J022146-034619, is
half the expected value given its mass.
This is caused by the normalization of the electron pressure profile
being lower than expected from the {\it Planck} SZE pressure profile
\citep{PlanckSZPprof}. The minor merger scenario we proposed agrees with the numerical
simulations. Since the model pressure outside the core of the minor
merger is higher than those of the other clusters, the integrated $Y$ of
this cluster is comparable though its peak signal-to-noise ratio is lower.  
However, the $Y$ parameters of the major merger are $7.7_{-2.6}^{+8.3}$ times
higher than the scaling relation suggests, where we do not use the intrinsic scatter from the numerical simulation \citep{Gupta17}.
\citet{Poole07} have investigated using numerical simulations the time evolution of the cylindrical
Compton $Y$ parameter within $R_{2500}$ and found that it increases by
about a factor four just after
first pericentre of major merger with a 1:3 mass ratio.
\citet{Wik08} have shown that the simulated spherical Compton $Y$ parameter over
the entire region of clusters increases by about $50$ percent times in major mergers with a 1:3 mass ratio.

Although the measurement methods are different, 
the increase in the $Y$ parameter provides similar trends to our results. 
However, \citet{Yu15} have shown in cosmological hydrodynamic simulations that the spherical $Y$ parameter of the thermal component
has no significant merger boost at $0.15$ Gyr after core passage and that the
scatter of the $Y$ parameter in the scaling relation is at most about 12 per
cent. This is not supported by our data.  
They also found that the normalization of the $Y$ parameter for the thermal component is $\sim20-30\%$ lower than that obtained when non-thermal pressure is included. Similarly,
\citet{Krause12} have studied merger-induced scatter and bias in the $M-Y$ scaling relation using cosmological hydrodynamic simulations. They found that the $Y$ parameter of major mergers within a Gyr after core-passage is below the baseline of the scaling relation and the $Y$ parameter increases more slowly during mergers than expected from the overall scaling relation. This is not supported by our data, either. 

There seems to be a discrepancy in results between cosmological and non-cosmological simulations.
The discrepancy would depend on how much the thermal energy or the non-thermal pressure is increased by cluster mergers. 
In cosmological simulations \citep{Krause12,Yu15}, the level of non-thermal pressure support such as the bulk motion and turbulence is more dominant. Non-cosmological simulations \citep[e.g.][]{Ricker01,Poole07,Wik08} studied gas heating induced by supersonic motions in the major merger regime.  The temperature and density enhancements are correlated in the shock region and thereby the $Y$ parameter increases. 
From optical cluster samples, we in principle can find major mergers in various stages, thanks to the long lifetime of   galaxy subhalos (see details in the introduction of \cite{Okabe19}). A combination of optical surveys and follow-up observations 
is therefore a powerful approach to understand gas physics in clusters that are outliers in cosmological simulations.

\subsubsection*{$M-M_{\rm gas}$ relation}

\begin{figure}
 \begin{center}
 \includegraphics[width=\hsize]{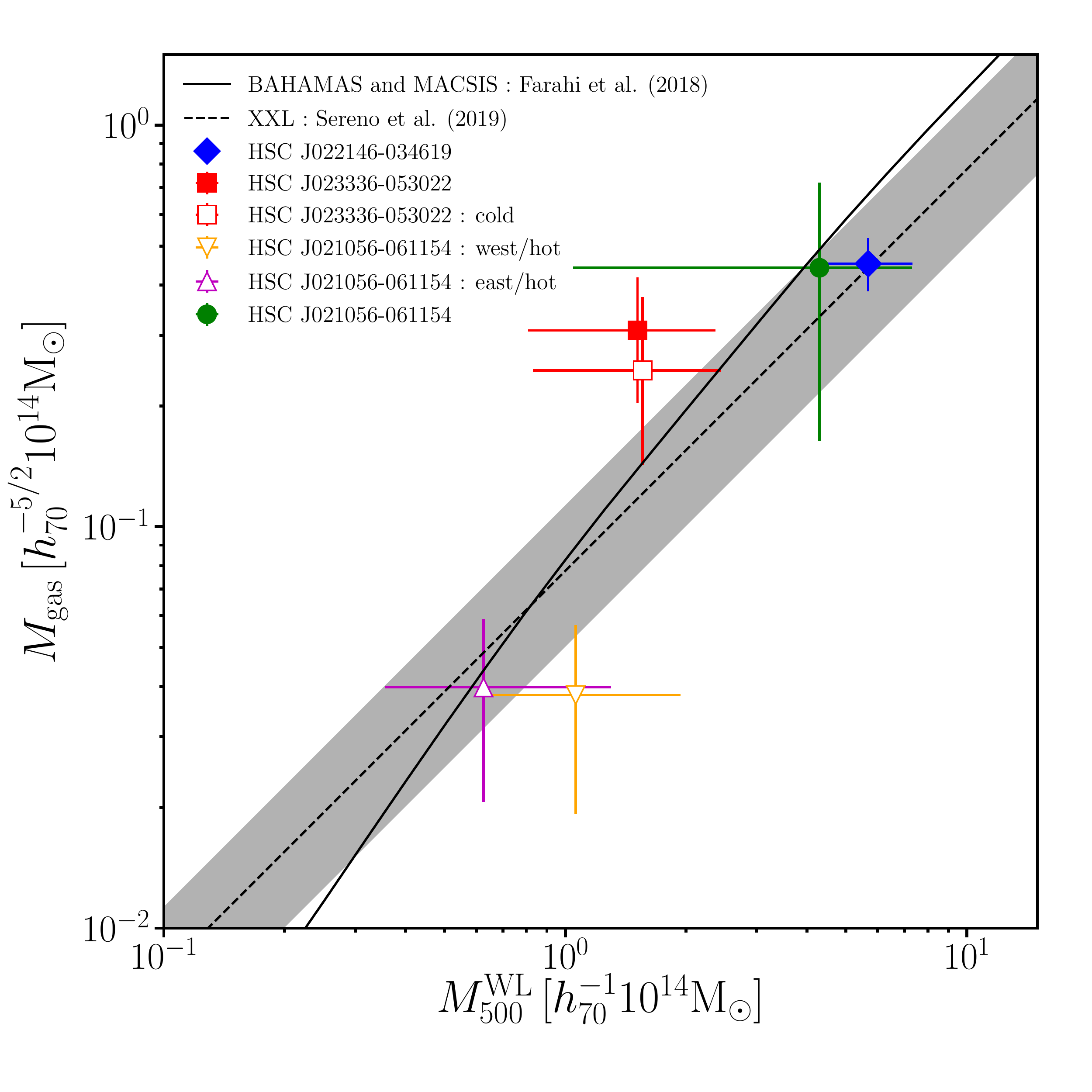}
 \end{center}
 \caption{Relation between the WL mass and gas mass within spherical radius, $r_{500}$.
 The solid and dashed lines denote the best-fit scaling
  relations from numerical simulations \citep{Farahi18} and the XXL
 clusters \citep{Sereno19}, respectively.
 Colour symbols and gray region are the same as Figure \ref{fig:M-T}.
 }
 \label{fig:Mgas}
 \end{figure}

We also investigate the relation between total mass, inferred from WL
analysis, and gas mass within each cluster's spherical radius $r_{500}$, as shown in Figure \ref{fig:Mgas}.
The cases of the major and minor mergers use the results of the multi-component analyses.
In three-dimensional dimensions we calculate the off-centering effect of
each component, as follows
\begin{eqnarray}
 \rho^{\rm off}(r)&=&\frac{1}{2\pi^2}\int_0^{\pi} d\theta \int_0^{2\pi}
  d\phi \nonumber \\ 
  &&\rho(\sqrt{d_{\rm off}^2+r^2-2rd_{\rm off}\cos\theta\sin\phi}), \label{eq:off} 
\end{eqnarray}
where $\rho$ is the mass density, $d_{\rm off}$ is the off-centering
radius and $r$ is the distance from the center in the three-dimensional space.
We ignore the separation between components along the line of sight.
Since the central component dominates for the gas mass,
this assumption does not significantly change the result.
We propagate errors of the WL-determined $r_{500}$ into gas mass estimates.
We also plot a theoretical scaling relation from cosmological
hydrodynamical simulations \citep{Farahi18} and the scaling
relation \citep{Sereno19} derived for the XXL sample (assuming $M_{\rm gas} \propto M$ and no
evolution) based on the HSC-WL \citep{Umetsu19} and XXL survey data.
We find that the total gas mass within the major merger (red square) and
the gas mass for the western hot component of the major merger (orange
downward-triangle) are slightly higher and lower than the baseline scaling relations, respectively. If the gas mass follows the scaling relations before the merger, 
the feature suggests a possibility that a small fraction of the gas mass of the main cluster is moved to the region of the subcluster. Indeed, the X-ray core is composed of the two components (Sec \ref{subsec:hsc2}). The surface-brightness weighted center close to the secondary X-ray peak is at the intermediate position of the two $y$ peaks and the main X-ray peak is close to the eastern component. Some fraction of the X-ray core could be the remnant of the main cluster. A difference between collisional gas and collisionless dark matter distributions is reported by previous studies. For instance, \citet{Okabe08} found that the gas distribution for on-going mergers is completely different from the dark matter distribution, while the two distributions before mergers are similar. 
The deviation of the total gas mass from the baseline is $3.6_{-0.8}^{+1.4}\sigma$, where we consider the error correlation through WL-determined radius. The deviation may be affected by cluster mergers, though we cannot rule out the possibility that the total gas mass before the merger is intrinsically higher than the baseline.

\subsubsection*{Time evolution}

\begin{figure}
 \begin{center}
 \includegraphics[width=\hsize]{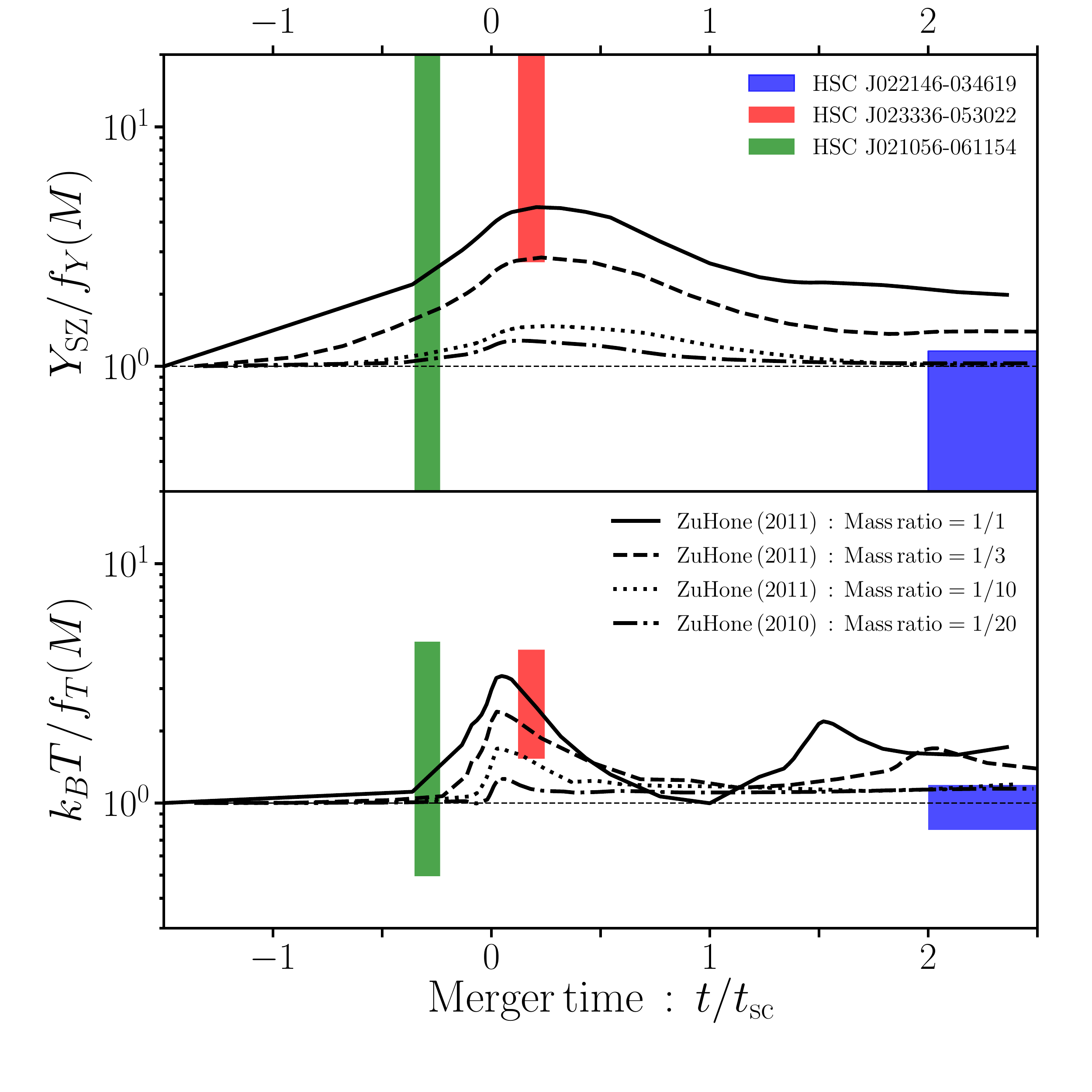}
 \end{center}
 \caption{Time evolution of the merger boost. The $x$-axis is the time
 in units of the sound-crossing time. The function, $f(M)$, is the
 temperature or the integrated Compton $y$ parameter expected from
 the WL masses on the basis of the scaling relations
 \citep{2016A&A...592A...4L,Gupta17}. From left to right, rectangles
 mark HSC J021056-061154, HSC J023336-053022, and HSC J022146-034619,
 respectively. The errors in $y$-axes take into account the $1\sigma$
 uncertainties of both gas measurements and WL masses.  The solid,
 dashed and dotted lines are results of numerical simulations
 \citep{ZuHone11} of mass ratio $1:1$, $1:3$ and $1:10$ respectively, assuming head-on mergers.
 The dash-dotted line shows the result of numerical simulations \citep{ZuHone10} of mass ratio $1:20$ with an impact parameter of $200$ kpc. }
 \label{fig:time}
 \end{figure}

We finally investigate how much the temperature and the integrated Compton $y$ parameter change with dynamical time during cluster mergers, as discussed above. 
Assuming that the gas properties before cluster mergers follow the
scaling relations \citep{2016A&A...592A...4L,Gupta17}, we compute the
ratio between the observable and the expectation of WL masses via the
scaling relations, $f(M)$ .
The range of plausible dynamical times is estimated by spatial separations between WL-determined centres (Sec. \ref{subsec:hsc3}) or galaxy clumps (Secs. \ref{subsec:hsc2} and \ref{subsec:hsc1}) and inferred velocity. 
The dynamical time is normalised by the sound-crossing time to allow
comparisons from different mass systems.
We ignore the distance along the line of sight. 
We assume that the dynamical time ranges are $0.1$ Gyr for the pre- and post mergers and $1$ Gyr for the sloshing phase as uncertainties. 
The result is shown in Figure \ref{fig:time}. The errors in the scaled
$Y_{\rm SZ}$ and $k_BT$ values for each cluster take into account the $1\sigma$ uncertainties of both spherically symmetric gas models and WL masses. 
 We also plot results of N-body and hydrodynamic numerical simulations
 of binary mergers \citep{ZuHone11} and sloshing \citep{ZuHone10} which
 are computed from the publicly-available catalogue 
 \citep{ZuHone18}. The solid, dashed and dotted lines are the time
 evolution for head-on mergers of mass ratios $1:1$, $1:3$ and $1:10$ from \citet{ZuHone11}, respectively.  The dash-dotted line is for mass ratio of $1:20$ with impact parameter of $200$ kpc, retrieved from \citet{ZuHone10}. 
 The first two and last one are our references for the major and minor mergers, respectively. 
 We normalise the simulated gas properties by the initial states. 
 The temperature and the integrated Compton $y$ parameter of
 \citet{ZuHone11} are shown as would be observed for a merger in the plane of
 the sky, and calculated within $300$ kpc for the temperature and $r_{500}$ for the integrated Compton $y$ parameter from the X-ray surface brightness peaks. 
 The off-axis merger of the sloshing simulation \citep{ZuHone10} are
 calculated for the same direction as Figure \ref{fig:ZuHone_sloshing}.
  The overall trend of the time evolution from numerical simulations do not conflict with our results. Although the current sample of clusters is only three, 
future follow-up studies will significantly increase the number of clusters. 
Both the $k_B T$ and $Y$ parameters for the major merger are boosted from the baseline, which suggests that the scatter in the temperature and $Y$ parameter, that is, the temperature and gas density, is correlated. Numerical simulations have shown negative \citep{Kravtsov06} and positive \citep{Stanek10} correlations between temperature and gas mass scatter.  \citet{Gaspari14} have discussed the origin of this pressure fluctuations. The subsonic motions mainly drive isobaric turbulence and entropy index perturbations, while high velocity motions with $\mathcal{M}>0.5$ trigger compressive pressure fluctuations. Super-sonic motion in the major merger regime supports the latter case. Previous observational studies 
\citep{Okabe14a,Ghirardini19} found that intrinsic scatter of entropy profiles is lower than that of pressure profiles in several local cluster samples. It indicates that the intrinsic scatter of the electron number density is correlated with that of the temperature \citep[see also;][]{Okabe10c}. When temperature and density fluctuations are correlated, the $Y$ parameter is affected and major mergers can give a systematic bias cluster cosmology. Although significant merger boosts were found in previous studies using non-cosmological simulations \citep[e.g.][]{Ricker01,Poole07,Wik08} and this paper, the merger boost in the $Y$ parameter is not significant in some cosmological simulations \citep[e.g.][]{Krause12,Yu15}. This discrepancy is an open question theoretically. Based on the observational approach, follow-up multi-wavelength observations for a large sample of clusters are essential to answer this question and understand the thermodynamics of the gas.  As pointed out by \cite{Okabe19}, the galaxy distribution contains unique and ideal information to construct a
homogeneous sample of cluster mergers, in particular because of long life-time of galaxy subhalos, a similarity between galaxy and dark matter distributions, and insensitivity to the ICM merger boosts. Although the sample can cover from pre- to post- mergers, it is difficult to distinguish between pre- and post- mergers and thus, follow-up observations are important to measure the gas properties and characterize the impact of merging phenomena on the cluster evolution.  We do not use any X-ray information for the follow-up sample definition in this paper;  nevertheless, we found evidence for a merger boost, which is promising for future follow-up observations.
Furthermore, a large sample of merging clusters will fill the parameter space of the dynamical time and the mass ratio, and statistically  
overcome uncertainties in the values of the impact parameter of cluster mergers, inclination angle of cluster mergers relative to the plane of sky, and intrinsic scatter of initial states.

\subsection{Central entropy index profiles and centroid offsets}

The distribution of galaxies provides a strong indication of whether a
cluster is undergoing a major merger, and the phase of that interaction,
but cannot inform us about the status of the cluster atmospheres.
Further, optical searches are not good at identifying low-mass subhalos
and mergers where the halo barycentres are closely spaced.
Since gas properties are more or less changed by both major and minor
mergers, and the duration of the change continues over several
sound-crossing times \citep[e.g.][]{Ricker01,Poole07,Wik08,ZuHone11},
and repeated interactions with small subhalos can sustain X-ray
perturbation \citep{Ascasibar06,ZuHone10}, the gas properties provides
us with the essential information of the activity of gas, which is
complementary to the optical galaxy distributions.

\begin{figure*}
 \begin{center}
 \includegraphics[width=\hsize]{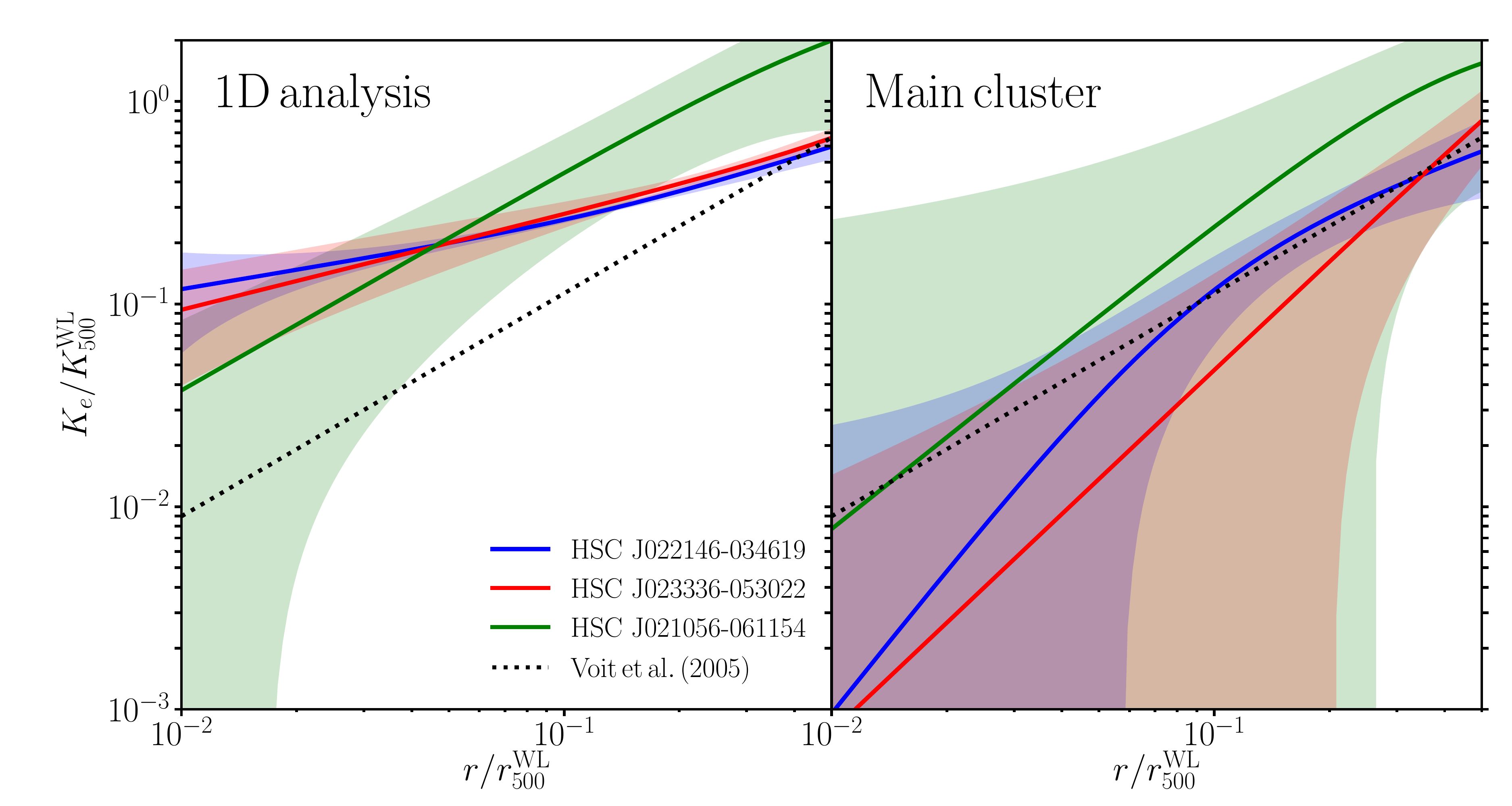}
 \end{center}
   \caption{Central entropy index profiles normalised by weak-lensing
 radii and masses. {\it Left:} results from one-dimensional
 analysis. {\it Right}: results for the main cluster components (gNFW
 model) derived from the multicomponent analysis.  Blue, red and green
 solid lines are HSC J022146-034619, HSC J023336-053022 and HSC
 J021056-061154, respectively. The colour shaded regions represent the
 $1\sigma$ uncertainty bounds. The black dotted line is $K/K_{500}=1.41(r/r_{500})^{1.1}$ of \citet{Voit05}. $K_{500}=375{\rm keV\,cm^2}\left(M_{500}^{\rm WL}/10^{14}h_{70}^{-1}M_\odot E(z)\right)^{2/3}(f_b/0.15)^{-2/3}$ from eq (3) of \citet{Pratt10}, where $f_b$ is the baryon fraction.} 
 \label{fig:Knorm}
\end{figure*}

\citet{Pratt10} have investigated cluster dynamical properties using X-ray based entropy index profiles in
the inner regions of 31 nearby clusters ($z\simlt0.15$) from the
representative XMM-Newton cluster structure survey (REXCESS). They
assumed a spherically symmetric model and found that the central entropy
for morphologically disturbed clusters is higher than the baseline expected from cosmological simulations \citep{Voit05}, but that for cool core clusters follows the baseline. 
That feature would be explained by the scenario that merging subhalos penetrate into central regions of the main clusters, and then heat and disturb the central gas.
We therefore investigate central entropy profiles. Following
\citet{Pratt10}, we first assume a spherically-symmetric model from
one-dimensional analysis. We normalised the radius by the WL-determined
$r_{500}$ and the entropy by the characteristic entropy computed by WL
masses. The characteristic entropy is specified by eq. (3) of
\citet{Pratt10}. We here ignore WL measurement errors for simplicity.
The resulting profiles of electron entropy index are higher than the baseline from numerical simulations \citep{Voit05}, as shown in the left panel of Figure \ref{fig:Knorm}. The electron entropy index at $r\sim0.01r_{500}$ is $\sim0.1K_{500}$, which is similar to the case of the morphologically disturbed clusters of \citet{Pratt10}. 
Indeed, the $y$ and $S_X$ maps for the three clusters are complexly distributed.  
Since we have carried out multi-component analyses, we computed the corresponding profiles for the main cluster component of the gNFW model. 
The right panel of Figure \ref{fig:Knorm} shows that the entropy for the
main cluster component follows the baseline. That suggests that, when we
interpret the entropy profile in the three dimensional space from the
projected information, the geometrical assumption and the number of
components are both important. If the hot component is spatially offset
from the cluster center, the entropy index close to the centre for the
spherical model would be overestimated due to the low-density in this region.

\begin{figure*}
 \begin{center}
 \includegraphics[width=\hsize]{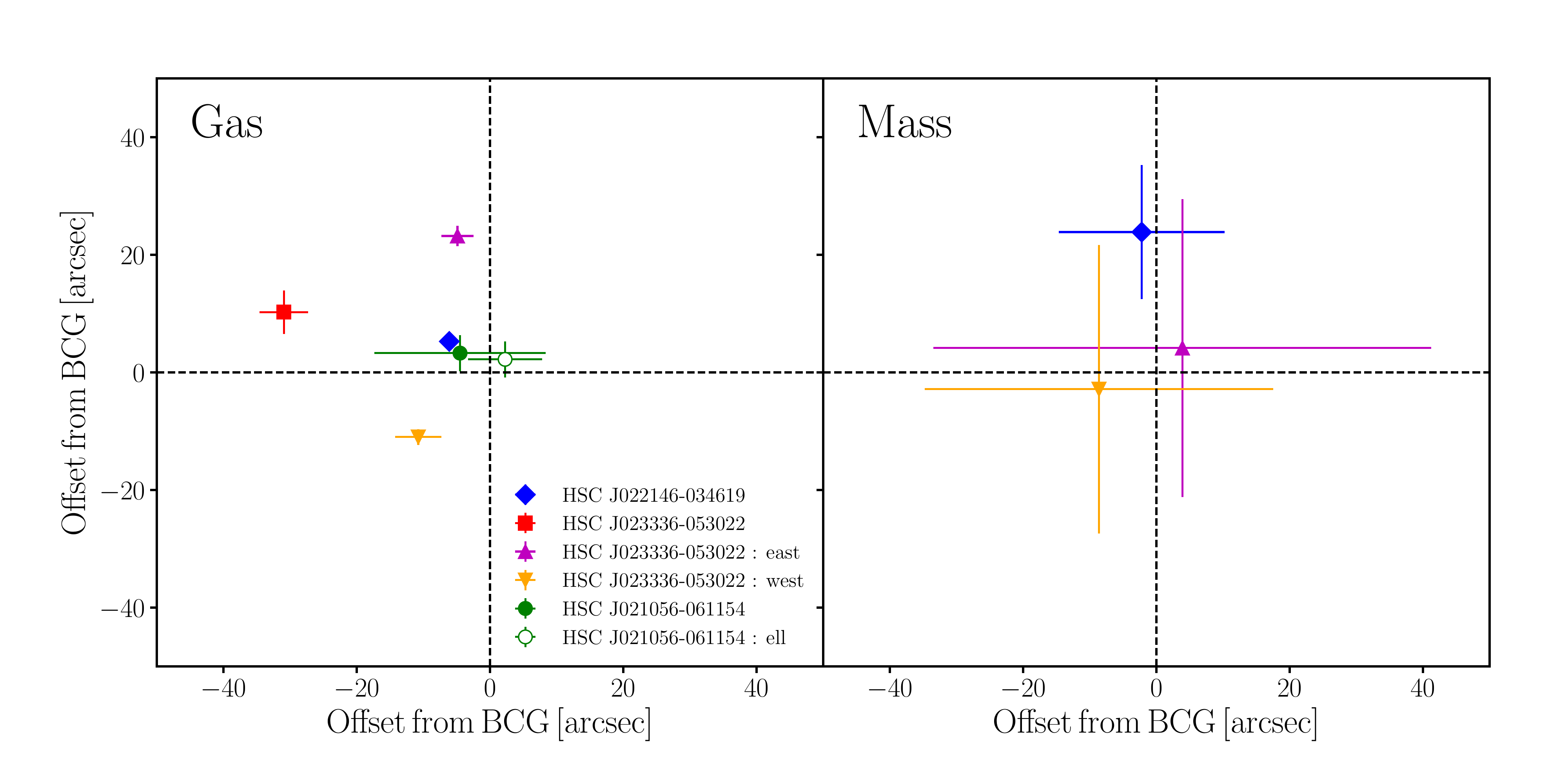}
 \end{center}
 \caption{{\it Left}: offsets between gas centers obtained by the two-dimensional joint SZE and X-ray analysis and the BCGs. {\it Right} offsets between mass centers determined by the two-dimensional analysis and the BCGs. Colour symbols are the same as Figure \ref{fig:M-T}.　}
 \label{fig:offset}
 \end{figure*}

We next compare centroids determined by the two-dimensional analyses
(Table \ref{tab:center}) with the BCG positions. The left and right
panels in Figure \ref{fig:offset} show gas center and mass center
offsets. The gas centers are obtained by a forward modeling method, and
thus differ from the peak positions of the SZE and/or X-ray
distributions. The mass center for the pre-minor merger, HSC
J021056-061154, cannot be estimated since there is no shape catalogue in the central region (Sec. \ref{subsec:wl}).  
The gas offsets for the sloshing cluster, HSC J022146-034619 and the pre-minor merger are smaller than those of the major-merger, HSC J023336-053022. We adopted the single gNFW model for HSC J022146-034619.
The WL-determined centroids for the major merger agree with the BCGs (Sec. \ref{subsec:hsc3_dis}), while that for the sloshing cluster is slightly offset (Sec. \ref{subsec:hsc2_dis}). 
Offset features in gas and mass distributions show different trends and have no correlation with central entropy profiles. Therefore, the offset distances, especially gas centers, will give complementary indicators about the dynamical state.

\subsection{Mass Comparison} \label{sub:mass_com}

\begin{figure*}
 \begin{center}
 \includegraphics[width=\hsize]{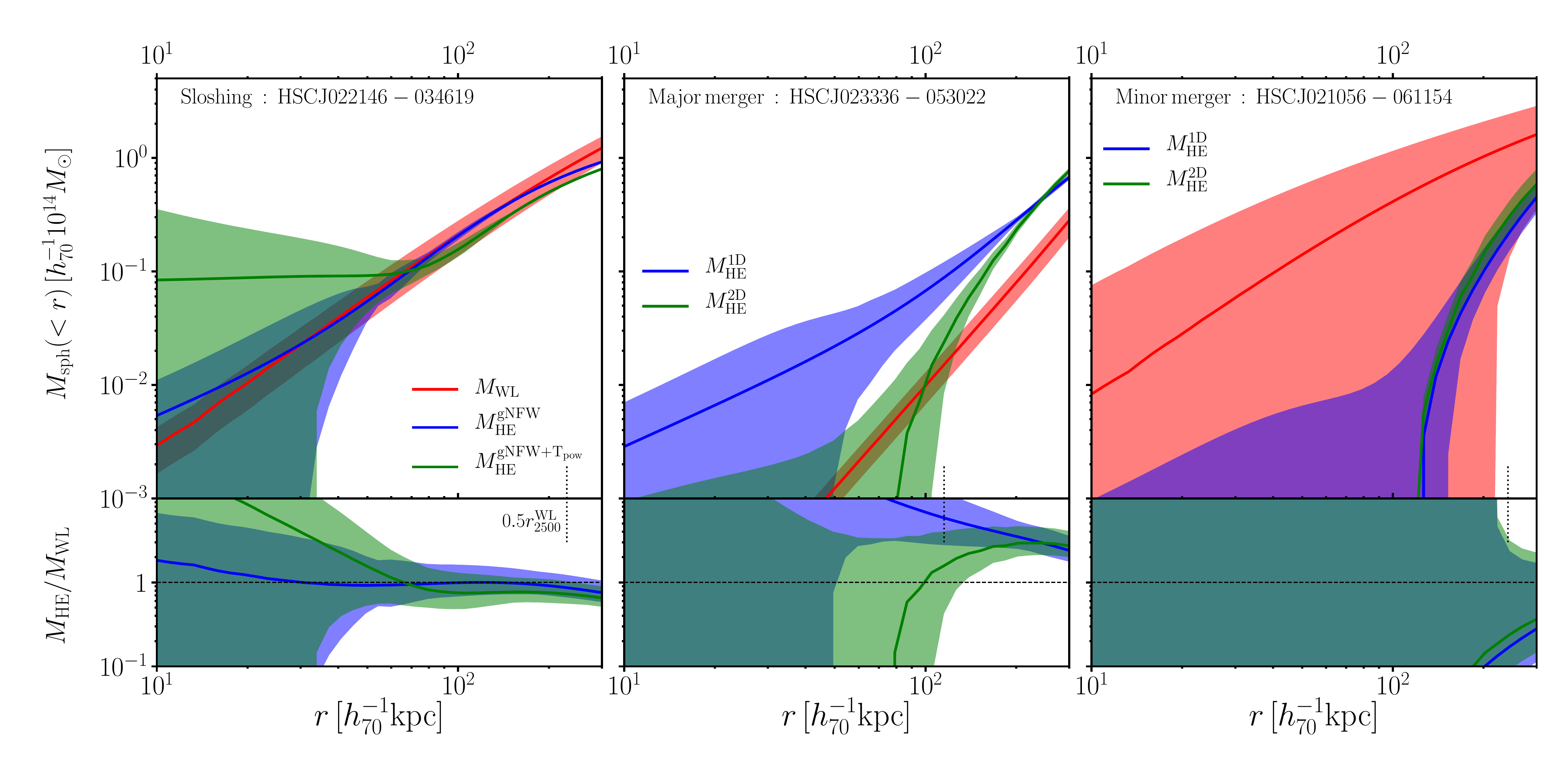}
 \end{center}
 \caption{{\it Top}: HE and WL enclosed masses as a function of 3-D radius (from left to right ; HSC J022146-034619, HSC J023336-053022 and HSC J021056-061154) . 
  The red and blue lines denote the WL and HE mass profiles, respectively. 
  The green lines are the HE mass profile derived from gNFW+$T_{\rm pow}$ model in the left panel and two-dimensional analysis in the middle and right panels. 
  The color transparent regions denote the $1\sigma$ uncertainty. 
   {\it Bottom}: HE to WL total mass ratios as a function of radius. The vertical dotted lines denote $0.5r_{2500}^{\rm WL}$.}
 \label{fig:Mratio}
 \end{figure*}

 We next compare hydrostatic equilibrium (HE) masses with WL masses as a function of radius.
 The HE masses are evaluated through the HE equation
 \begin{eqnarray}
  M_{\rm HE}(<r)=-\frac{r^2}{G\mu m_p n_e}\frac{d P_e}{d r}
 \end{eqnarray}
where $\mu=0.62$ is the mean molecular weight. For the multi-component
gas analysis, we take account of the off-centering effect (eq. \ref{eq:off}) for the
electron number density and pressure profiles but ignore the offsets along the line of sight. Similarly, the two WL mass
estimates by the two-dimensional analysis are converted into
one-dimensional radial profiles through the off-centering effect in the mass density. 
Figure \ref{fig:Mratio} shows the spherical mass profiles derived by the
joint SZE and X-ray analysis and by WL analysis. The figure also shows the mass ratio $M_{\rm HE}/M_{\rm WL}$ out
to $300\,h_{70}^{-1}{\rm kpc}$ which is comparable to the maximum radii
($\sim 1$ arcmin) of our positive $y$ measurements.
Since we cannot use the shape catalogue in the central region, 
the WL measurement errors for HSC J021056-061154 are large and the mass ratio cannot be constrained well.
We find for the single-peaked cluster, HSC J022146-034619, that the HE mass
is consistent with the WL one, while for the merging cluster, HSC
J023336-053022, that the HE mass exceeds the WL one because of the
merger-boost.

Although it is difficult to make a fair comparison with the literature
because of the small set of clusters 
and differences in the radial range ($r\simlt r_{2500}$; Figure \ref{fig:Mratio}), 
we first compare results from a large compilation of clusters with masses measured at $\Delta=2500-500$, because the mass bias at central regions is not yet well studied. 
Agreement of the HE and WL mass estimations of are reported by Local Cluster Substructure Survey \citep[LoCuSS;][]{Smith16} and the XXL Survey \citep{Sereno19} at $\Delta=500$, and weighing the Giants \citep[wtG;][]{Applegate16} at $\Delta=2500$. Their differences are at most of the order of $\simlt10\%$. 
The LoCuSS \citep{Smith16} uses {\it Chandra} and {\it XMM-Newton} data with X-ray temperature measurements from the two satellites \citep{Martino14} and Subaru/Suprime-Cam WL analysis \citep{Okabe16b}.  
The XXL Survey \citep{Sereno19} uses {\it XMM-Newton} X-ray measurements
and assumed the universal pressure profile  and Subaru/HSC-SSP WL data
\citep{Umetsu19}. The wtG \citep{Applegate16} uses {\it Chandra} X-ray
measurements and Subaru/Suprime-Cam WL analysis \citep{Applegate16} for
12 relaxed clusters. All the X-ray and WL mass measurement techniques are different \citep[see also][]{Pratt19}, nevertheless the comparisons only found a discrepancy $\sim30\%$. 
However, the Canadian Cluster Comparison Project \citep[CCCP;][]{Mahdavi13} with their new WL mass measurement \citep{Hoekstra15} have shown that the HE mass is on average $\sim 25\%$ lower than the WL mass \citep[see also][]{Smith16}. \citet{Pratt19} have summarized that the CCCP WL masses are similar to those of LoCuSS and Cluster Lensing And Supernova survey with Hubble \citep[CLASH;][]{Umetsu16} and thus, the discrepancy would be caused by a difference between X-ray analyses. \citet{Siegel18} have carried out a joint analysis of Chandra
X-ray observations, Bolocam thermal SZ observations, HST
strong-lensing data, and Subaru/Suprime-Cam weak-lensing data for 6 regular CLASH clusters, and constrained that the non-thermal pressure fraction at $r_{2500}-r_{500}$ is $<10\%$. Thus, state-of-art analyses using good resolution data resolving the internal structure suggest only a minor contribution of non-thermal pressure.
However, interestingly, the {\it Planck} masses are $\sim30-40\%$ lower than the WL masses, even when we use the same WL masses from \citet[CLASH;][]{2017A&A...604A..89P} and \citet[wtG;][]{2014MNRAS.443.1973V}. It indicates 
that an observational discrepancy between WL and HE masses highly depends on how the HE masses are estimated.

A numerical simulation \citep{Nelson14} has found that the non-thermal pressure has a strong dependence of 
cluster-centric radius and a weak dependence of the mass accretion rate. The non-thermal pressure changes from $\sim 10\%$, $\sim15\%$, to $\sim40\%$ as the radius increases from $0.2r_{200m}\sim r_{2500}$, $0.4r_{200m}\sim r_{500}$, to $r_{200m}$. \citet{Biffi16} also found a similar radial dependence. The average HE mass biases for cool-core and non-cool-core clusters are $\simlt 5\%$ and $\sim 10\%$ at $r_{2500}$, and $\sim10\%$ for the both at $r_{500}$, respectively. Therefore, the simulated HE mass bias at cluster cores is likely to be small, which agrees with our results. Similar results in central regions are reported by \citet{Okabe16} and \citet{2016Natur.535..117H}.
\cite{Okabe16} have found that the central mass profile ($r < 300$ kpc) determined from the joint stellar kinematics and WL analysis is in excellent agreement with those from independent measurements, including dynamical masses estimated from the cold gas disc component, the HE mass profile, and the BCG stellar mass. The quiescent gas motion around the BCG in the Perseus cluster is directly observed by the {\it Hitomi} satellite \citep{2016Natur.535..117H}. 
Since the amplitude of non-thermal pressure varies from cluster-to-cluster, it is important to increase the number of clusters for further assessments.

The differences between the HE and WL masses in previous numerical simulations and observations are small ($\simlt 5-10\%$) at small radii. We found a similar result in the single-peaked cluster, but for the merging clusters the HE mass is higher than the WL one.

\section{Conclusions} \label{sec:con}

We performed GBT/MUSTANG-2 observations for three HSC-SSP CAMIRA clusters \citep{Oguri18} 
with different galaxy distributions: one single-peaked cluster; one double-peaked cluster; and one belonging to a supercluster.

We carried out the following analyses for each cluster. 
We compared the $y$ maps with X-ray images taken from the XXL survey and HSC-SSP optical galaxy distributions.
The gas distributions observed by the SZE and X-ray method provide different information.
We performed Bayesian forward modeling via simultaneous fits to MUSTANG-2 $y$ and XXL X-ray data and measured gas properties. 
We measured WL masses using the HSC-SSP shape catalog. We have looked through the library of simulations presented by \citet{ZuHone18} and identified systems with similar X-ray and $y$ properties for each of the three clusters. We summarize the main results for each cluster.

The results of the single-peaked cluster, HSC J022146-034619, are as follows :
\begin{itemize}
  \item we found that the SZE and X-ray distributions have regular morphology, but the galaxy distribution is elongated. 
  \item the temperature profile measured by the joint analysis agrees with the temperature profile based on the X-COP method \citep{Eckert17}. 
  \item the excess $y$ distribution from the best-fit gNFW profile is found at $3\sigma$ level at $\simlt4$ arcsec from the BCG. 
  \item the residual $y$ and $S_X$ patterns from the model-independent, azimuthally-averaged profiles are coherent, indicating that the cluster is likely to be in a sloshing phase. 
  \item  two subhalo candidates which plausibly drive the sloshing mode are found. The first candidate is the second brightest galaxy and the second one is at $1.3$ Mpc south of the BCG. 
  \item the coherent residual patterns is found in simulated $y$ and $S_X$ distributions \citep{ZuHone10}.
\end{itemize}

The results of the double-peaked cluster, HSC J023336-053022, are as follows:
\begin{itemize}
    \item a double-peaked $y$ morphology with each peak associated with a separate galaxy
concentration and a single X-ray core between the two $y$ peaks are found.
    \item such a double-peaked $y$ feature is not yet reported in previous studies.
    \item the multi-component analysis indicates two hot components with temperatures $28.4_{-6.0}^{+5.9}$ keV (west) and $20.2_{-3.4}^{+3.5}$ keV (east), where the temperatures are the X-ray-like emission-weighted temperatures measured within 300kpc from the best-fit centers.  
    \item the two-dimensional WL analysis indicates that the western component is the main cluster. The mass ratio is $0.54_{-0.28}^{+0.93}$.
    \item  the density and temperature distributions and the mass and galaxy distributions indicate that the cluster is likely to be a major merger after core crossing.
    \item some numerical simulations of merging systems \citep{ZuHone11} show a double-peaked $y$ distribution and single $S_X$ distribution. 
    \item we do not find significant levels of diffuse radio emission in the FIRST, GMRT and TGSS data with high angular resolutions. The absence of diffuse radio emission implies an efficiency of less than $1$ percent for conversion of kinetic energy into relativistic electrons, assuming a magnetic field strength $B>1\,\mu {\rm G}$. 
\end{itemize}

The results for the supercluster member, HSC J021056-061154, are as follows:
\begin{itemize}
\item an elongated $y$ distribution is offset from the X-ray main peak, which is around the BCG position.
\item the multi-component analysis indicates a hot component elongated perpendicular to the major-axis of the X-ray core. 
\item the anisotropic $y$ and temperature distributions indicates that the cluster is likely to be in a pre-merger phase.
\item from stellar mass estimates of member galaxies we suggest that the cluster is a minor merger with a total mass ratio of $\sim1:10$. 
\item distributions in $y$ and $S_X$ like the observed ones are also found in numerical simulations \citep{ZuHone11}.
\end{itemize}

We then studied cluster properties and their relationship with their dynamical dependence. 
One of the striking results is that the distributions of the gas properties (temperature, density and pressure) are more or less disturbed regardless of the global red galaxy distributions. 
The projected temperatures derived from the joint SZE and X-ray analysis are in a good agreement with those of X-ray measurements. 
We computed deviations from the mass scaling relations of the temperature, the integrated $Y_{\rm cyl}$ parameter, and the gas mass and the relationships with their dynamical dependence. 
We find a merger-driven boost in the $M-T$ \citep{Lieu16} and $M-Y_{\rm cyl}$ \citep{Gupta17} relations, which is in good agreement with numerical simulations \citep{ZuHone11}. 
Although the $y$ and $S_X$ distributions of all the three clusters are disturbed and the central entropy index profiles are higher than the baseline from numerical simulations \citep{Voit05}, the global $k_BT$ and $Y_{\rm cyl}$ are 
changed only for the major merger just after core-passage.
The WL mass profiles for the sloshing and minor merger agree with the HE mass profiles at $r\simlt 300$ kpc, while the HE mass for the major merger is higher than the WL one.

The joint analysis of the high-angular resolution SZE and X-ray data enables us to simultaneously determine the three-dimensional profiles of the temperature and the density and their centers. 
It can spatially resolve hot components at temperatures of tens of keV, which are not well measured by existing X-ray satellites. The spatial resolution of the projected temperature distribution is $\sim0.05\,{\rm arcmin}^2$ and higher than those of X-ray spectroscopic measurement (the order of the arcmin$^2$).
Therefore, such analyses can overcome the problems of poor angular
resolution of X-ray temperature measurements and provide a tool for
studying the hottest components of clusters and cluster mergers.
In the future,  systematic follow-up observations for optical clusters
in various dynamical stages will play an important role in cluster
physics.

\section*{Acknowledgments}

We thank the anonymous referee for helpful comments.

MUSTANG2 is supported by the NSF award number 1615604 and by the Mt.\ Cuba Astronomical Foundation.
The National Radio Astronomy Observatory is a facility of the National
Science Foundation operated under cooperative agreement by Associated
Universities, Inc.

 This work was in part supported by the Funds for the Development of Human
 Resources in Science and Technology under MEXT, Japan and
  Core Research for Energetic Universe in Hiroshima University
 and in-house grant for international conferences under the MEXT's
 Program for Promoting the Enhancement of Research Universities, Japan.
 
This paper is supported in part by JSPS KAKENHI Grant Number JP20K04012
(N. O.), JP18K03704 (T. K.), JP15H05892 (M.O.) and JP18K03693 (M. O.) .
SRON is supported financially by NWO, the Netherlands Organization for Scientific Research. 
MS acknowledges financial contribution from contract ASI-INAF n.2017-14-H.0 and INAF `Call per interventi aggiuntivi a sostegno della ricerca di main stream di INAF'. KU acknowledges support from the Ministry of Science and Technology of
Taiwan (grant MOST 106-2628-M-001-003-MY3) and from the Academia Sinica Investigator Award (grant AS-IA-107-M01).

 We acknowledge Lucio Chiappetti for his editorial comments on the manuscript.

\section*{Data availability}

GBT data was taken under the project ID AGBT17B\_101.

The Hyper Suprime-Cam (HSC) collaboration includes the astronomical communities of Japan and Taiwan, and Princeton University. The HSC instrumentation and software were developed by the National Astronomical Observatory of Japan (NAOJ), the Kavli Institute for the Physics and Mathematics of the Universe (Kavli IPMU), the University of Tokyo, the High Energy Accelerator Research Organization (KEK), the Academia Sinica Institute for Astronomy and Astrophysics in Taiwan (ASIAA), and Princeton University. Funding was contributed by the FIRST program from Japanese Cabinet Office, the Ministry of Education, Culture, Sports, Science and Technology (MEXT), the Japan Society for the Promotion of Science (JSPS), Japan Science and Technology Agency (JST), the Toray Science Foundation, NAOJ, Kavli IPMU, KEK, ASIAA, and Princeton University. 

This paper makes use of software developed for the Vera C. Rubin Observatory (VRO). We thank the VRL Project for making their code available as
 free software at \url{https://www.lsst.org/about/dm}.

 The Pan-STARRS1 Surveys (PS1) have been made possible through contributions of the Institute for Astronomy, the University of Hawaii, the Pan-STARRS Project Office, the Max-Planck Society and its participating institutes, the Max Planck Institute for Astronomy, Heidelberg and the Max Planck Institute for Extraterrestrial Physics, Garching, The Johns Hopkins University, Durham University, the University of Edinburgh, Queen’s University Belfast, the Harvard-Smithsonian Center for Astrophysics, the Las Cumbres Observatory Global Telescope Network Incorporated, the National Central University of Taiwan, the Space Telescope Science Institute, the National Aeronautics and Space Administration under Grant No. NNX08AR22G issued through the Planetary Science Division of the NASA Science Mission Directorate, the National Science Foundation under Grant No. AST-1238877, the University of Maryland, and Eotvos Lorand University (ELTE) and the Los Alamos National Laboratory.

Based on data collected at the Subaru Telescope and retrieved from the
 HSC data archive system, which is operated by Subaru Telescope and Astronomy Data Center at National Astronomical Observatory of Japan.

XXL is an international project based around an XMM Very Large Programme surveying two 25 deg$^2$ extragalactic fields at a depth of $\sim6 \times 10^{-15}\, {\rm erg\,cm^{-2}\,s^{-1}}$ in the [0.5-2] keV band for point-like sources. The XXL website is http://irfu.cea.fr/xxl. Multi-band information and spectroscopic follow-up of the X-ray sources are obtained through a number of survey programmes, summarised at http://xxlmultiwave.pbworks.com/.

\bibliographystyle{mnras}
\bibliography{references}


\bsp
\label{lastpage}
\end{document}